\documentclass[%
 reprint,
superscriptaddress,
 amsmath,amssymb,
 aps,
]{revtex4-1}
\usepackage{ragged2e} 
\usepackage{booktabs}
\usepackage{multirow}
\usepackage{rotating}
\usepackage{subcaption}

\usepackage{float}
\usepackage{caption}
\usepackage{bm} 
\usepackage{graphicx}
\usepackage{dcolumn}
\usepackage{bm}
\usepackage{hyperref}

\begin{document}

\preprint{APS/123-QED}

\title{Radiative properties and optical appearance of a thin accretion disk around a charged-PFDM black hole}
\author{Taiyang Zhang}
\affiliation{%
 College of Physics,Guizhou University,Guiyang,550025,China
}%

\author{Zongyuan Qin}
\affiliation{%
	College of Physics,Guizhou University,Guiyang,550025,China
}%

\author{Qian Feng}
\affiliation{%
	College of Physics,Guizhou University,Guiyang,550025,China
}%

\author{Zheng-Wen Long}%
\email{zwlong@gzu.edu.cn (corresponding author)}
\affiliation{%
 College of Physics,Guizhou University,Guiyang,550025,China
}%


\begin{abstract}
Accretion onto magnetically charged black holes in a perfect fluid dark matter (PFDM) background opens a new window for testing strong-field gravity. This paper investigates the radiative properties and optical appearance of a thin accretion disk surrounding a charged-PFDM black hole. We numerically compute the radiative energy flux, temperature distribution, and radiative efficiency, and employ a ray-tracing method to construct the direct images, secondary images, redshift distribution, and observed flux. By comparing with the Schwarzschild and pure PFDM cases, we find that the thin-disk efficiency of the charged-PFDM black hole lies between the two, with the magnetic charge partially counteracting the dark-matter-induced efficiency enhancement; for M87*, the efficiency is estimated to be 7\%--8\%. The optical appearance is predominantly governed by the PFDM parameter, while the magnetic charge plays only a marginal role. Larger inclination angles give rise to stronger Doppler asymmetry, producing the characteristic ``hat-like'' shape. These results provide falsifiable predictions for future high-resolution observations, such as those by ngEHT, to distinguish dark-matter environments from magnetic-charge effects.
\end{abstract}

\maketitle


\section{Introduction}

Black holes, as one of the most fascinating predictions of general relativity, have transitioned from theoretical constructs to observable astrophysical objects over the past decade. The groundbreaking detections of gravitational waves from black hole mergers by LIGO and the first image of the supermassive black hole M87* captured by the Event Horizon Telescope (EHT) have ushered in a new era of black hole astrophysics \cite{LIGOScientific:2016emj, EventHorizonTelescope:2019dse}. These observations not only confirm the existence of black holes but also provide unprecedented opportunities to test gravitational theories in the strong-field regime. However, general relativity predicts the existence of spacetime singularities within black holes \cite{EventHorizonTelescope:2019ths}, where classical physics breaks down, signaling the need for more fundamental theories.

Nonlinear electrodynamics (NED), originally proposed by Born and Infeld in 1934 to resolve the infinite self-energy problem of point charges in classical electrodynamics, has emerged as a promising framework for addressing black hole singularities \cite{Gibbons:2001gy,Born:1934ji, Born:1934gh}. Coupling nonlinear electrodynamics with general relativity makes it possible to construct black hole solutions in which the central singularity is regularized and replaced by a finite curvature core \cite{Gunasekaran:2012dq, Bronnikov:2000vy, Bronnikov:2017sgg, Rincon:2021hjj, Bronnikov:2022ofk}. This line of thought regained widespread attention in the context of string theory, when it was realized that the effective action of open strings on D branes can be precisely described by nonlinear electrodynamics \cite{Gibbons:2001gy, Fradkin:1985qd, Seiberg:1999vs}. Over the past decades, numerous regular black hole solutions have been constructed within the framework of general relativity coupled to NED, with applications ranging from cosmological models to astrophysical phenomena \cite{DeLorenci:2002mi, Novello:2003kh, Novello:2008ra, Novello:2008xp, Xiang:2013sza, Balart:2014cga, Culetu:2014lca}. Extending these studies to rotating black holes, Lombardo et al. \cite{CiriloLombardo:2004qw} investigated Kerr-Newman-type solutions using the Newman-Janis algorithm and related methods \cite{Newman:1965tw, Azreg-Ainou:2014aqa, Azreg-Ainou:2014nra, Azreg-Ainou:2014pra}. Subsequently, Ghosh \cite{Ghosh:2021clx} further considered spacetimes with reflection symmetry and scalar polynomial singularities at $r=0$ \cite{Hawking:1973uf}, demonstrating that for certain ranges of the NED charge parameter, the black hole possesses both inner and outer horizons.

Perfect fluid dark matter (PFDM) provides an effective phenomenological description of dark matter that successfully reproduces the asymptotically flat rotation curves observed in spiral galaxies  \cite{Kiselev:2003ah, Kiselev:2002dx, Guzman:2000zba, Rahaman:2010xs, Potapov:2016obe}. Although dark matter may not strictly satisfy the perfect fluid condition, this model has been widely adopted due to its simplicity and consistency with observational data \cite{Rizwan:2018rgs, Ndongmo:2021how, Xu:2018wow}. Recently, Vachher et al. \cite{Vachher:2024ldc} investigated strong gravitational lensing by a static spherically symmetric magnetically charged black hole coupled to PFDM, while quasinormal modes of such black holes have also been studied \cite{Tan:2025usr}. The EHT observations of M87* and Sgr A* have imposed stringent constraints on the parameter space of these black hole models \cite{EventHorizonTelescope:2022wkp, EventHorizonTelescope:2021dqv}, providing a solid foundation for further theoretical investigations. 

The accretion process around black holes stands as one of the most powerful probes for testing general relativity and its modifications. Among various accretion models, the thin-disk framework developed by Shakura-Sunyaev and Novikov-Thorne \cite{Shakura:1972te, Thorne:1974ve, Page:1974he, Novikov:1973kta} has become the standard paradigm for studying black hole accretion systems, due to its observational viability and analytical tractability. The innermost stable circular orbit (ISCO), serving as the inner boundary of the thin disk, determines the radiative efficiency—i.e., the fraction of rest-mass energy converted into radiation. It has been demonstrated that properties such as the energy flux, temperature profile, and emission spectrum of the accretion disk can reveal the fundamental characteristics of the central black hole and test alternative theories of gravity \cite{Li:2004aq, Pun:2008ua, Harko:2010ua, Chakraborty:2014eha, Bambi:2015kza, He:2022lrc}. With the EHT's recent success in imaging the shadows of M87* and Sgr A*, the optical appearance of thin accretion disks—comprising direct images, secondary images, photon rings, redshift distributions, and observed fluxes—has become a powerful means of distinguishing between gravitational theories. Investigations along this line have been conducted in a variety of modified gravity scenarios \cite{Zheng:2024ftk,Wang:2025buh,Chen:2025wut,Nieto:2025apz,Darvishi:2024ndu,Yang:2024nin,Zeng:2025pch,Cai:2025rst}. They reveal that accretion-disk images are highly responsive to departures of the spacetime geometry from general relativity, thereby offering falsifiable predictions for strong-field gravity tests.

In this paper, we investigate several aspects of the charged-PFDM black hole proposed in \cite{Vachher:2024ldc}. In Section \ref{sec:2}, we impose constraints on the magnetic charge and PFDM parameters using EHT shadow data from M87*. In Section \ref{sec:3}, we compute the effective potential, ISCO, specific energy, specific angular momentum, and angular velocity for massive test particles around the thin accretion disk. In Sections \ref{sec:4}, we discuss the radiation flux, temperature profile, radiation efficiency, observed luminosity, and emission efficiency of a thin accretion disk. In Sections \ref{sec:5}, we focus on null geodesics and examine how the PFDM parameter modifies photon trajectories around the black hole. In Sections \ref{sec:6}, we analyze the direct and secondary images, redshift distribution, and observed flux of the thin accretion disk, and discuss the influence of the PFDM parameter on the black hole’s optical appearance. The final section provides a summary of our findings. In this study, we adopt a unit system in which $8\pi G = c = 1$ and the black hole mass $M=1$. Under this convention, the Einstein field equations reduce to $G_{\mu\nu} = T_{\mu\nu}$.

\section{Black hole shadow constraints from M87*}
\label{sec:2}

The line element of charged-PFDM black hole metrics is provided by \cite{Vachher:2024ldc}:

\begin{equation}
ds^2 = -f(r)\,dt^2 + \frac{1}{f(r)}\,dr^2 + r^2\left(d\theta^2 + \sin^2\theta\,d\phi^2\right),
\label{eq:1}
\end{equation}
where 
\begin{equation}
f(r) = 1 - \frac{2M}{\sqrt{r^2 + a^2}} + \frac{\zeta}{r} \log\frac{r}{|\zeta|}.
\label{eq:2}
\end{equation}
Here, $M$ denotes the black hole mass, $a$ is a parameter characterizing the magnetic charge, and $\zeta$ is the PFDM parameter, in the limit where $a \to 0$ and $\zeta \to 0$, it degenerates to the Schwarzschild black hole.

According to the strong-field limit theory of gravitational lensing \cite{Bozza:2002zj}, an extremal black hole is characterized by the coincidence of its event horizon and the innermost unstable photon orbit(photon sphere). In this study, the critical curve corresponding to extremal black holes in the parameter space $(a/M,\zeta/M)$ is determined by simultaneously solving $f(r)=0$ (the horizon condition) and $f'(r)=0$ (the photon-sphere condition). This curve delineates the physically admissible region where black hole solutions exist, as shown in Fig.1, where $a/M$ is the dimensionless magnetic charge parameter and $\zeta/M$ is the dimensionless PFDM parameter. The solid blue curve represents the trajectory of extremal black holes, corresponding to the coincidence of the event horizon and the inner horizon. The gray shaded region indicates the parameter range where regular black holes with two horizons exist, while the white region denotes the parameter space where no black hole solution (i.e., no physical horizon) is present. Furthermore, for the convenience of subsequent numerical analysis, the parameters $a$ and $\zeta$ appearing throughout this paper are dimensionless, i.e., $a \equiv a/M$ and $\zeta \equiv \zeta/M$.

To constrain the parameters of the charged-PFDM black hole using the shadow data observed by the EHT, we follow the methodology outlined in \cite{EventHorizonTelescope:2022wkp, Perlick:2021aok}. For a static spherically symmetric black hole, the radius of the photon ring $r_{\text{ph}}$ is determined by:
\begin{equation}
	   r f'(r) = 2 f(r), 
\end{equation}
for an asymptotically flat metric background, the shadow radius $R_s$ of  black hole as observed at infinity is defined by \cite{EventHorizonTelescope:2022wkp}
\begin{equation}
R_s = \frac{r_{\mathrm{ph}}}{\sqrt{f(r_{\mathrm{ph}})}}.
\label{eq:4}
\end{equation}
\begin{figure}[t]
\centering
\includegraphics[width=0.8\columnwidth]{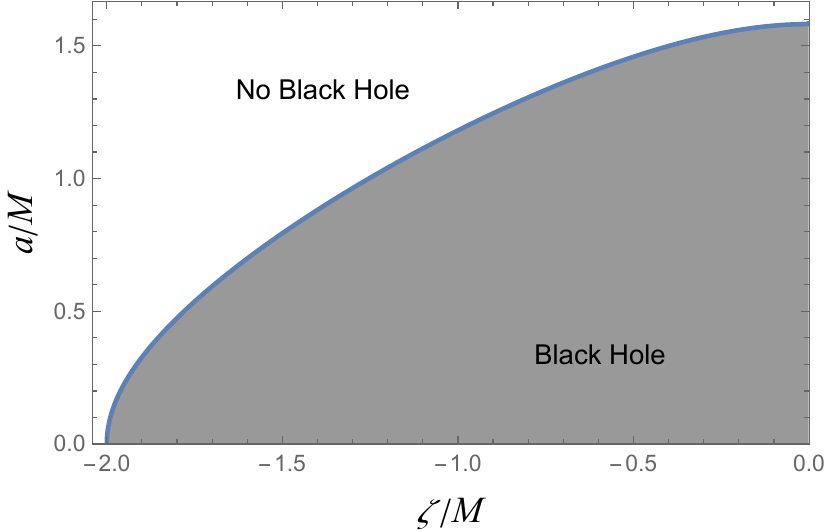}
\caption{\justifying The phase diagram for the parameter space $(a/M,\zeta/M)$ of charged-PFDM black hole}
\label{fig:1}
\end{figure}

In 2019, the Event Horizon Telescope (EHT) Collaboration successfully captured the first shadow image of the supermassive black hole M87* at the center of the elliptical galaxy M87 \cite{EventHorizonTelescope:2019dse, EventHorizonTelescope:2019ths, EventHorizonTelescope:2021dqv, EventHorizonTelescope:2019ggy}. The observations yielded an angular shadow diameter of $\theta_{\mathrm{sh}} = (42 \pm 3)\,\mu\mathrm{as}$, a distance of $D = (16.8 \pm 0.8)\,\mathrm{Mpc}$, and a mass estimate of  $(6.5 \pm 0.7) \times 10^9 \, M_\odot$. This landmark achievement provided the first strong constraints on theoretical models from event-horizon scales \cite{Vachher:2024ldc}. Subsequently, in 2022, the EHT reported the shadow observation of Sgr A*, the black hole at the Galactic Center \cite{EventHorizonTelescope:2022wkp, EventHorizonTelescope:2022urf, EventHorizonTelescope:2022xqj}. The measured angular diameter is $\theta_{\mathrm{sh}} = (48.7 \pm 7)\,\mu\mathrm{as}$, the distance is $D = 8\,\mathrm{kpc}$, and the mass is about$(4.0_{-0.6}^{+1.1}) \times 10^6 \, M_\odot$. These high-precision data establish a solid foundation for testing various black hole models \cite{Saurabh:2020zqg, Vagnozzi:2022moj}.

\begin{figure}[h]
\centering
\includegraphics[width=0.8\columnwidth]{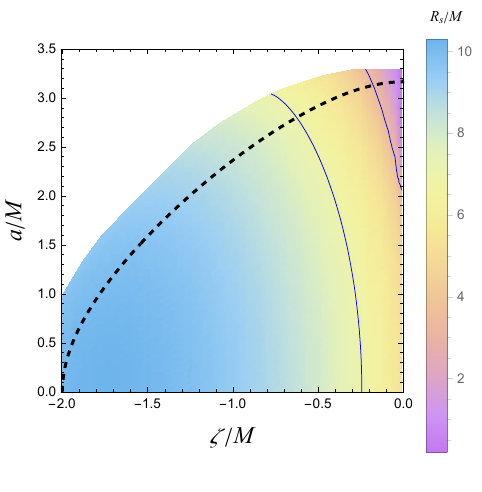}
\caption{\justifying EHT-constrained parameter space of M87* candidate charged black holes in PFDM background. The horizontal axis is PFDM parameter $\zeta/M$, left vertical axis is magnetic charge parameter $a/M$, and the right color bar links to $R_s/M$. The black dashed line marks extreme black holes (black holes exist below it); the area between two curves is constrained by EHT observations of M87*.}
\label{fig:2}
\end{figure}

Here, we utilize the shadow measurement of M87*, with a reported radius of $3.430 \leq R_s \leq 6.963 \; (2\sigma)$ \cite{EventHorizonTelescope:2022wkp}, to constrain the parameter space of the charged-PFDM black hole. Fig 2 illustrates the parameter space constrained by EHT observations when the black hole is considered a candidate for M87*. The allowed region lies between two characteristic curves and below the critical line for extremal black holes (black dashed line). The analysis shows that when the magnetic charge parameter $a/M$ is small, the effective range of the dark matter parameter $\zeta/M$ is approximately between $-0.8$ and $0$. As a/M increases, this range gradually extends toward more negative values, eventually covering $-1.7$ to $-0.8$. This work focuses on the constraints imposed by EHT observations in a dark matter dominated background. Synthesizing the above results, the parameter ranges adopted in the subsequent analysis are $0 < a \leq 1$ and $-0.25\leq \zeta<0$. In this work, the above constraint has been consistently adopted throughout our analysis.

\section{Time-like circular geodesics and orbital properties of the accretion disk}
\label{sec:3}

In this section, we establish the kinematic foundation for the thin accretion disk. The Lagrangian that describes the motion of a test particle orbiting a charged PFDM black hole takes the form

 \begin{equation}
   \mathcal{L} = \frac{1}{2} g_{\mu\nu} \frac{dx^\mu}{ds} \frac{dx^\nu}{ds} = \frac{1}{2} \varepsilon,
   \label{eq:5}
\end{equation}

The Lagrangian presented in Eq.\eqref{eq:5} describes massive particles for $\varepsilon=1$ and photons for $\varepsilon=0$. Our analysis is confined to equatorial orbits with $\theta=\frac{\pi}{2}$.The conserved energy E and angular momentum L are obtained as:

\begin{equation}
E = -g_{tt} \frac{dt}{ds} 
= \left(1 - \frac{2M}{\sqrt{r^2 + a^2}} + \frac{\zeta}{r} \log\frac{r}{|\zeta|}\right) \frac{dt}{ds},
\label{eq:6}
\end{equation}

\begin{equation}
	L = g_{\phi\phi} \frac{d\phi}{ds} 
	= r^2 \frac{d\phi}{ds}.
	\label{eq:7}
\end{equation}

The geodesic equations for a massive particle in the spacetime \eqref{eq:1} take the following forms:

\begin{align}
	\left(\frac{dr}{ds}\right)^2 &= E^2 - \left(1 - \frac{2M}{\sqrt{r^2 + a^2}} + \frac{\zeta}{r} \log\frac{r}{|\zeta|}\right) \left(1 + \frac{L^2}{r^2}\right), \label{eq:8} \\
	\left(\frac{dr}{d\phi}\right)^2 &= \frac{r^4}{L^2} \left[ E^2 - \left(1 - \frac{2M}{\sqrt{r^2 + a^2}} + \frac{\zeta}{r} \log\frac{r}{|\zeta|}\right) \left(1 + \frac{L^2}{r^2}\right) \right], \label{eq:9} \\
	\begin{split}
		\left(\frac{dr}{dt}\right)^2 &= \frac{1}{E^2} \left(1 - \frac{2M}{\sqrt{r^2 + a^2}} + \frac{\zeta}{r} \log\frac{r}{|\zeta|}\right)^2 \\
		&\quad \times \left[ E^2 - \left(1 - \frac{2M}{\sqrt{r^2 + a^2}} + \frac{\zeta}{r} \log\frac{r}{|\zeta|}\right) \left(1 + \frac{L^2}{r^2}\right) \right]. \label{eq:10}
	\end{split}
\end{align}

The system's dynamics are completely governed by Eqs.\eqref{eq:8}\eqref{eq:9}\eqref{eq:10}, with the effective gravitational potential given directly by Eq.\eqref{eq:8}:
\begin{equation}
 V_{\text{eff}}(r)=-g_{tt}(1+\frac{L^2}{r^2})=(1 - \frac{2M}{\sqrt{r^2 + a^2}} + \frac{\zeta}{r} \log\frac{r}{|\zeta|})(1+\frac{L^2}{r^2}).
\label{eq:11}
\end{equation}
When a test particle travels along a circular orbit, its radius must satisfy the following constraints:

\begin{equation}
V_{\text{eff}} = E^2, \quad V_{\text{eff},r} = 0.
\label{eq:12}
\end{equation}

Through the solution of the above two equations, the explicit dependence of $\Omega$, $E$, $L$ on the circular orbit radius $r$ can be obtained:

\begin{equation}
	\begin{aligned}
		\Omega &= \frac{d\phi}{dt} = \sqrt{\frac{-g_{tt,r}}{g_{\phi\phi,r}}} 
		= \sqrt{\frac{M}{(r^2 + a^2)^{3/2}} + \frac{\zeta\left(1 - \log\frac{r}{|\zeta|}\right)}{2r^3}}, 
	\end{aligned}
	\label{eq:13}
\end{equation}

\begin{equation}
	\begin{aligned}
		E &= \frac{-g_{tt}}{\sqrt{-g_{tt} - g_{\phi\phi}\Omega^2}}  \\
		&= \frac{1 - \dfrac{2M}{\sqrt{r^2 + a^2}} + \dfrac{\zeta}{r}\log\dfrac{r}{|\zeta|}}
		{\sqrt{1 - \dfrac{2M}{\sqrt{r^2 + a^2}} - \dfrac{Mr^2}{(r^2 + a^2)^{3/2}} + \dfrac{\zeta}{2r}\left(3\log\dfrac{r}{|\zeta|} - 1\right)}},
	\end{aligned}
	\label{eq:14}
\end{equation}

\begin{equation}
	\begin{aligned}
		L &= \frac{g_{\phi\phi}}{\sqrt{-g_{tt} - g_{\phi\phi}\Omega^2}} \\
		&= \sqrt{\frac{
				\dfrac{2Mr^4}{(r^2 + a^2)^{3/2}} + \zeta r\left(1 - \log\dfrac{r}{|\zeta|}\right)
			}{
				2 - \dfrac{4M}{\sqrt{r^2 + a^2}} - \dfrac{2Mr^2}{(r^2 + a^2)^{3/2}} + \dfrac{\zeta}{r}\left(3\log\dfrac{r}{|\zeta|} - 1\right)
		}},
	\end{aligned}
	\label{eq:15}
\end{equation}

In general, not all circular orbits are stable. An interesting critical circular orbit is the innermost stable circular orbit (ISCO), which is determined by the equations $V_{\text{eff}}=0$, $\quad V_{\text{eff},r} = 0$, $\quad V_{\text{eff},rr} = 0$. Therefore, the radius $r_{isco}$ of the ISCO satisfies the following equation:

\begin{equation}
r_{isco} = \frac{2r f''(r) f(r) - 4r f'(r)^2 + 6 f(r) f'(r)}{2r f(r) - r^2 f'(r)}.
	\label{eq:16}
\end{equation}

\begin{figure*}
	a) \includegraphics[width=8.1 cm]{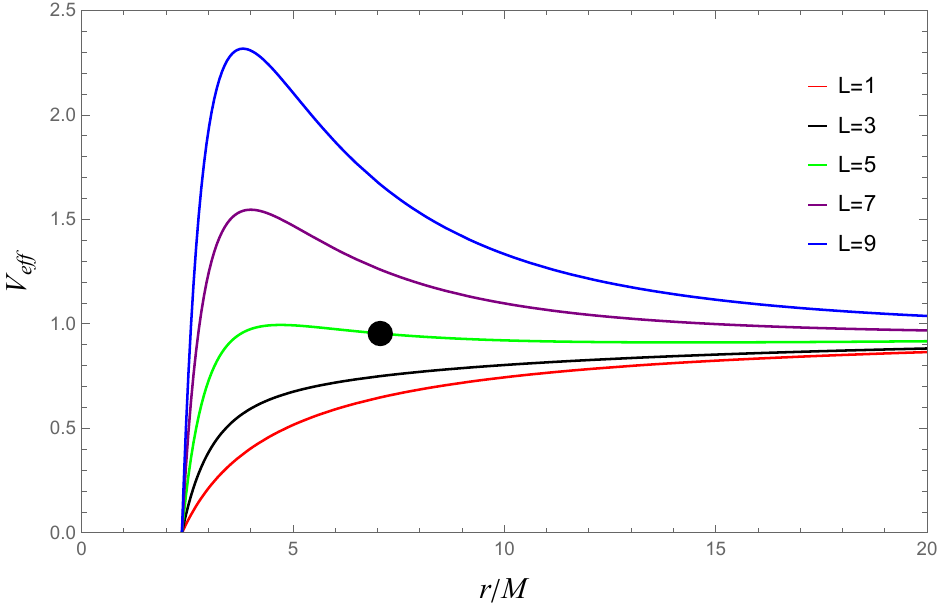}
	b) \includegraphics[width=8.1 cm]{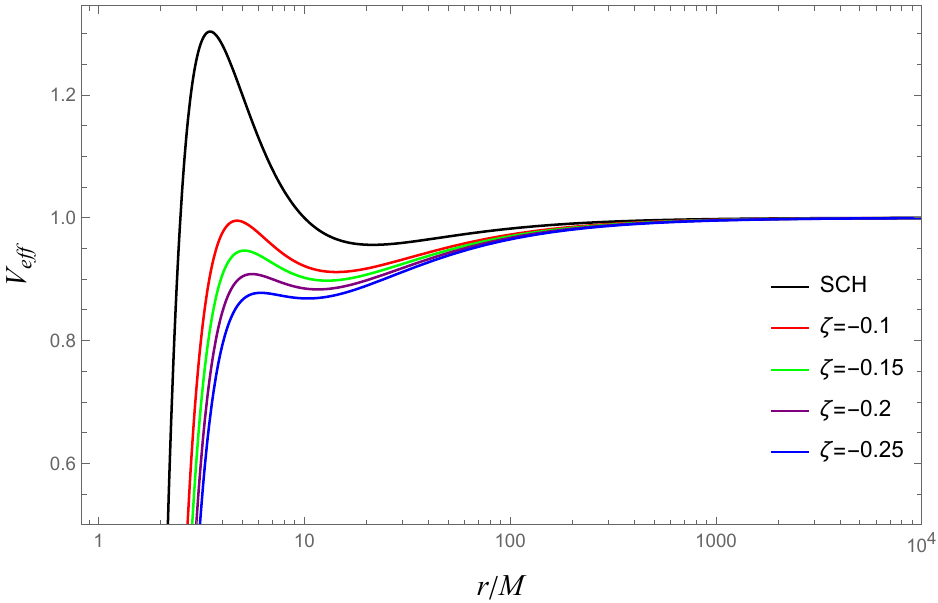}\\
	c) \includegraphics[width=8.1 cm]{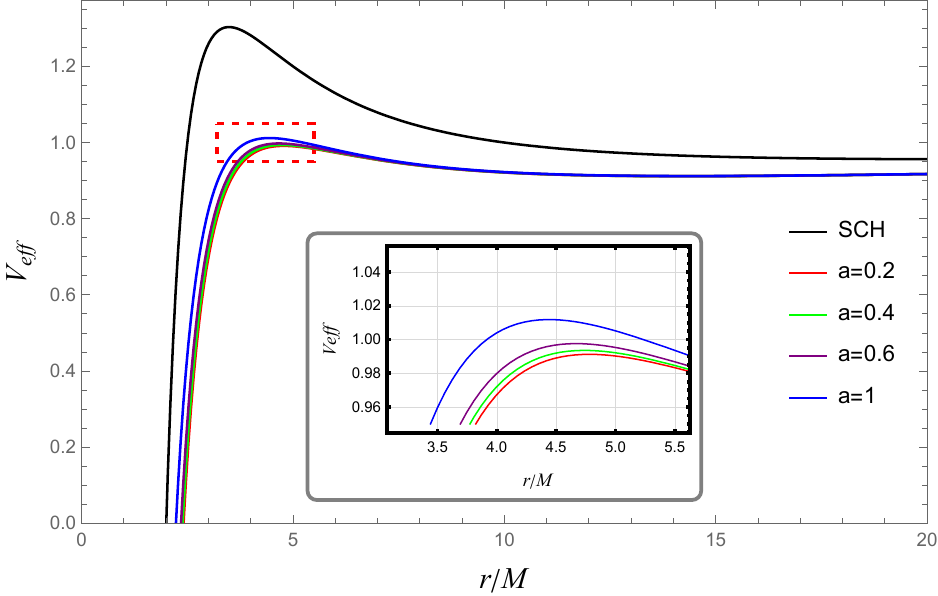}
	\caption{ The effective potential $V_{eff}$ depends on the radial coordinate $r$, and its variation characteristics under different parameter combinations are as follows (a) for $\zeta=-0.15$, $a=0.5$, with different values of $L$ (b) for $L=5$, $a=0.5$ with different values of $\zeta$ (c) for $L=5$, $\zeta=-0.15$ with different values of $a$.} \label{fig: 3}
\end{figure*}

Assuming the magnetic charge parameter $a=0.5$, PFDM parameter $\zeta=-0.15$, and black hole mass $M=1$, the ISCO radius $r_{isco}\approx 7.06414$ is obtained by solving Eq.\eqref{eq:16}. The radial profile of the effective potential $V_{\text{eff}}$ for a test particle is presented in Fig.3a reveals the first extremum occurring at $L=5$, with no additional extrema for $L<5$. The effective potential is further observed to increase with the angular momentum $L$. The ISCO, marked by a black dot in Fig.3a, is located at $r=7.06414$. In Fig.3b, a larger magnetic charge parameter $a$ leads to a higher effective potential, indicating that the trajectories of both unstable and stable circular orbits are situated farther from the central mass. As shown in Fig.3c, the effective potential decreases as the PFDM parameter $\zeta$ increases. Plotted on a logarithmic scale, a secondary minimum at larger $r$ becomes apparent; for particles with sufficiently large $L$,  $V_{\text{eff}}$ thus exhibits two extrema—a maximum (unstable circular orbit) and a minimum (stable circular orbit). Furthermore, comparison between Fig.3b and Fig.3c indicates that the effective potential of particles around the charged-PFDM black hole remains consistently lower than that in the Schwarzschild case.

Fig.4 shows the angular velocity $\Omega$ of particles on circular orbits as a function of orbital radius $r$. The angular velocity decreases monotonically with increasing $r$. The influence of the PFDM parameter $\zeta$ and the magnetic charge parameter $a$ on the orbital angular velocity is analyzed. The results indicate that $\Omega$ decreases with increasing values of $a$ or $\zeta$. However, outside the event horizon, it is consistently greater than that in the case of the Schwarzschild black hole. This implies that, at the same orbital radius, the gravitational field of the charged-PFDM black hole is stronger than that of a Schwarzschild black hole.

\begin{figure*}
	a)\includegraphics[width=8.1 cm]{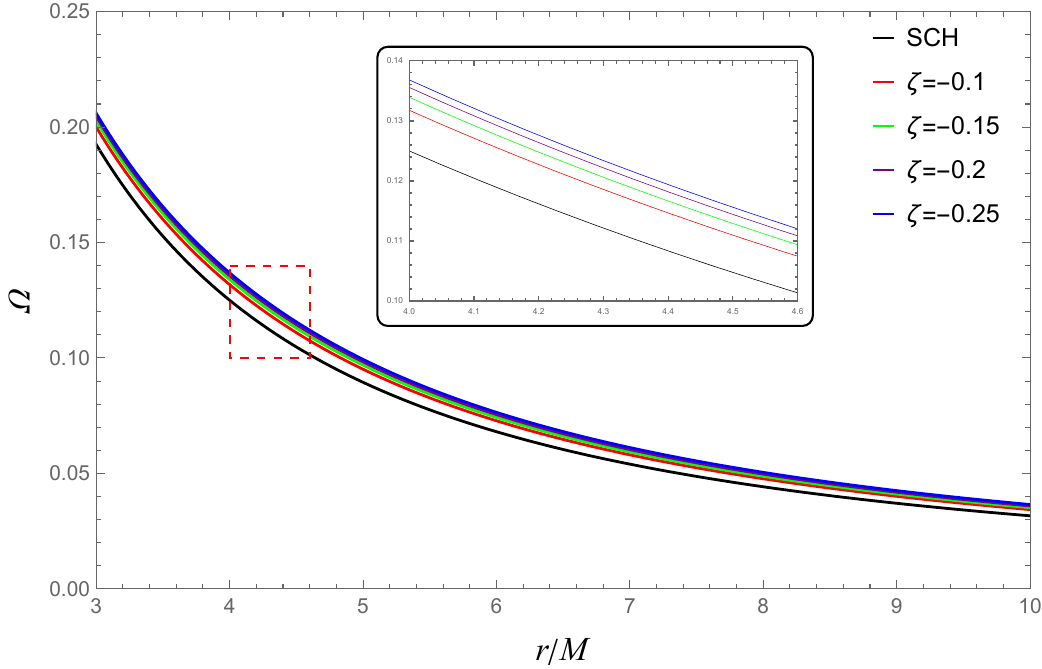}
	b)\includegraphics[width=8.1 cm]{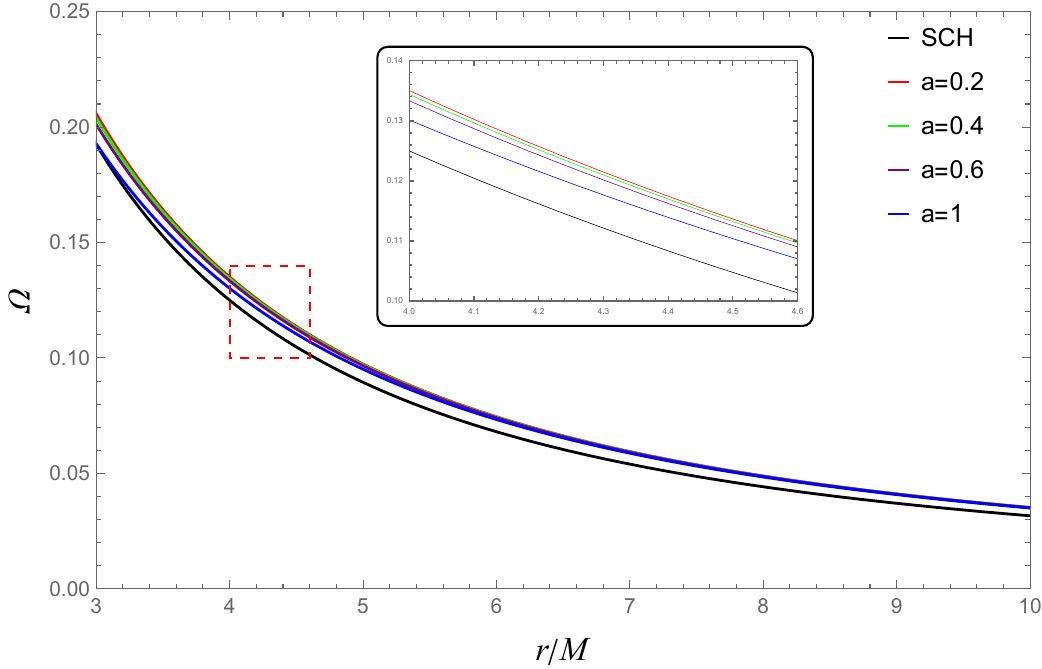}
	\caption{Angular velocity $\Omega$ versus $r$ (a) for $a=0.5$, varying $\zeta$, (b) for $\zeta=-0.15$, varying $a$.} \label{fig: 4}
\end{figure*}

The radial profiles of specific energy and specific angular momentum are displayed in Fig.5. From the left panel it is clearly seen that both specific energy and specific angular momentum decrease as the PFDM parameter $\zeta$ increases. The right panel shows that they also decline with rising magnetic charge $a$. Notably, at the same radius $r$, the angular momentum of the charged-PFDM black hole remains consistently higher than that of the Schwarzschild black hole, while its specific energy drops below the Schwarzschild value beyond a certain radial distance.

\begin{figure*}
	(a)\includegraphics[width=8.1 cm]{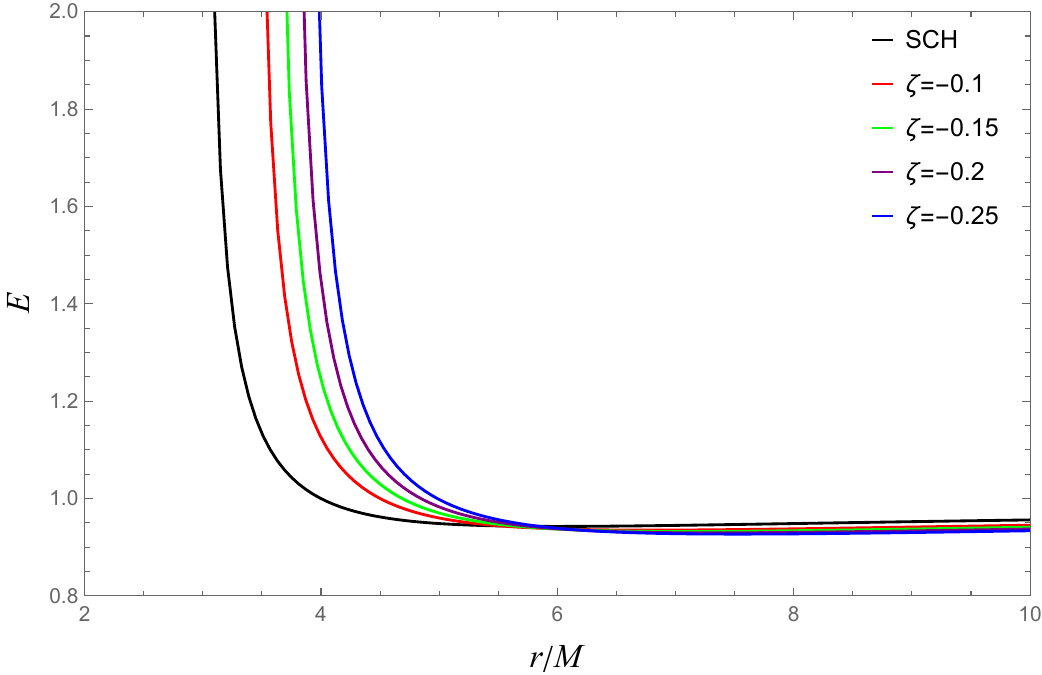}
	(b)\includegraphics[width=8.1 cm]{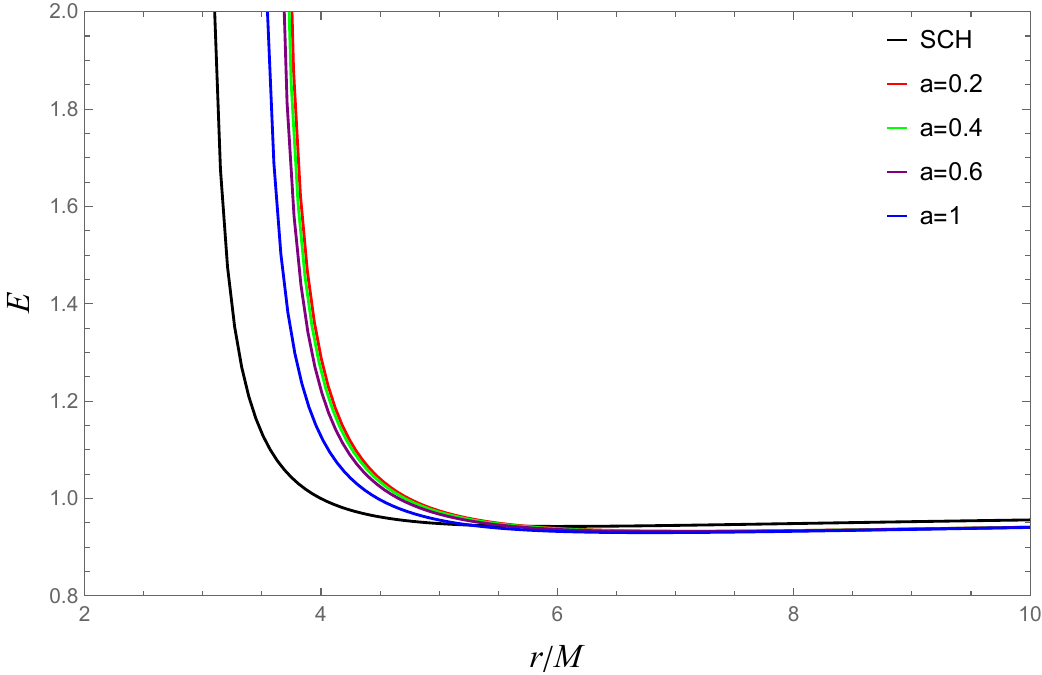}
	(c)\includegraphics[width=8.1 cm]{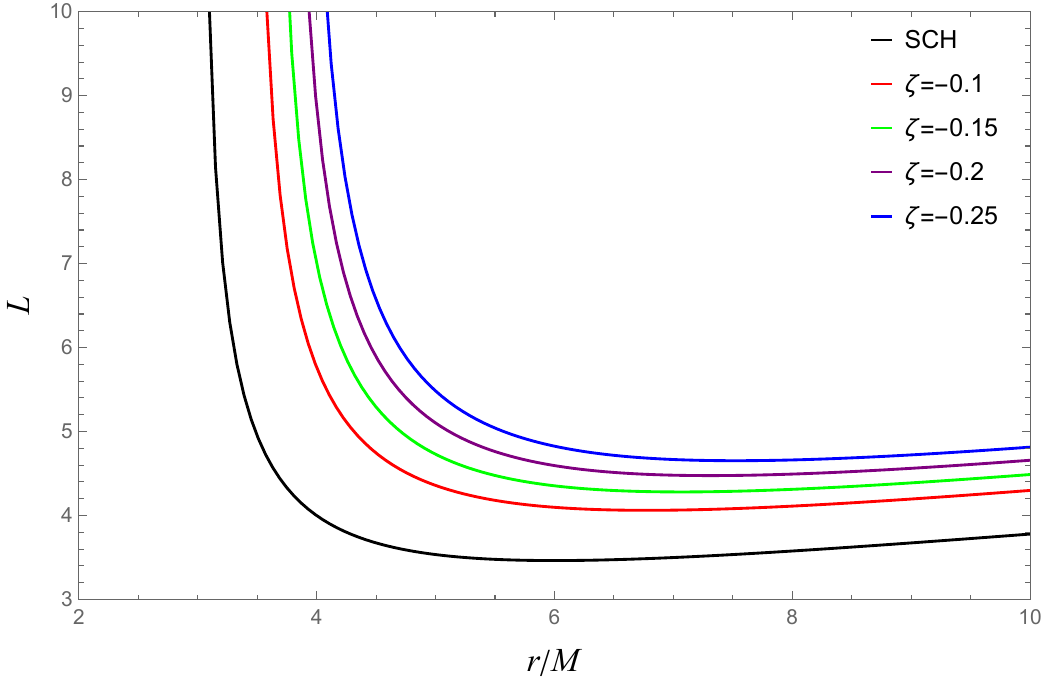}
	(d)\includegraphics[width=8.1 cm]{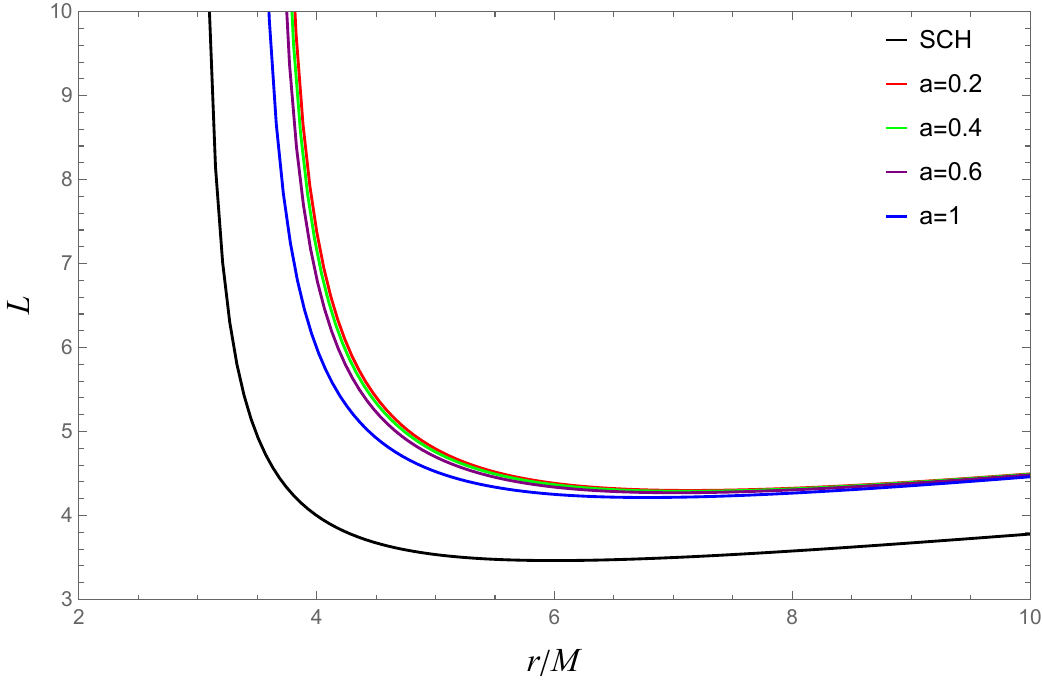}
	\caption{ The distributions of specific energy and specific angular momentum are shown as functions of the radial coordinate $r$. The left panel displays results for different PFDM parameter $\zeta$ values with $a = 0.5$ , while the right panel corresponds to different magnetic charge parameter a values with $\zeta=-0.15$.}. \label{fig: 5}
\end{figure*}

\section{Properties of thin accretion disk}
\label{sec:4}

In this section, we investigate the physical properties based on the Novikov-Thorne model of a thin accretion disk \cite{Novikov:1973kta}. Within a stationary, axisymmetric, and asymptotically flat spacetime background, the thickness of a thin accretion disk is typically negligible, meaning its vertical half-thickness H is always much smaller than its characteristic radial distance $r$ $(H \ll r)$. The thin disk is in a state of hydrodynamic and thermodynamic equilibrium. Pressure gradients and vertical entropy gradients within the accreting material are negligible, and the heat generated by viscous stresses is efficiently radiated away from the disk surfaces. This efficient radiative cooling prevents the disk from puffing up and maintains its slender vertical structure. The inner edge of the disk is defined by the ISCO within the gravitational potential of the central compact object, located at $r_{isco}$. Furthermore, the mass accretion rate $\dot{M}_0$ is assumed to be constant over time.

In the following subsection, we take the central black hole of M87* to be described by the charged-PFDM model and conduct a detailed investigation of its radiative energy flux, temperature profile, observed luminosity, and other key quantities associated with a thin accretion disk. Since the Novikov-Thorne model requires an asymptotically flat spacetime, we primarily focus on the case of a vanishing cosmological constant. The specific values of the physical constants and the adopted thin-disk parameters are summarized in \textbf{Table I}.

In investigating the radiative flux from the outer region of the accretion disk within the equatorial plane, we employ the quantities $E$, $L$, and $\Omega$. The radiant energy flux across the disk can be determined using the following relation \cite{Novikov:1973kta, Page:1974he, Collodel:2021gxu}:

\begin{align}
	F(r) = -\frac{\dot{M}_0 \Omega_{,r}}{4\pi \sqrt{-g/g_{\theta\theta}} (E - \Omega L)^2} 
	\int_{r_{\mathrm{isco}}}^{r} (E - \Omega L) L_{,r} \, dr,
	\label{eq:17}
\end{align}

where $\dot{M}_0$ is the mass accretion rate. Fig.6 presents the profiles for the radiative energy flux $F(r)$ from accretion disks surrounding the charged-PFDM black hole for different values of $\zeta$ or $a$. For comparison, the flux distribution for a disk around a Schwarzschild black hole is also included. It is observed that the energy flux in the charged-PFDM geometry is consistently lower than that in the standard Schwarzschild case. Furthermore, the flux increases with larger values of the parameters $\zeta$ or $a$, indicating that the deviation from the standard Schwarzschild result diminishes as these black-hole parameters grow. In addition, the peak (maximum) of the flux shifts toward smaller radii as the black hole parameters increase.

\begin{figure*}
	a)\includegraphics[width=8.1 cm]{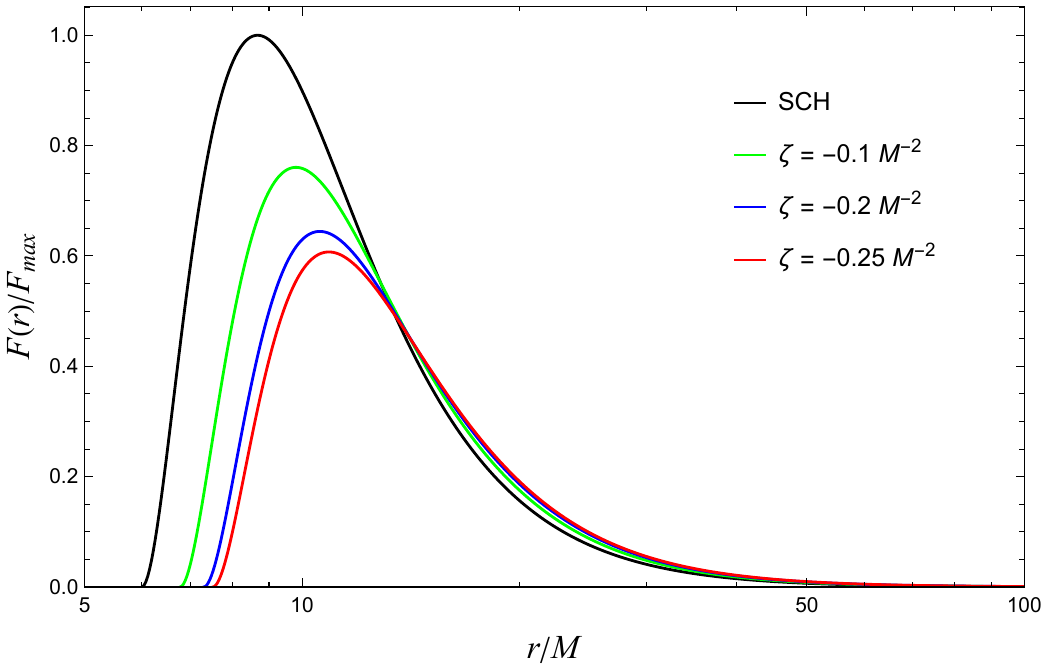}
	b)\includegraphics[width=8.1 cm]{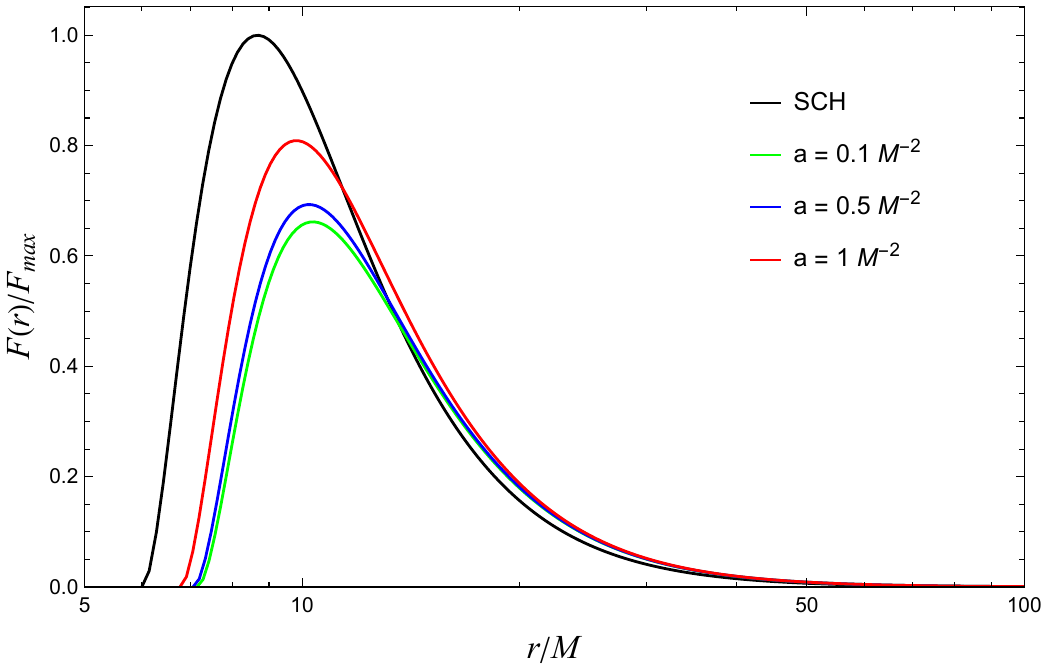}
	\caption{Radiative energy comparison between accretion disks around charged-PFDM and Schwarzschild black holes of equal total mass (a) different $\zeta$ values with fixed $a=0.5$ (b) different $a$ values with fixed $\zeta=-0.15$. Flux values are normalized to the maximum flux of the Schwarzschild case, $F_{\max} = 1.37 \times 10^{-5} \, \dot{M}_0/M^2$.} \label{fig:6}
\end{figure*}

Within the Novikov-Thorne framework, the accreting matter maintains thermodynamic equilibrium. Consequently, the radiation emanating from the disk can be treated as perfect blackbody radiation. The disk's radiative temperature $T(r)$ is connected to its energy flux $F(r)$ through the Stefan-Boltzmann law:

\begin{equation}
	F(r) = \sigma_{\mathrm{SB}} T^4,
	\label{eq:18}
\end{equation}

where $\sigma_{\mathrm{SB}}$ is the Stefan-Boltzmann constant. This indicates that the radial profile of the temperature $T(r)$ follows a trend similar to that of the energy flux $F(r)$. As shown in Fig.7, the temperatures are normalized to the maximum value for a standard Schwarzschild black hole. In Fig.7a (with fixed $a=0.5$), the disk temperature increases for larger values of the PFDM parameter $\zeta$, and the temperature peak shifts toward the inner edge of the disk. Similarly, Fig.7b (with fixed $\zeta=-0.15$) shows that the temperature rises with increasing magnetic charge parameter $a$, accompanied by an inward shift of the peak. Overall, the accretion disk around the charged-PFDM BH is cooler to the disk around a Schwarzschild BH.

\begin{figure*}
	a)\includegraphics[width=8.1 cm]{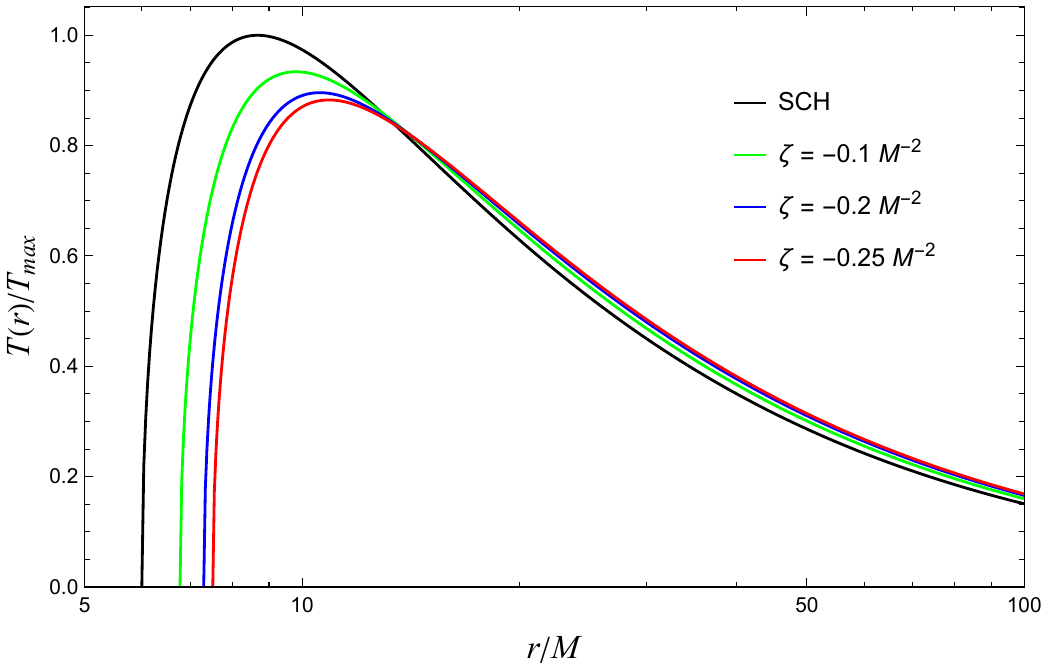}
	b)\includegraphics[width=8.1 cm]{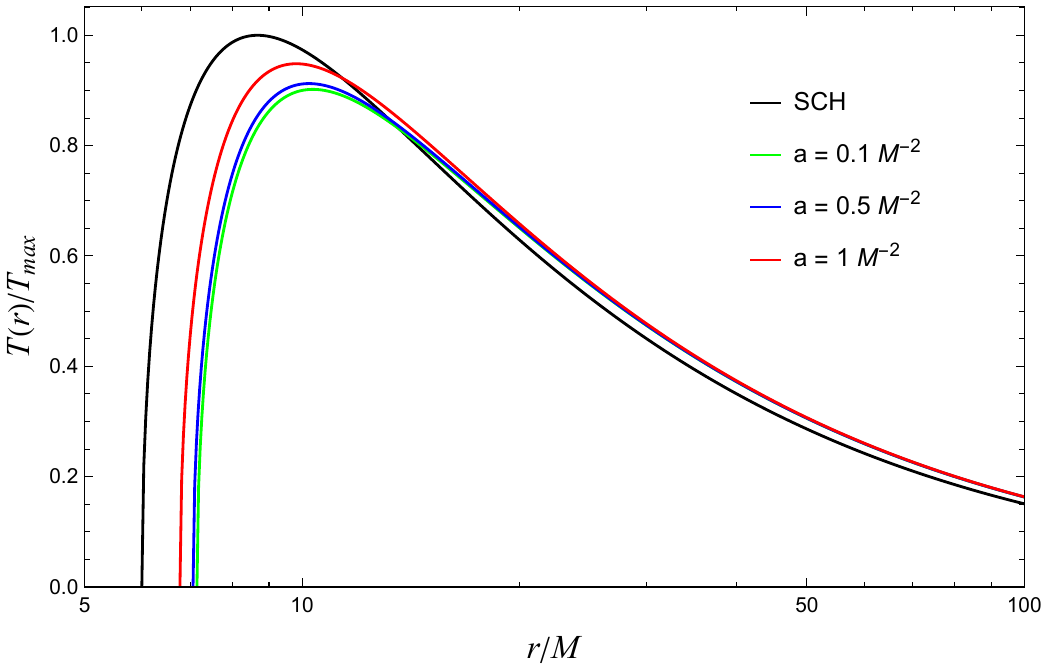}
	\caption{The disk temperature for charged-PFDM and Schwarzschild black holes of equal total mass (a) different $\zeta$ values with fixed $a=0.5$ (b) different $a$ values with fixed $\zeta=-0.15$. Temperature values are normalized to the maximum temperature of the Schwarzschild case, $T_{\max} = 0.7015 \, (\dot{M}_0/M^2)^{1/4}$.} \label{fig:7}
\end{figure*}

Under the assumption that disk-emitted radiation has a blackbody spectrum, the disk’s monochromatic luminosity $L(\nu)$ can be calculated by:

\begin{align}
	L(\nu) &= 4\pi d^2 I(\nu) \nonumber \\
	&= 8\pi h \cos\theta \int_{r_i}^{r_f} \int_0^{2\pi} 
	\frac{\nu_e^3 r}{e^{h\nu_e/(k_B T)} - 1} \, dr \, d\phi,
	\label{eq:19}
\end{align}

where $d$ is the distance to the disk center, $\nu$ is the frequency of emitted photons, $I(\nu)$ stands for the thermal energy flux corresponding to a specific frequency, $h$ is the Planck constant and $k_B$ is the Boltzmann constant. $\theta$ denotes the disk’s inclination angle which we set it to zero. $r_i$ and $r_f$ denote the positions of the inner and outer edges of the disk, respectively. To ensure that the flux over the disk surface vanishes at $r_f \to \infty$ for any asymptotically flat geometry, we set $r_i =r_{isco}$ and take $r_f \to \infty$. Also, $\nu_e= \nu(1+z)$ corresponds to the emitted photon frequency, with $z$ being the redshift factor. The redshift factor $z$ is computed without accounting for light bending effects\cite{Luminet:1979nyg, Bhattacharyya:2000kt, Salahshoor:2018plr}:

\begin{equation}
1+z = \frac{1 + \Omega r \sin\theta \sin\phi}{\sqrt{-g_{tt} - \Omega^2 g_{\phi\phi}}},
	\label{eq:20}
\end{equation}

Employing Eqs.\eqref{eq:19} and \eqref{eq:20}, we calculate the spectral energy distribution (SED) of the thin accretion disk around the black hole, the results are presented in Fig.8. This figure compares the radiation spectra of the charged-PFDM black hole with that of the Schwarzschild black hole. For all parameter choices, the SED of the charged-PFDM black hole lies systematically above the Schwarzschild case, in agreement with the higher radiative efficiency $\eta^*$ shown in \textbf{Table II}, which implies a higher total luminosity for the same accretion rate. Specifically, in Fig.8a (fixed $a=0.5$), a more negative PFDM parameter $\zeta$ leads to a larger ISCO radius (see \textbf{Table II}), an increase in radiative efficiency and total luminosity, and consequently an upward shift of the entire SED. In Fig.8b (fixed $\zeta=-0.15$), as the magnetic charge parameter $a$ increases, the deepening gravitational potential raises the radiative efficiency and simultaneously shifts the spectral peak toward higher frequencies (blueshift). 

\begin{figure*}
	a)\includegraphics[width=8.1 cm]{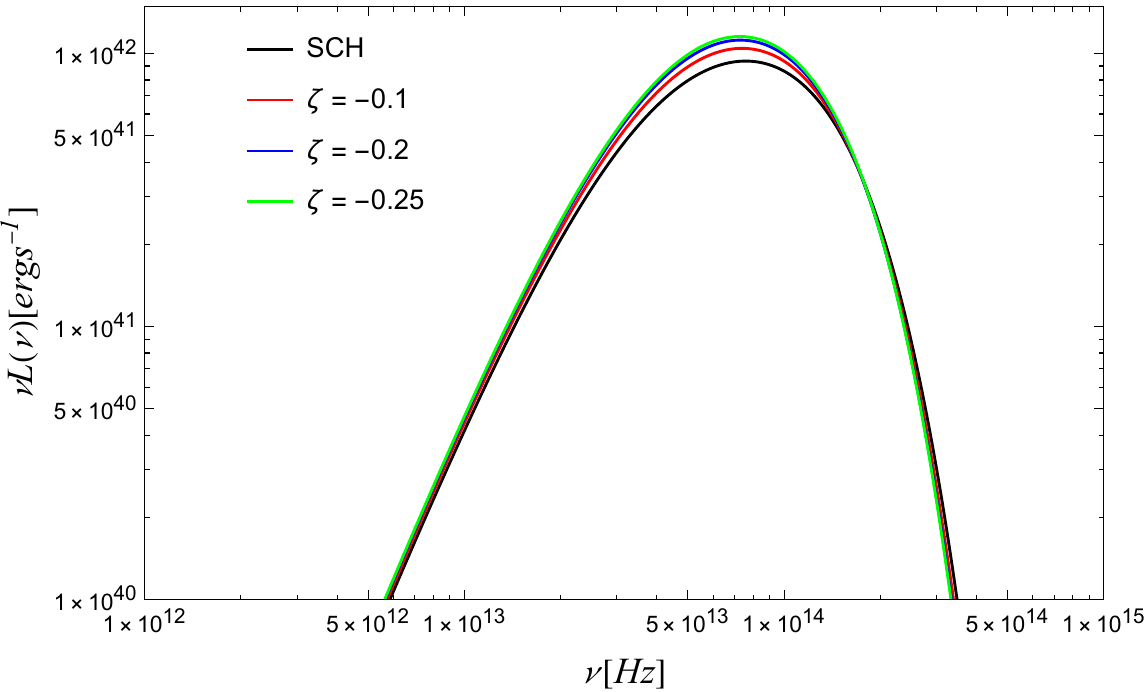}
	b)\includegraphics[width=8.1 cm]{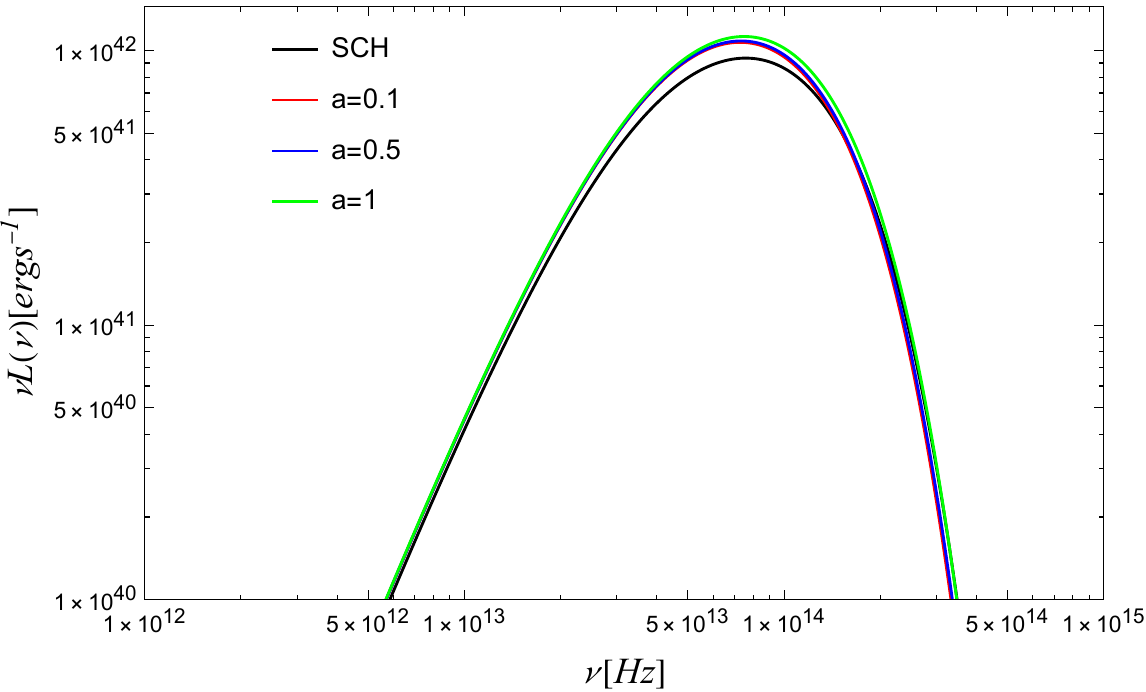}
	\caption{The disk spectra for charged-PFDM and Schwarzschild black holes of same total mass (a) different $\zeta$ values with fixed $a=0.5$ (b) different $a$ values with fixed $\zeta=-0.15$.} \label{fig:8}
\end{figure*}

A key physical quantity associated with mass accretion is the radiation efficiency of the black hole. This efficiency is defined as the ratio of the photon energy radiated to infinity from the disk surface to the rate at which rest mass energy is supplied to the central compact object, with both quantities measured at infinity \cite{Novikov:1973kta, Page:1974he}. Considering a test particle of unit mass falling from infinity down to the ISCO, and assuming that all the lost gravitational energy is converted into radiation that reaches infinity, the radiation efficiency $\eta^*$ is given by:

\begin{equation}
	\eta^* = 1 - E_{isco},
	\label{eq:21}
\end{equation}

where $E_{isco}$ is the specific energy at the ISCO. \textbf{Table II} lists the ISCO radius and the mass to radiation conversion efficiency for the charged-PFDM black hole under different parameter choices. Interestingly, when $\zeta$ is fixed at $–0.15$, the ISCO radius decreases with increasing $a$, whereas the radiative efficiency rises. As seen in Figs.6 and 7, both the flux $F(r)$ and the temperature $T(r)$ increase with larger $a$. This indicates that, although a smaller ISCO reduces the emitting area, the enhancement in the local flux and temperature dominates the increase in efficiency.

Conversely, with $a$ fixed at $0.5$, a more negative $\zeta$ leads to a higher efficiency. In this case, even though the local $F(r)$ and $T(r)$ are lower, the larger ISCO radius – and consequently the greater overall emitting area – results in a higher total radiated power. On the whole, the variation of the ISCO radius (and thus the disk area) appears to influence the thin disk behavior more significantly; the wider spread of the curves in Figs.6 and 7 for fixed $a$ and different $\zeta$ reflects this point.

For comparison, a Schwarzschild black hole has an efficiency of about $6\%$ (whether or not photon capture is included) and an ISCO at $r_{isco}=6M$ \cite{Page:1974he}. Therefore, although the local $F(r)$ and $T(r)$ of the charged-PFDM black hole are lower than those of the Schwarzschild case, its larger ISCO radius and a higher radiative efficiency and, for a fixed accretion rate, a higher total observed luminosity. However, a pure PFDM black hole can achieve a radiative efficiency as high as $20\%$ for similar dark matter parameters \cite{Narzilloev:2022avv}, which is substantially larger than in the charged case. This is because a more negative $\zeta$ pushes the ISCO outward, whereas a larger $a$ pulls it inward, resulting in a trade off effect between the two parameters. Within the parameter range allowed by the EHT constraints, the outward shift induced by PFDM dominates, but the presence of the magnetic charge partially suppresses the efficiency enhancement that would otherwise be present in the pure PFDM scenario. Consequently, the net efficiency of the charged-PFDM black hole is markedly lower than that of its pure PFDM counterpart. This difference provides a testable discriminant for future high precision spectroscopic observations.

\begin{table}[htbp]
	\centering
	\footnotesize 
	\setlength{\heavyrulewidth}{0.12ex} 
	\setlength{\lightrulewidth}{0.08ex} 
	\setlength{\tabcolsep}{3pt}
	\resizebox{\columnwidth}{!}{
		\begin{tabular}{cc}
			\toprule 
			parameters & values  \\
			\midrule %
			$G$  & $6.67430 \times 10^{-11} \, \mathrm{m}^3 \, \mathrm{kg}^{-1} \, \mathrm{s}^{-2}$   \\
			$c$  & $2.99792458 \times 10^8 \, \mathrm{m/s}$   \\
			$h$  & $6.62607015 \times 10^{-27} \, \mathrm{erg\,s}$   \\
			$k_B$  & $1.380649 \times 10^{-16} \, \mathrm{erg\,K^{-1}}$   \\
			$M_\odot$  & $1.989 \times 10^{33} \, \mathrm{g}$   \\
			$1yr$  & $3.156 \times 10^{7} \, \mathrm{s}$  \\
			$\sigma_{\mathrm{SB}}$  & $5.670374419 \times 10^{-5} \, \mathrm{erg\,s^{-1}\,cm^{-2}\,K^{-4}}$  \\
			$M (M87^*)$  & $6.5 \times 10^{9} \, M_{\odot}$  \\
			$\dot{M}_0 (M87^*)$  & $3 \times 10^{-4} \, M_{\odot} \, \mathrm{yr^{-1}}$  \\
			
			\bottomrule 
		\end{tabular}
	}
	\caption{\justifying Physical constants and M87* baseline parameters adopted in this study} 
	\label{tab:1}
\end{table}

\begin{table*}[htbp]
	\centering
	\footnotesize
	\setlength{\heavyrulewidth}{0.12ex} 
	\setlength{\lightrulewidth}{0.08ex} 
	\setlength{\tabcolsep}{20pt} 
	\resizebox{\columnwidth}{!}{
		\begin{tabular}{cccc}
			\toprule
			$\zeta$ & $a$ & $r_{isco}$ & $\eta^*$  \\
			\midrule %
			$-0.1$ & $0.5$ & $6.7768$ & $0.0647$   \\
			$-0.2$ & $0.5$ & $7.3089$ & $0.0703$   \\
			$-0.25$ & $0.5$ & $7.5233$ & $0.0730$      \\
			$-0.15$ & $0.1$ & $7.1531$ & $0.0669$     \\
			$-0.15$ & $0.5$ & $7.0641$ & $0.0676$   \\
			$-0.15$ & $1$ & $6.7753$ & $0.0699$   \\
			
			\bottomrule
		\end{tabular}
	}
	\caption{The ISCO radius $r_{isco}$ and radiative efficiency $\eta^*$ for different values of $\zeta$ or $a$.}
	\label{tab:2}
\end{table*}

\section{Null geodesics and photon deflection}
\label{sec:5}
We now shift our focus to null geodesics around a magnetically charged black hole immersed in perfect fluid dark matter. Confining the photon motion to the equatorial plane \(\theta=\pi/2\) and setting $\varepsilon=0$ for photons, the normalization condition derived from Eq.\eqref{eq:5} becomes

\begin{equation}
	-f(r)\dot{t}^2 + \frac{\dot{r}^2}{f(r)} + r^2 \dot{\phi}^2 = 0, 
	\label{eq:22}
\end{equation}

substituting the conserved quantities from  Eqs.\eqref{eq:6} and \eqref{eq:7} into the above expression, we obtain

\begin{align}
	-\frac{E^2}{f(r)} + \frac{\dot{r}^2}{f(r)} + \frac{L^2}{r^2} &= 0 \nonumber \\
	\Rightarrow \quad \dot{r}^2 &= E^2 - f(r)\frac{L^2}{r^2},
	\label{eq:23}
\end{align}

with the impact parameter defined as \(b = L/E\) and the affine parameter rescaled via \(\lambda' = L\lambda\), one arrives at
 
\begin{equation}
\frac{dt}{d\lambda'} = \frac{1}{b f(r)}, \qquad 
\frac{d\phi}{d\lambda'} = \frac{1}{r^2},
\label{eq:24}
\end{equation}
\begin{equation}
\left(\frac{dr}{d\lambda'}\right)^2 = \frac{1}{b^2} - \frac{f(r)}{r^2}.
\label{eq:25}
\end{equation}

Combining Eqs.\eqref{eq:24} and \eqref{eq:25}, we obtain

\begin{equation}
\left(\frac{dr}{d\phi}\right)^2 = r^4 \left(\frac{1}{b^2} - \frac{f(r)}{r^2}\right),
\label{eq:26}
\end{equation}

here, \(V_{\text{eff}} = \frac{1}{b^2} - \frac{f(r)}{r^2}\) denotes the effective potential for photons. The critical impact parameter \(b_c\) can be determined from the conditions \(V_{\text{eff}} = 0\) and \(V_{\text{eff}}' = 0\), and its expression is already given by Eq.\eqref{eq:4}. By introducing \(u = 1/r\), Eq.\eqref{eq:26} may be reformulated in the form

\begin{equation}
\left(\frac{du}{d\phi}\right)^2 = \frac{1}{b^2} - u^2 f\left(\frac{1}{u}\right) \equiv G(u),
	\label{eq:27}
\end{equation}

substituting the metric function f(r) from Eq.\eqref{eq:2} into Eq.\eqref{eq:27}, we obtain the explicit form of $G(u)$

\begin{equation}
G(u) = \frac{1}{b^2} - u^2 + \frac{2M u^3}{\sqrt{1+a^2 u^2}} - \zeta u^3 \ln\left(\frac{1}{u|\zeta|}\right).
	\label{eq:28}
\end{equation}

Figure 9 displays \(G(u)\) for various PFDM parameters with \(a=0.5\). 
When \(b>b_c\), \(G(u)\) possesses two roots; at \(b=b_c\), one root; 
and for \(b<b_c\), none. It is found that with increasing \(\lvert\zeta\rvert\), 
\(u_{\mathrm{ph}}\) and \(u_{\mathrm{min}}\) both move to smaller \(u\), 
and \(b_c\) becomes larger.

\begin{figure*}
	a)\includegraphics[width=8.1 cm]{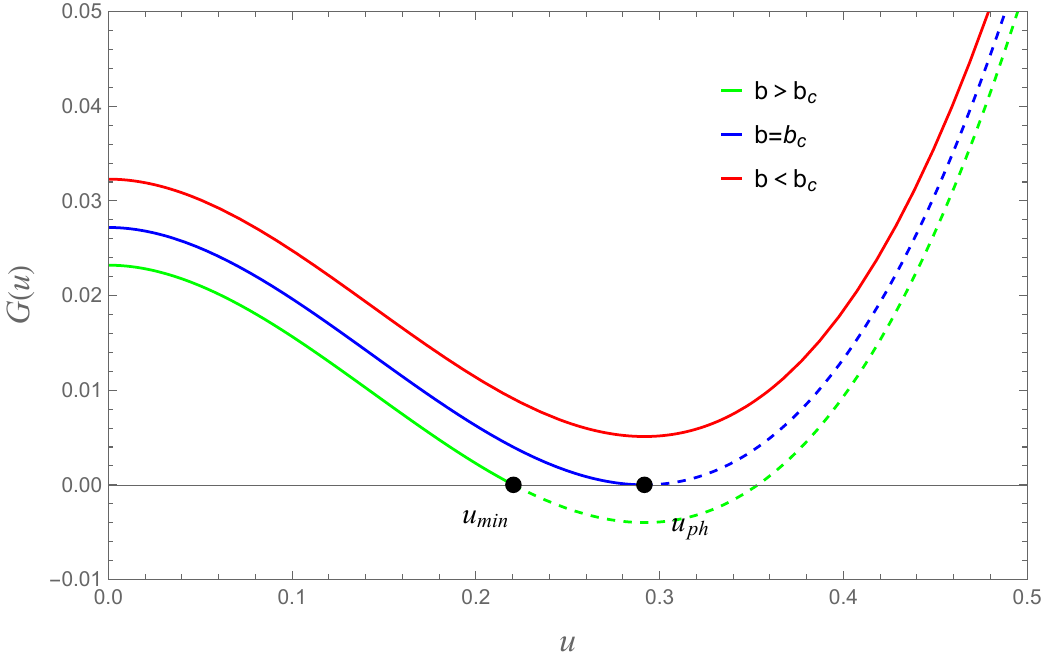}
	b)\includegraphics[width=8.1 cm]{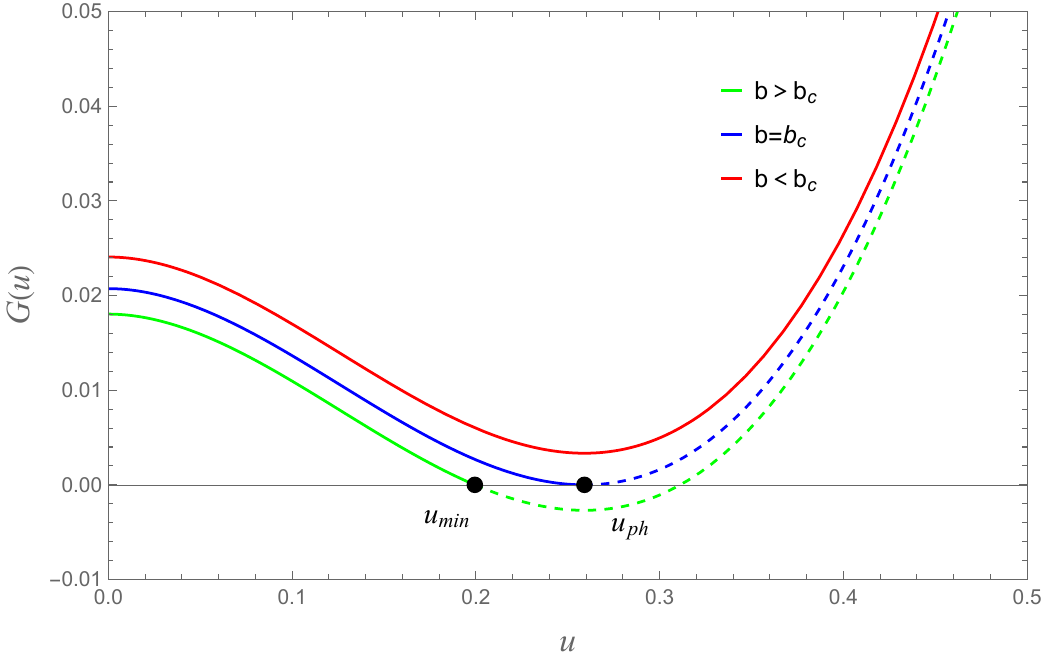}
	\caption{The function \(G(u)\) is plotted as a function of \(u\) for different values of \(\zeta\) at \(a=0.5\). The left panel corresponds to \(\zeta=-0.1\), and the right panel to \(\zeta=-0.25\). In these plots, \(u_{\mathrm{ph}}=1/r_{\mathrm{ph}}\), while \(u_{\mathrm{min}}\) denotes the smallest positive root of \(G(u)\) when \(b>b_c\).} \label{fig:9}
\end{figure*}

For a given impact parameter $b$, the net change in the azimuthal angle of a photon during its passage through the black hole spacetime is determined by the following integral:

\begin{equation}
	\varphi = 
	\begin{cases}
		\displaystyle \int_{0}^{u_h} \frac{du}{\sqrt{G(u)}}, & b < b_c \\[8pt]
		\displaystyle 2\int_{0}^{u_{\min}} \frac{du}{\sqrt{G(u)}}, & b > b_c
	\end{cases}
	\label{eq:29}
\end{equation}

where \(u_h = 1/r_h\), with \(r_h\) denoting the radius of the outermost horizon. The integral above gives the total azimuthal deflection \(\varphi(b)\), from which the winding number is defined as \(n(b) = \varphi/(2\pi)\), i.e., the number of full revolutions completed by the photon before arriving at the observer. Highly sensitive to the impact parameter \(b\), this quantity perfectly characterizes the three types of photon trajectories---direct emission, lensing rings, and photon rings---around a black hole \cite{Gralla:2019xty}. The winding number \(n(b)\) is defined by \cite{Peng:2020wun}

\begin{equation}
    n(b) = \frac{2m-1}{4}, \quad m = 1,2,3,\cdots,
	\label{eq:30}
\end{equation}

here, \(m\) denotes the number of intersections between the light ray and the accretion disk. For a given \(m\), the corresponding impact parameters are denoted by \(b_m^\pm\), which satisfy \(b_m^- < b_c < b_m^+\). On this basis, all trajectories can be classified as follows:

\begin{itemize}
	\item \textbf{direct:} The case \(m = 1\) corresponds to \(\frac{1}{4} < n < \frac{3}{4} \Rightarrow (b_1^-, b_2^-) \cup (b_2^-, \infty);\)
	\item \textbf{Lensed:} The case \(m = 2\) corresponds to \(\frac{3}{4} < n < \frac{5}{4} \Rightarrow (b_2^-, b_3^-) \cup (b_3^+, b_2^+);\)
	\item \textbf{Photon ring:} The case \(m \ge 3\) corresponds to \(n > \frac{5}{4} \Rightarrow (b_3^-, b_3^+).\)
\end{itemize}

Figure~10 shows the influence of \(\zeta\) and \(a\) on the winding number \(n(b)\). 
For larger \(\lvert\zeta\rvert\), the whole curve moves toward larger \(b\), with \(b_c\), \(b_1^-\), \(b_2^\pm\), and \(b_3^\pm\) all shifting outward; this signals that PFDM enhances the gravitational field and expands the photon ring. 
By contrast, when \(\zeta=-0.15\) is fixed, the curves for \(a=0.2\) and \(a=1\) differ only marginally. 
\textbf{Table III} confirms this quantitatively: at fixed \(\zeta=-0.15\), changing \(a\) from \(0.2\) to \(1\) alters \(b_c\) merely from \(6.4351\) to \(6.2276\) (about \(3.2\%\)), while changing \(\zeta\) from \(-0.1\) to \(-0.25\) changes \(b_c\) from \(6.0640\) to \(6.9452\) (about \(14.5\%\)). 
The physical reason is that the magnetic charge modifies the metric only near the horizon, whereas the PFDM term is a long-range logarithmic correction that accumulates along the photon path. 
Hence, the optical appearance of the charged-PFDM black hole is dominated primarily by \(\zeta\).

Figure~11 illustrates how the photon trajectories vary with the impact parameter \(b\). According to the winding number \(n(b)\), the gray, orange, and red curves correspond to \(n(b) < \frac{3}{4}\), \(\frac{3}{4} < n(b) < \frac{5}{4}\), and \(n(b) > \frac{5}{4}\), respectively. As \(|\zeta|\) increases, the boundaries of these three regions shift toward larger \(b\), while the black hole horizon and the photon sphere also expand in tandem. The range of \(b\) associated with the red photon-ring region extends outward, indicating that PFDM strengthens the gravitational field.

\begin{figure*}
	a)\includegraphics[width=8.1 cm]{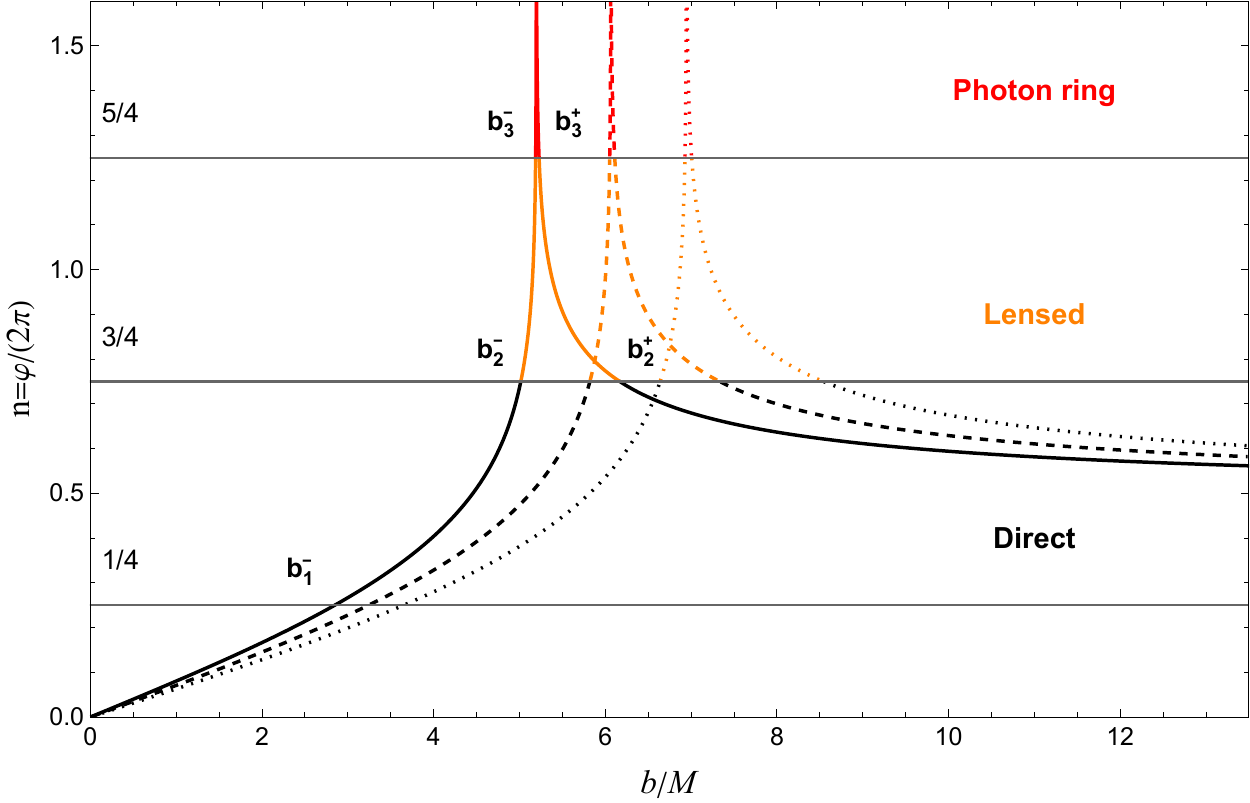}
	b)\includegraphics[width=8.1 cm]{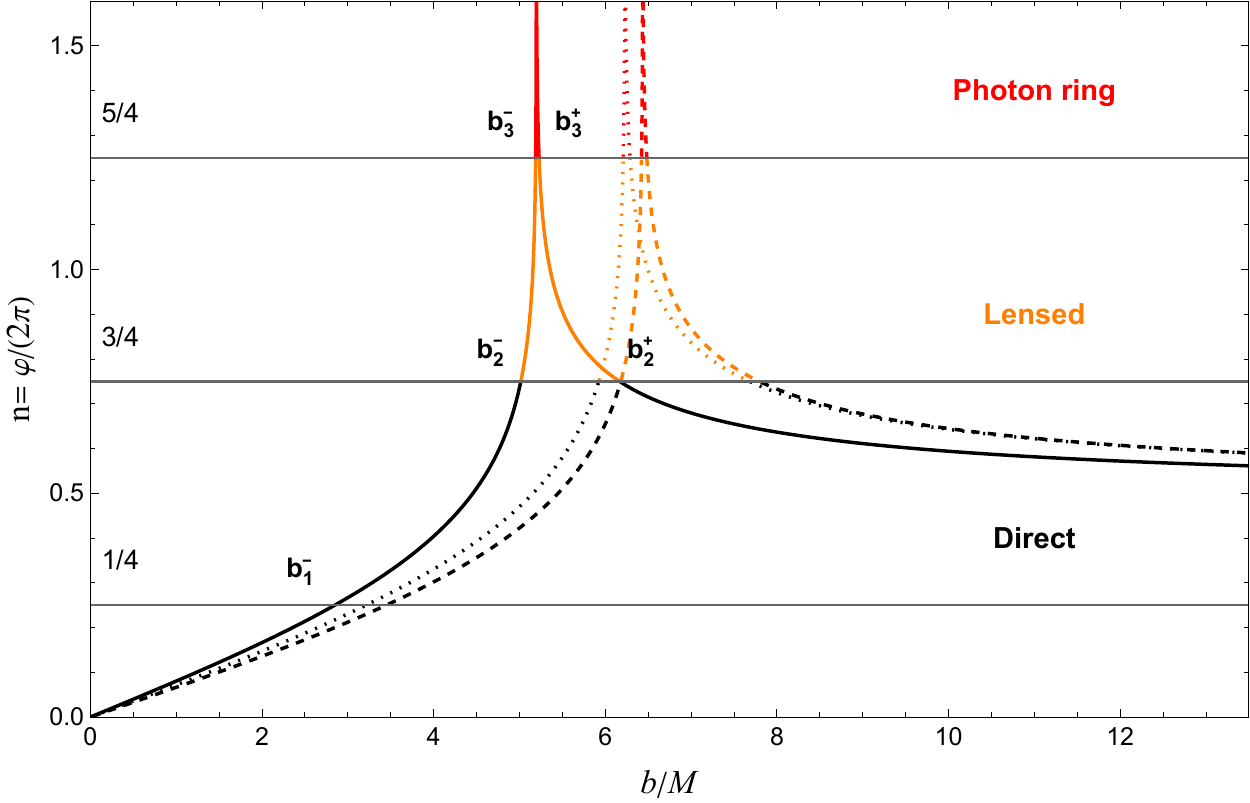}
	\caption{This figure illustrates the dependence of the total winding number \(n(b)\) on the impact parameter \(b\). In panel (a), the solid, dashed, and dotted lines correspond to the Schwarzschild case, and to the cases with \(a=0.5\) fixed and \(\zeta=-0.1\) and \(-0.25\), respectively. In panel (b), the solid, dashed, and dotted lines represent the Schwarzschild case, and the cases with \(\zeta=-0.15\) fixed and \(a=0.2\) and \(1\), respectively.} \label{fig:10}
\end{figure*}

\begin{table*}[htbp] 
	\centering
	\small 
	\renewcommand{\arraystretch}{1.1} 
	\setlength{\heavyrulewidth}{0.12ex} 
	\setlength{\lightrulewidth}{0.08ex} 
	\setlength{\tabcolsep}{6pt} 
	\resizebox{\textwidth}{!}{ 
		\begin{tabular}{cccccccccc}
			\toprule 
			\(\zeta\) & \(a\) & \(r_h/M\) & \(r_{ph}/M\) & \(b_c/M\) & \(b_1^-/M\) & \(b_2^-/M\) & \(b_2^+/M\) & \(b_3^-/M\) & \(b_3^+/M\) \\
			\midrule 
			\(-0.1\)  & \(0.5\) & \(2.2647\) & \(3.4282\) & \(6.0640\) & \(3.2415\) & \(5.8201\) & \(7.3310\) & \(6.0513\) & \(6.1098\) \\
			\(-0.25\)  & \(0.5\) & \(2.5422\) & \(3.8602\) & \(6.9452\) & \(3.6503\) & \(6.6391\) & \(8.5398\) & \(6.9279\) & \(7.0062\) \\
			\(-0.15\) & \(0.2\) & \(2.4096\) & \(3.6347\) & \(6.4351\) & \(3.4471\) & \(6.1796\) & \(7.7955\) & \(6.4219\) & \(6.4836\) \\
			\(-0.15\) & \(1\)   & \(2.2297\) & \(3.4349\) & \(6.2276\) & \(3.2125\) & \(5.9241\) & \(7.6803\) & \(6.2089\) & \(6.2880\) \\
			\(0\)     & \(0\)   & \(2\)      & \(3\)      & \(5.1962\) & \(2.8477\) & \(5.0151\) & \(6.1676\) & \(5.1878\) & \(5.2279\) \\
			\bottomrule 
		\end{tabular}
	}
	\caption{Characteristic radii and critical impact parameters of the charged-PFDM black hole for selected \(\zeta\) and \(a\). Schwarzschild: \(\zeta=0\), \(a=0\).} 
	\label{tab:3}
\end{table*}

\begin{figure*}
	a)\includegraphics[width=5.1 cm]{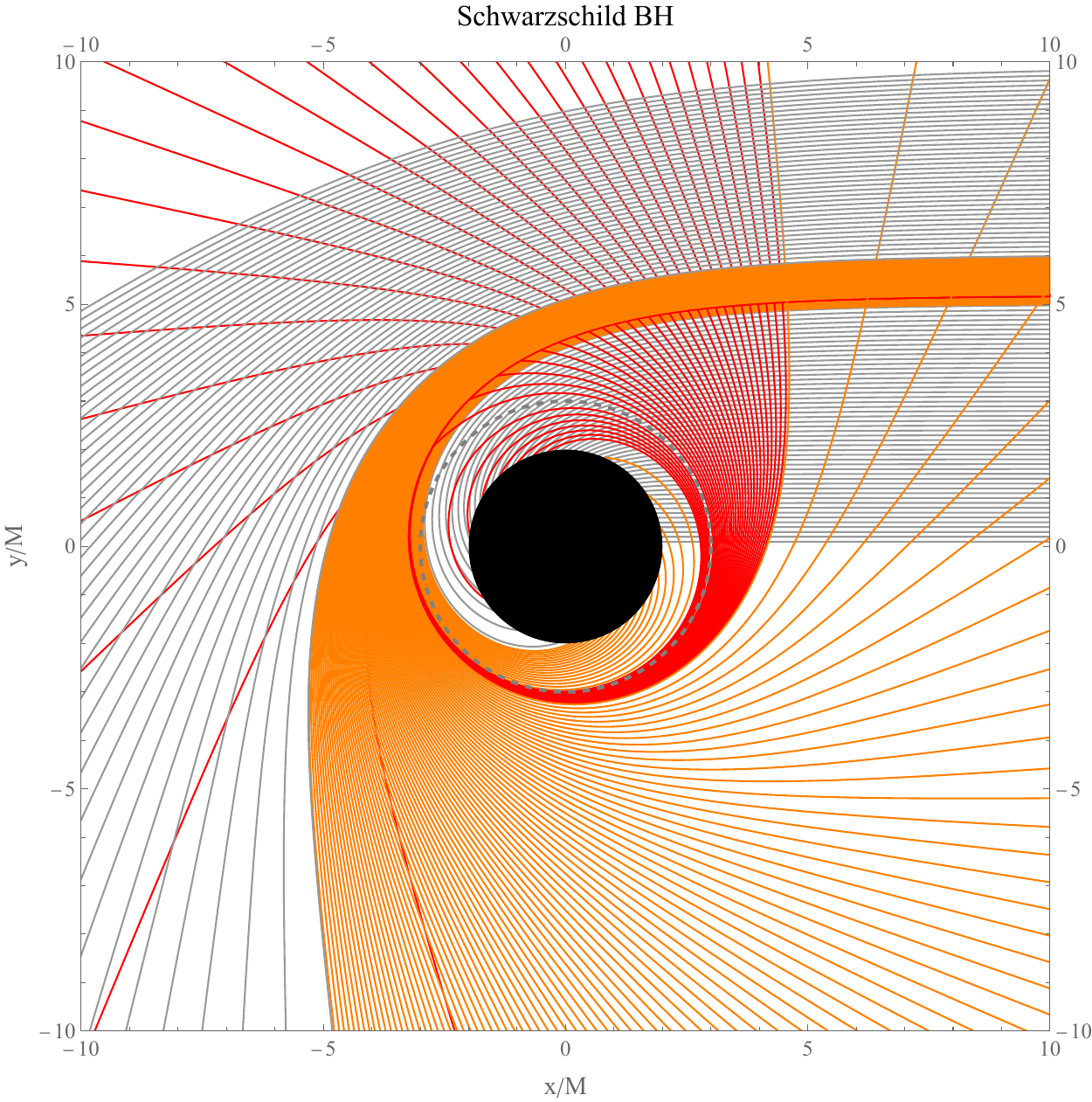}
	b)\includegraphics[width=5.1 cm]{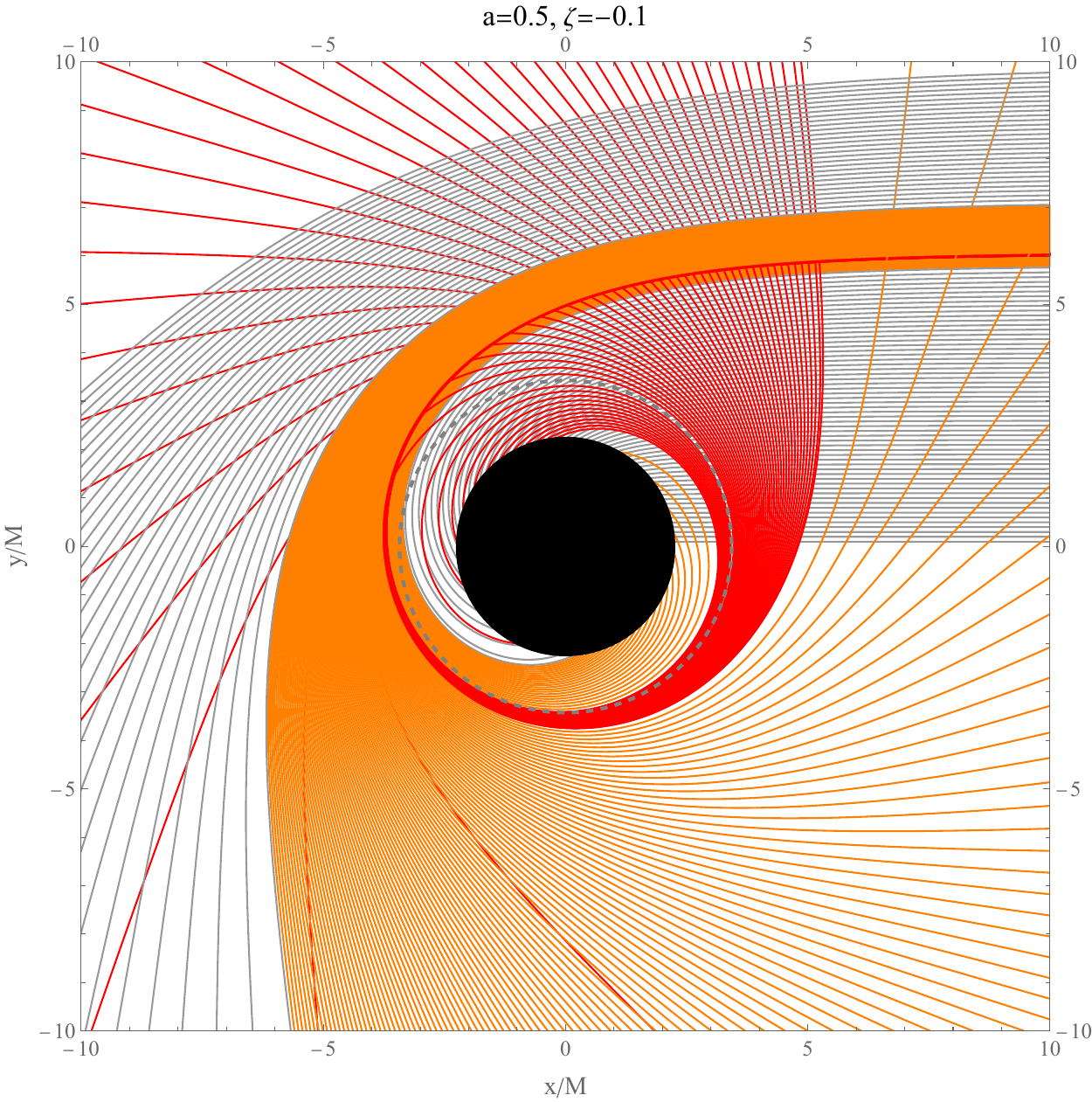}
	c)\includegraphics[width=5.1 cm]{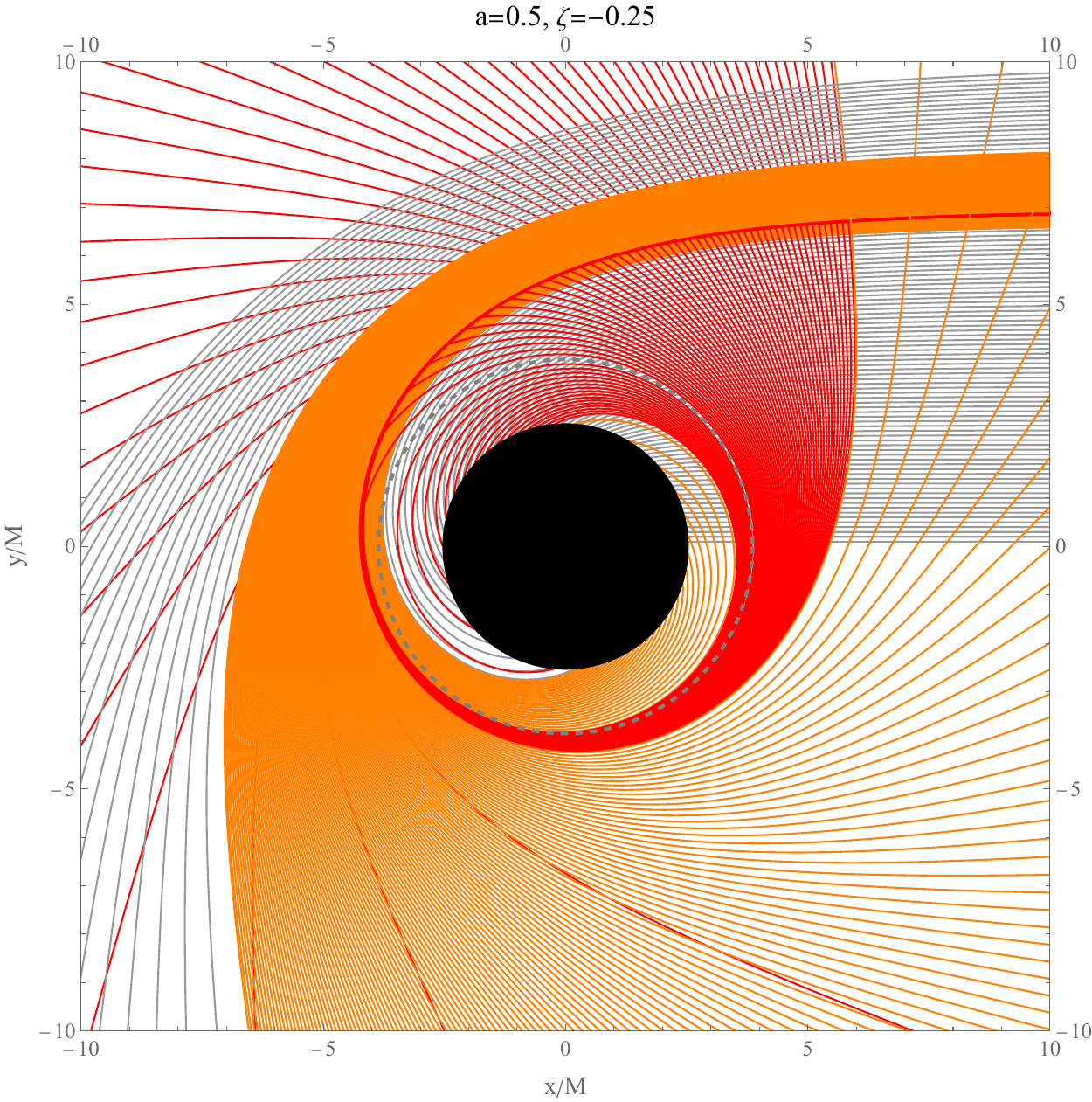}
	\caption{Photon trajectories as functions of \(b\) for Schwarzschild and charged-PFDM black holes. Red, orange, and black lines denote photon-ring, lensing-ring, and direct trajectories, respectively. The black disk is the black hole; the dashed black curve indicates the photon sphere. (a) Schwarzschild, (b) \(a=0.5\), \(\zeta=-0.1\), (c) \(a=0.5\), \(\zeta=-0.25\).} \label{fig:11}
\end{figure*}

\section{Images of the thin accretion disk around a charged-PFDM black hole}
\label{sec:6}

To construct images of the thin accretion disk, we introduce the observational coordinate system illustrated in Fig.12. The observer is assumed to be located at infinity, with its spatial position described by spherical coordinates \((r,\theta,\phi)\), where \(r\to\infty\), \(\theta\) is the observation inclination angle (i.e., the angle between the line of sight and the black hole spin axis), and \(\phi\) denotes the azimuthal angle. The black hole resides at the coordinate origin, and the accretion disk lies in the equatorial plane \(\theta=\pi/2\). As shown in Fig.12, we introduce an observer-plane coordinate system \(O'X'Y'\), with \(O'\) being the center of the observation plane. A photon is emitted from a point \(M(r,\pi/2,\phi)\) on the disk and propagates along a geodesic to an image point \(m(b,\alpha)\) on the observation plane, where \(b\) is the impact parameter (i.e., the perpendicular distance from the asymptotic line of the photon trajectory to the black hole center) and \(\alpha\) is the azimuthal angle of the image point on the observation plane.

Owing to the reversibility of light paths, a photon emitted from point \(M(r,\pi/2,\phi)\) on the accretion disk will follow a trajectory that eventually reaches the image point \((b,\alpha)\) on the observer plane. The entire trajectory is confined to a single plane, and the total azimuthal deflection angle \(\phi\) is related to the geometric parameters \(\theta\) and \(\alpha\). For a given emission radius \(r\), two symmetric points on the disk at \(\phi\) and \(\phi+\pi\) produce two images on the observation plane: one is the direct image (the photon reaches the observer directly), and the other is the secondary image (the photon reaches the observer after being deflected around the black hole) \cite{Liu:2021lvk}.

Using the law of sines for spherical triangles, the geometric relation between the disk azimuthal angle \(\phi\) and the image-plane angle \(\alpha\) can be derived. Consider the spherical triangle formed by the spin axis, the line of sight, and the radial vector \(OQ\). From the triangles \(\triangle Myy'\) and \(\triangle Mxx'\) in Fig.~12, we obtain

\begin{figure*}
	\includegraphics[width=12.1 cm]{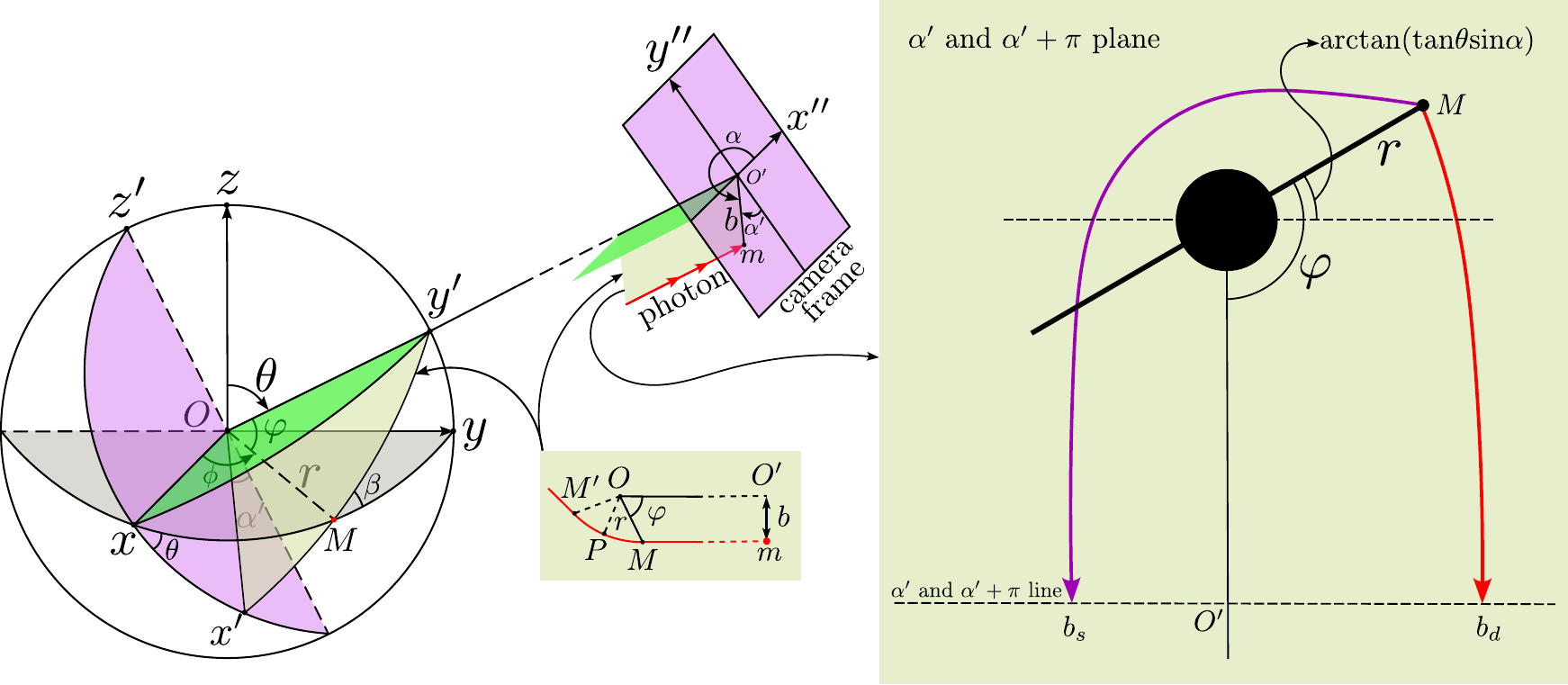}
	\caption{The coordinate system employed for the accretion-disk images is specified in Ref.~\cite{Xamidov:2025gcs}.} \label{fig:12}
\end{figure*}

\begin{equation}
	\frac{\sin\left(\frac{\pi}{2}\right)}{\sin\varphi} = \frac{\sin\beta}{\sin\left(\frac{\pi}{2} - \theta_0\right)}, \qquad
	\frac{\sin\theta_0}{\sin\left(\frac{\pi}{2} - \varphi\right)} = \frac{\sin\beta}{\sin\left(\frac{\pi}{2} - \alpha\right)},
	\label{eq:31}
\end{equation}

where \(\beta\) is an auxiliary angle. Eliminating \(\beta\) and using the relation \(\alpha+\alpha'=3\pi/2\) (with \(\alpha'\) being an intermediate angle), the explicit expression for the deflection angle \(\phi\) is obtained \cite{You:2024jeu, Sharipov:2025yfw}:

\begin{equation}
	\varphi=\frac{\pi}{2}+\arctan(\tan\theta_0\sin\alpha).
	\label{eq:32}
\end{equation}

For a photon emitted from the accretion disk at radius \(r\) and arriving at the image point \(q(b,\alpha)\), its total deflection angle is determined by the integral along the photon trajectory. Since the photon may cross the equatorial plane multiple times before reaching the observer, the deflection angle corresponding to the \(n\)-th order image can be expressed as \cite{You:2024uql}

\begin{widetext}
	\begin{equation}
		\varphi_n =
		\begin{cases}
			\dfrac{n}{2}2\pi + (-1)^n \left[ \dfrac{\pi}{2} + \arctan(\tan\theta_0\sin\alpha) \right], & \text{for } n \text{ even}, \\[10pt]
			\dfrac{n+1}{2}2\pi + (-1)^n \left[ \dfrac{\pi}{2} + \arctan(\tan\theta_0\sin\alpha) \right], & \text{for } n \text{ odd},
		\end{cases}
		\label{eq:33}
	\end{equation}
\end{widetext}

where \(n=0\) corresponds to the direct image, \(n=1\) to the secondary image, and higher-order images correspond to photons that reach the observer after winding multiple times around the black hole. Even-order images appear at the azimuthal angle \(\alpha\) on the same side as the source, while odd-order images appear on the opposite side at \(\alpha+\pi\). Once the deflection angle $\varphi_n$ is determined, the corresponding impact parameter \(b_n\) can be obtained by using the previously derived integral formula for the deflection angle. 

In Fig.~13, the deflection angle \(\varphi\) is plotted as a function of the impact parameter \(b\), with curves shown for different values of \(\zeta\) and different emission radii. As the disk radius increases, the direct image (\(n=0\)) varies appreciably, while the secondary images (\(n\geq 1\)) remain essentially unchanged. A larger \(\lvert\zeta\rvert\) gives rise to a larger critical impact parameter \(b_c\) and shifts the entire \(\varphi(b)\) curve upward and to the right. The purple dashed line in the figure denotes the periapsis curve \(\varphi_3(b)\), which asymptotically approaches the line \(\pi/2\) at infinity. Using this purple dashed line as a dividing boundary, the region above it corresponds to the incident branch \(\varphi_1(b)\), which describes photons traveling directly from infinity to a given radius; the region below it corresponds to the outgoing branch \(\varphi_2(b)\), which describes photons returning to the same radius after passing through periapsis. On this basis, we can define:

\begin{align}
	\varphi_1(b) &= \int_0^{u_{\min}} \frac{1}{\sqrt{G(u)}} \, du, \label{eq:34} \\
	\varphi_2(b) &= \int_0^{u_r} \frac{1}{\sqrt{G(u)}} \, du, \label{eq:35} \\
	\varphi_3(b) &= 2\int_0^{u_{\min}} \frac{1}{\sqrt{G(u)}} \, du - \int_0^{u_r} \frac{1}{\sqrt{G(u)}} \, du, \label{eq:36}
\end{align}

By numerically integrating Eqs.\eqref{eq:33}, \eqref{eq:34}, and \eqref{eq:35}, we obtain the coordinates \((B,\alpha)\) of the accretion disk in the observer plane \cite{Gyulchev:2020cvo,Guo:2024mij,Cai:2025rst}.Figure~14 displays the direct and secondary images of stable circular orbits surrounding a charged-PFDM black hole for various inclination angles; solid curves correspond to direct images, while dashed curves correspond to secondary images. The leftmost column represents the Schwarzschild black hole. In comparison with Schwarzschild, the overall image scale becomes larger as \(\lvert\zeta\rvert\) increases, which signals an enhanced gravitational lensing effect.

Owing to the combined gravitational redshift and Doppler effect, the energy flux detected by a distant observer differs from that emitted at the disk surface, and the observed flux \(F_{\mathrm{obs}}\) is given by \cite{Ellis:1971pg}

\begin{equation}
	F_{\text{obs}} = \frac{F(r)}{(1+z)^4},
	\label{eq:37}
\end{equation}

as in Refs.~\cite{Luminet:1979nyg,Bhattacharyya:2000kt,Cai:2025rst}, we neglect light bending for simplicity. In this case, the disk-plane coordinates and the observer-plane coordinates satisfy \(b\cos\alpha = r\sin\phi\), and Eq.\eqref{eq:20} can be recast as

\begin{equation}
	1+z = \frac{1 + \Omega b \sin\theta \cos\alpha}{\sqrt{-g_{tt} - g_{\phi\phi}\Omega^2}}.
	\label{eq:38}
\end{equation}

By combining Eqs.\eqref{eq:17}, \eqref{eq:37}, and \eqref{eq:38}, we compute and plot the distribution of the observed flux \(F_{\mathrm{obs}}\) from the accretion disk, spanning the radial range from \(r_{\mathrm{isco}}\) to \(r=25M\). The results are normalized to the observed flux of a Schwarzschild black hole at \(\theta=85^\circ\), which is set to \(100\%\), and are visualized using a thermal color map. Figure~15 presents the flux distributions for three representative spacetimes---Schwarzschild, \(a=0.5\) with \(\zeta=-0.1\), and \(a=0.5\) with \(\zeta=-0.25\)---at three inclination angles: \(\theta=17^\circ, 53^\circ\), and \(85^\circ\). As can be seen from Fig.~15, with increasing inclination angle \(\theta\), the brightness distribution becomes increasingly asymmetric due to the Doppler effect, with one side significantly brightened. For a fixed inclination, as \(|\zeta|\) increases, the overall observed flux \(F_{\mathrm{obs}}\) decreases and the accretion disk appears dimmer; this is consistent with the expectation that PFDM strengthens the gravitational field and thereby reduces the radiative efficiency. At \(\theta=85^\circ\) and \(\zeta=-0.25\), the peak observed flux is only a small fraction of that in the Schwarzschild case.

Figure~16 displays the redshift distributions of a thin accretion disk around a charged PFDM black hole for three representative spacetimes and three observation inclinations. The combined action of the Doppler effect and gravitational redshift shifts the frequency of the emitted light as seen by the observer, while the disk rotation renders the image asymmetric---one side appears blueshifted and the other redshifted. At low inclination (\(\theta=17^\circ\)), the redshift distribution is nearly axisymmetric, indicating that gravitational redshift dominates. As \(\theta\) increases, the Doppler effect becomes progressively more pronounced: the approaching side of the disk exhibits a strong blueshift (negative \(z\)), whereas the receding side shows a marked redshift (positive \(z\)). This asymmetry reaches its maximum at \(\theta=85^\circ\), where the disk appears as a highly asymmetric, elongated strip. The subpanels at the same inclination share an identical color scale to allow consistent comparison. The figure reveals that a larger \(|\zeta|\) leads to stronger both blueshift and redshift, and to a more pronounced asymmetry in the observed image. This trend is consistent with the PFDM-enhanced gravitational field, i.e., the overall gravitational redshift across the disk is strengthened. These differences suggest that the redshift distribution of the accretion disk can serve as a potential probe for future high-resolution observations to distinguish black holes in a PFDM environment from Schwarzschild black holes.

\begin{figure*}
	a)\includegraphics[width=8.1 cm]{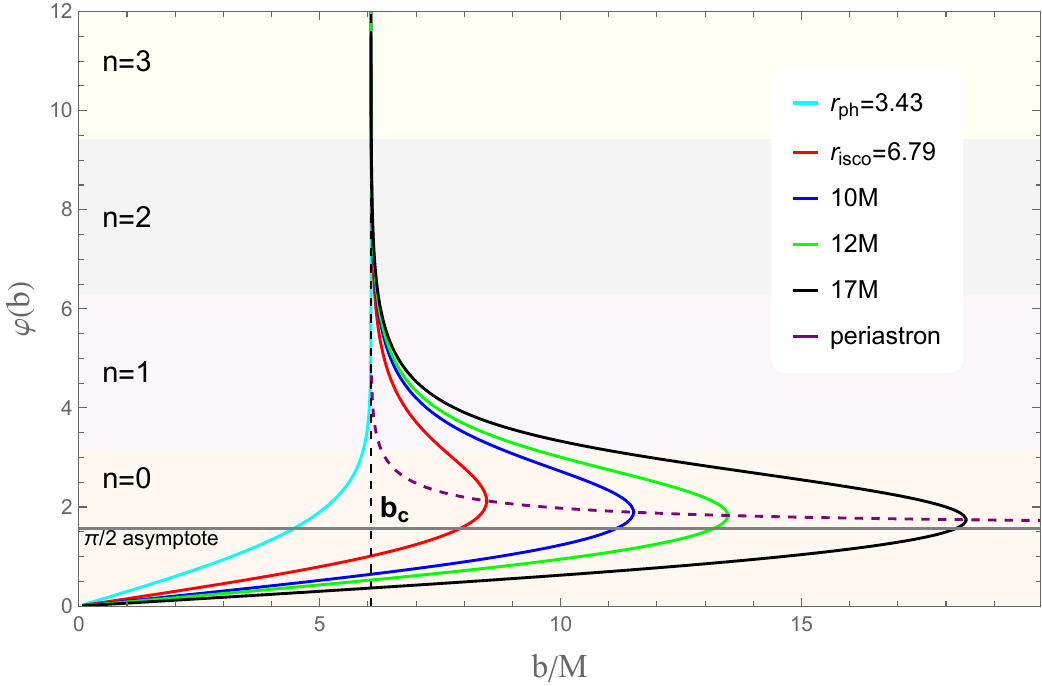}
	b)\includegraphics[width=8.1 cm]{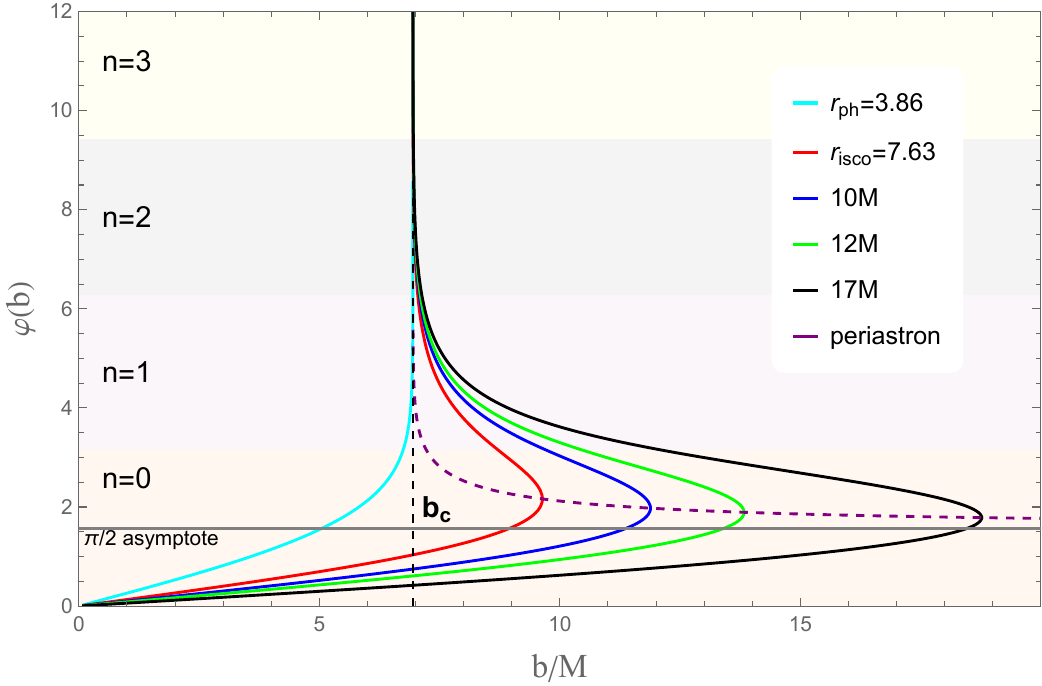}
	\caption{Deflection angle \(\varphi(b)\) versus impact parameter \(b\) for constant-radius photon orbits around a charged-PFDM black hole. (a) \(a=0.5\), \(\zeta=-0.1\), (b) \(a=0.5\), \(\zeta=-0.25\).} \label{fig:13}
\end{figure*}

\begin{figure*}
	\includegraphics[width=5.1 cm]{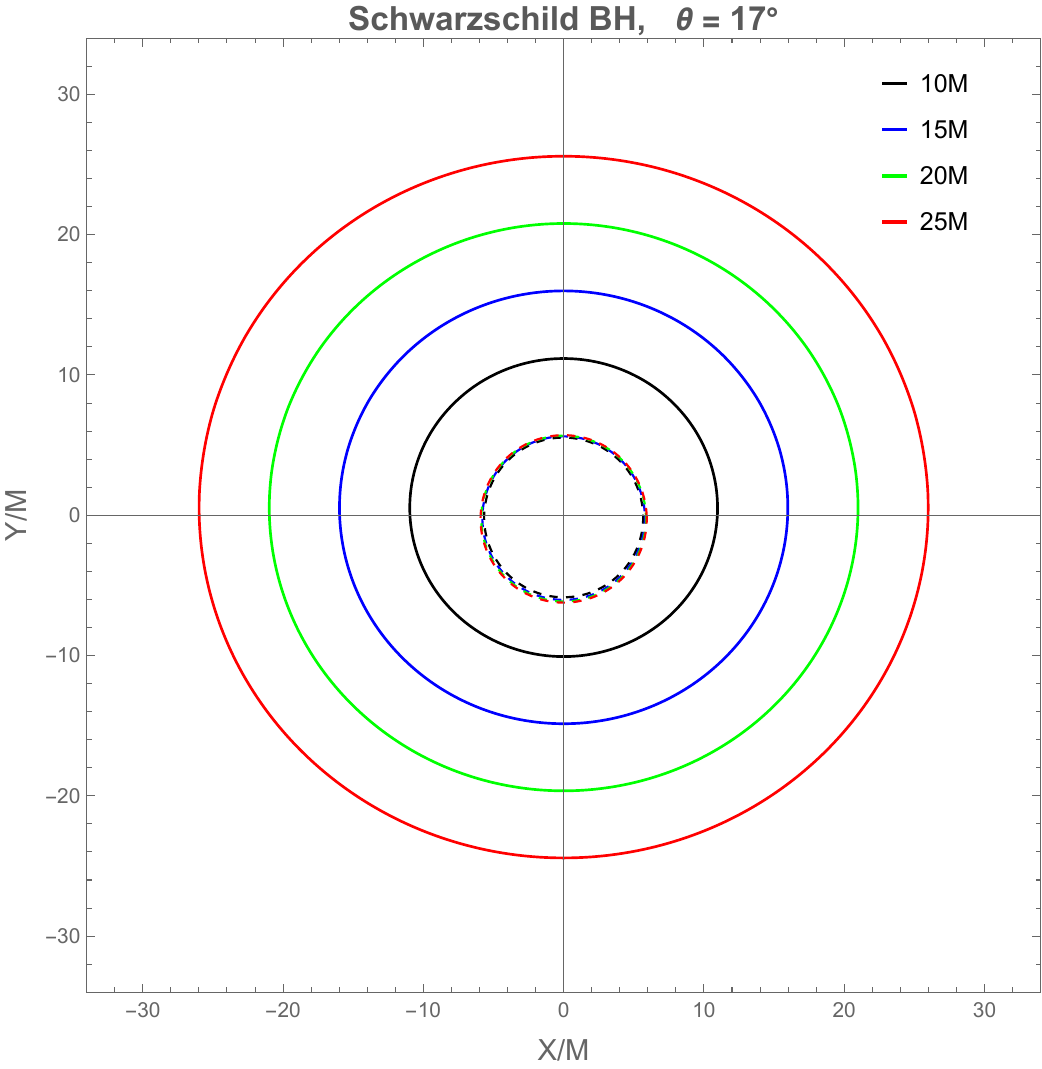}
	\includegraphics[width=5.1 cm]{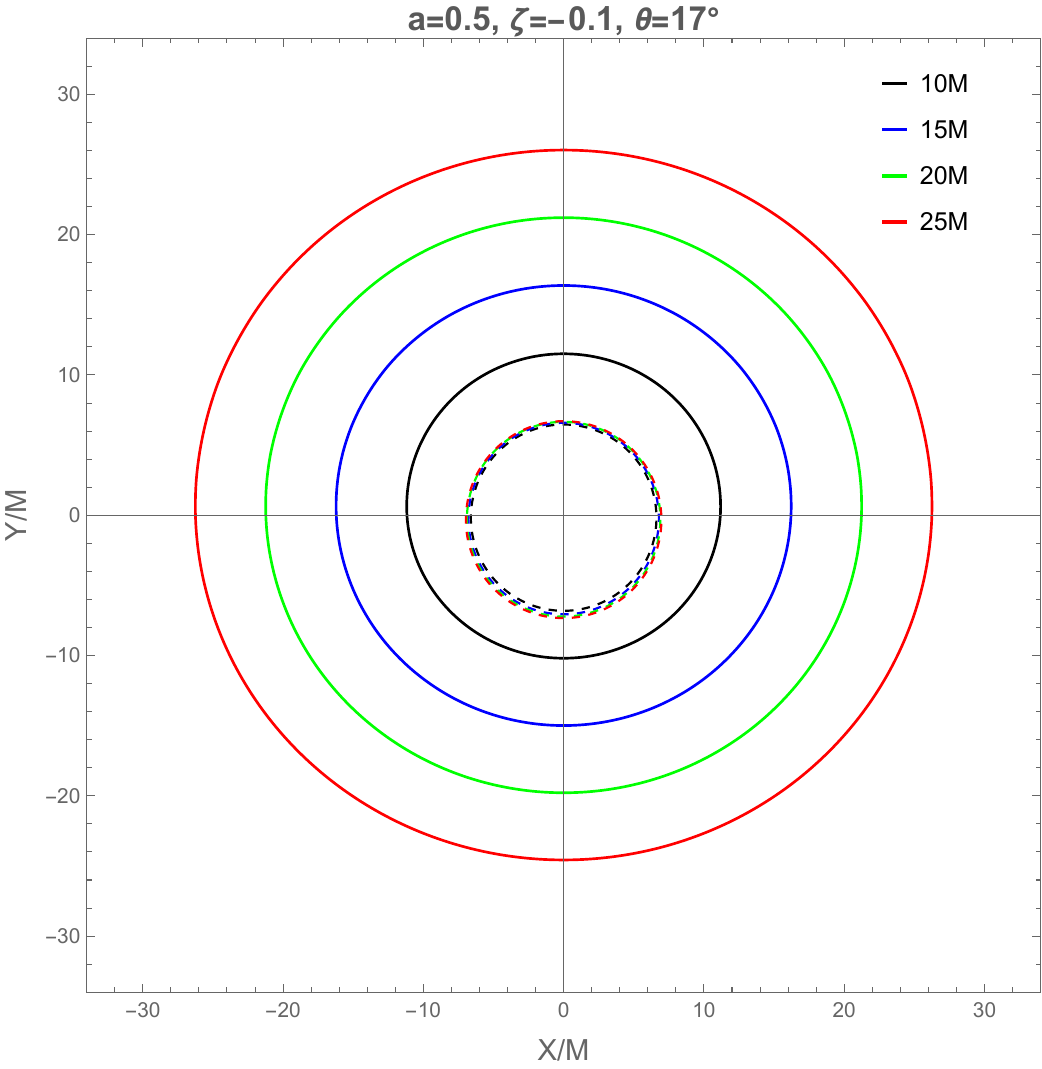}
	\includegraphics[width=5.1 cm]{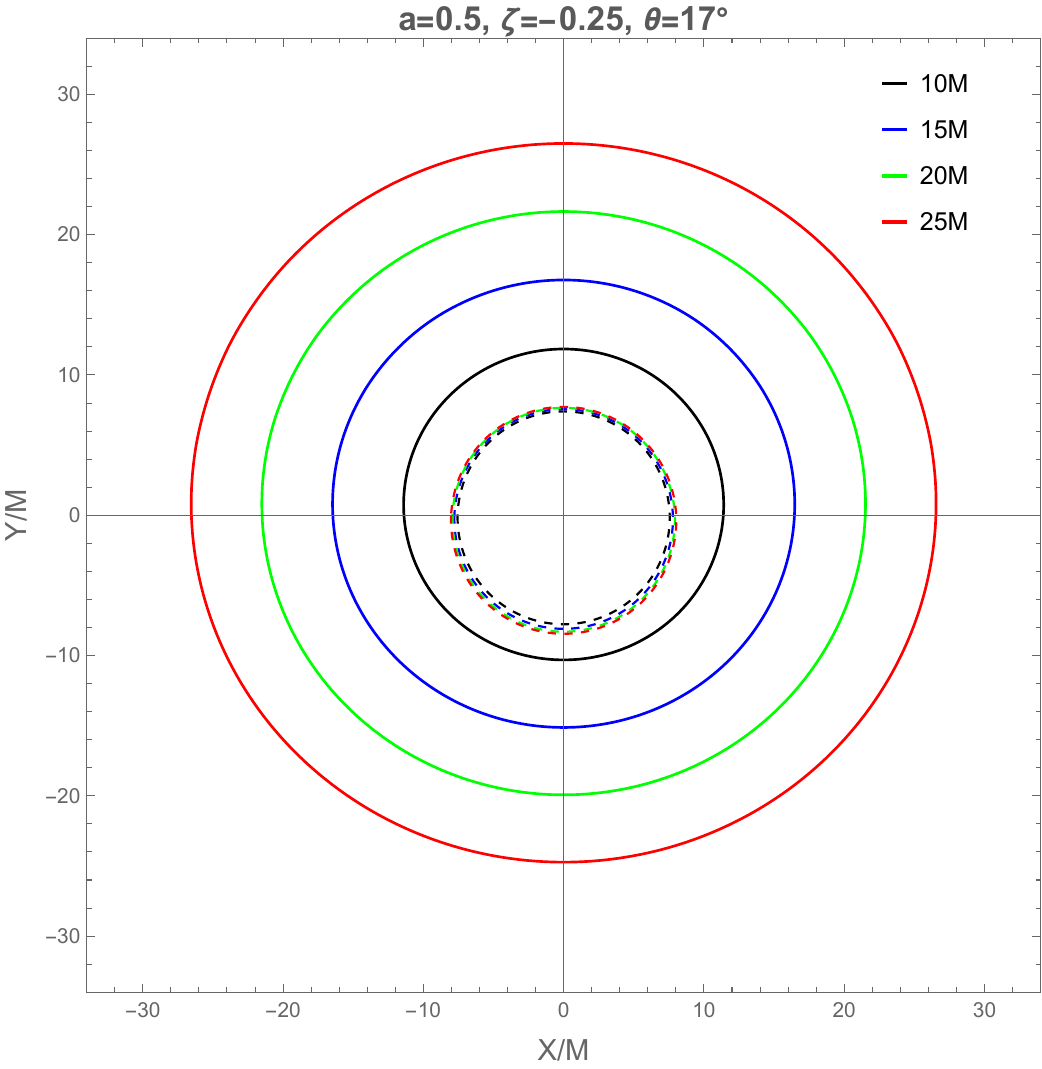}
	\includegraphics[width=5.1 cm]{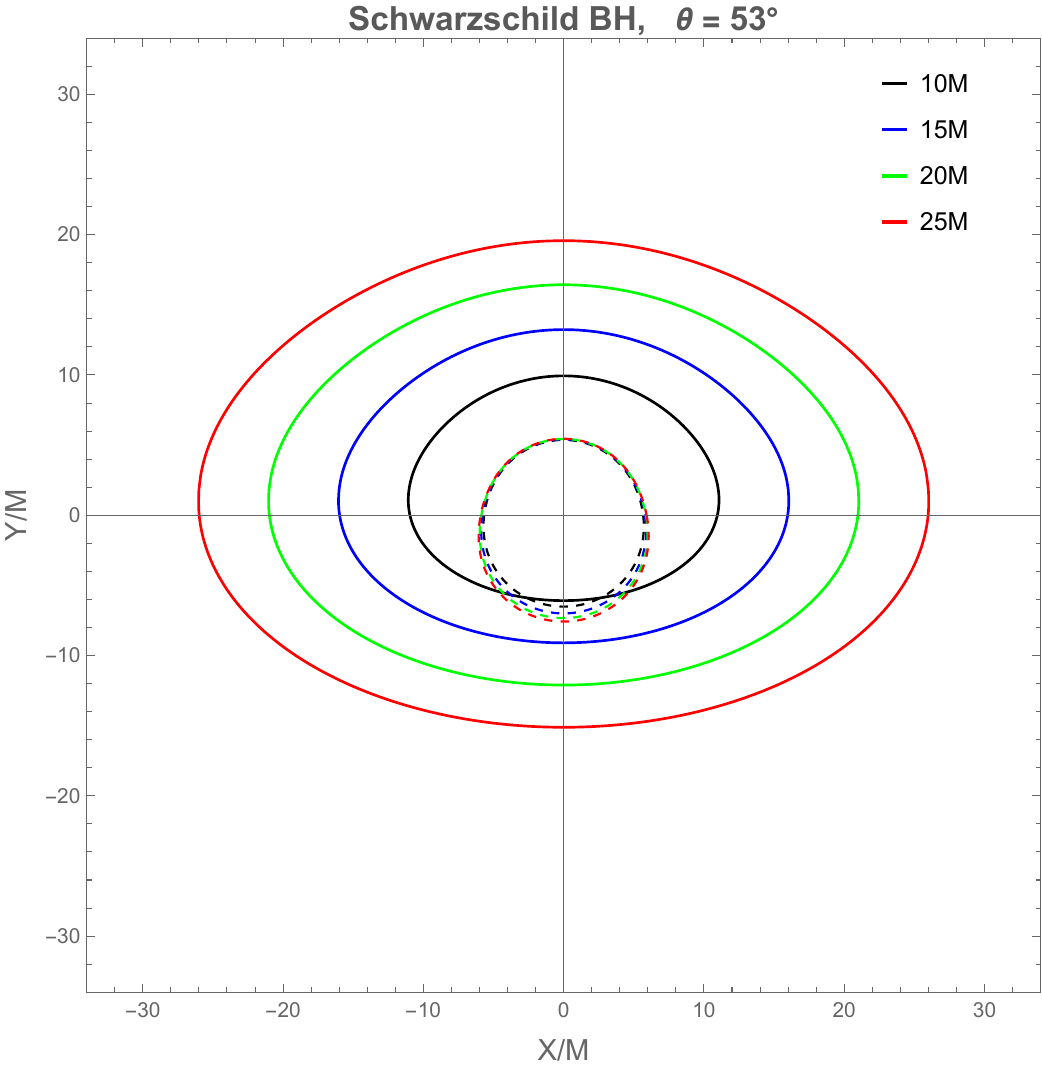}
	\includegraphics[width=5.1 cm]{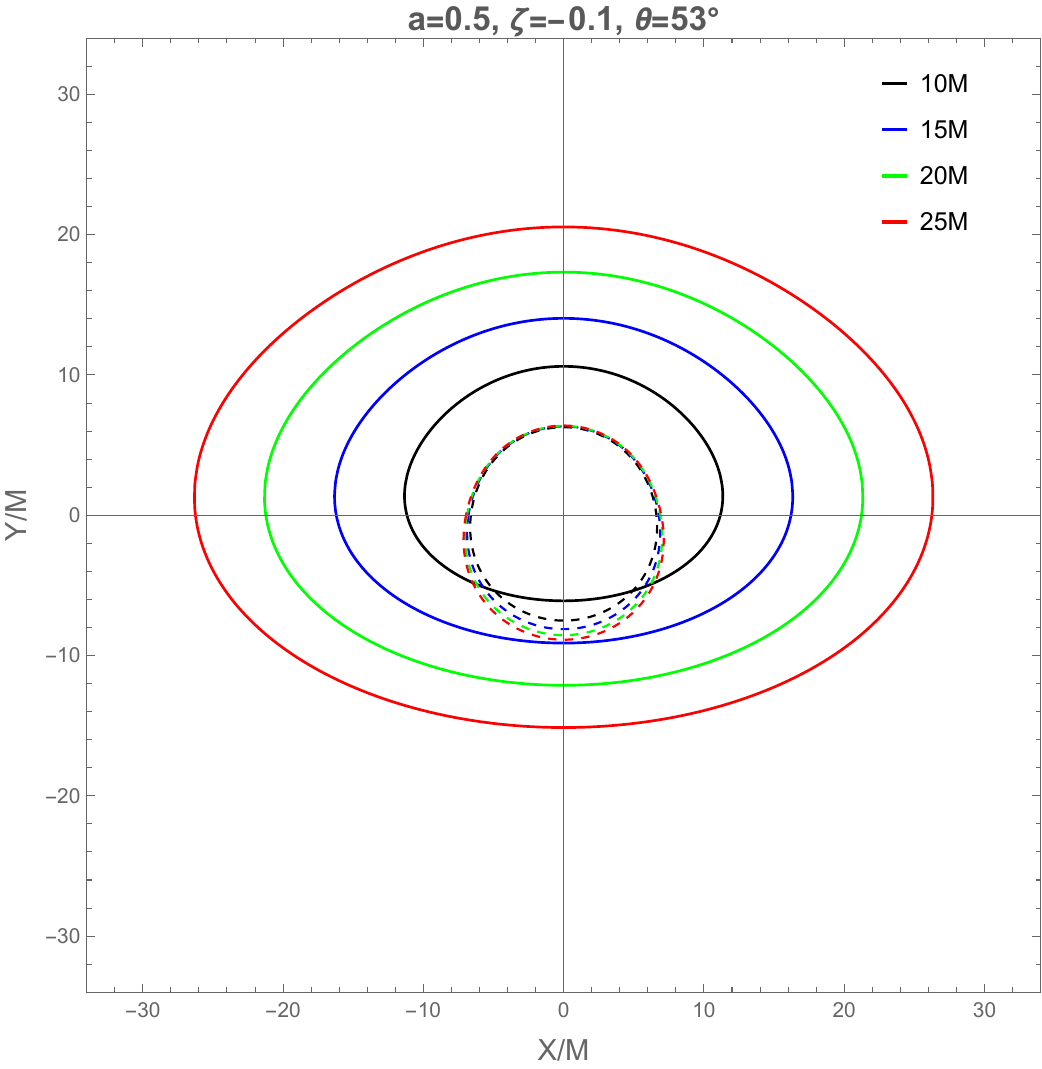}
	\includegraphics[width=5.1 cm]{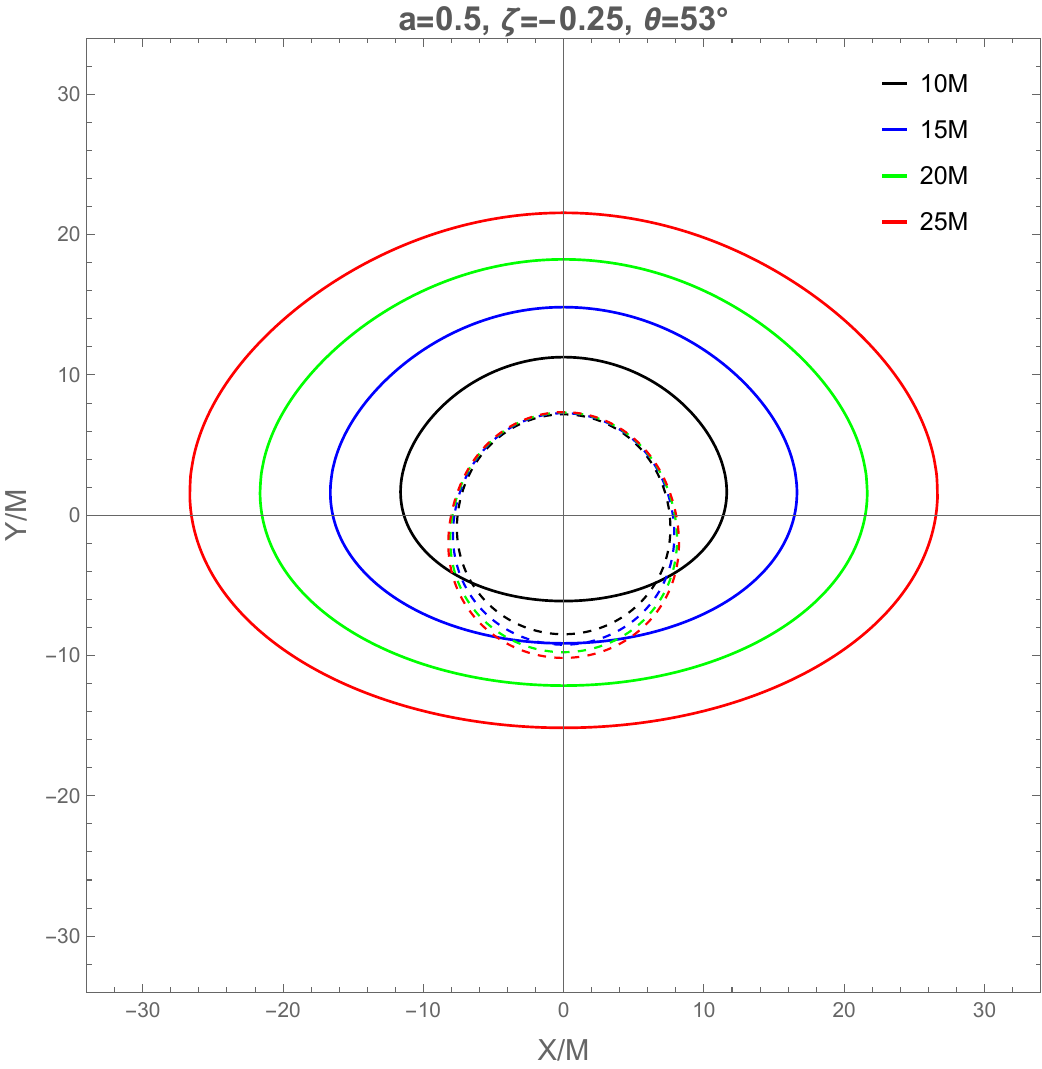}
	\includegraphics[width=5.1 cm]{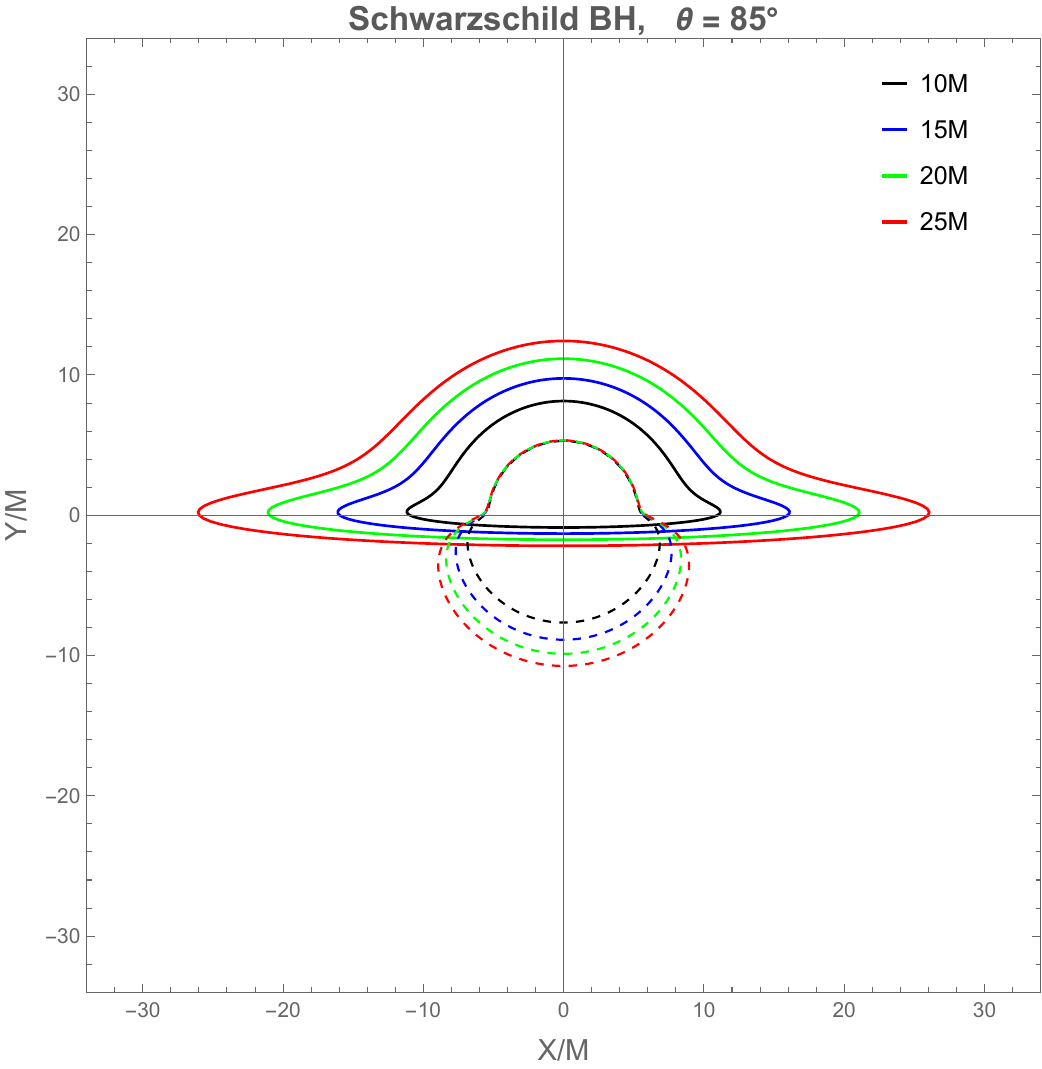}
	\includegraphics[width=5.1 cm]{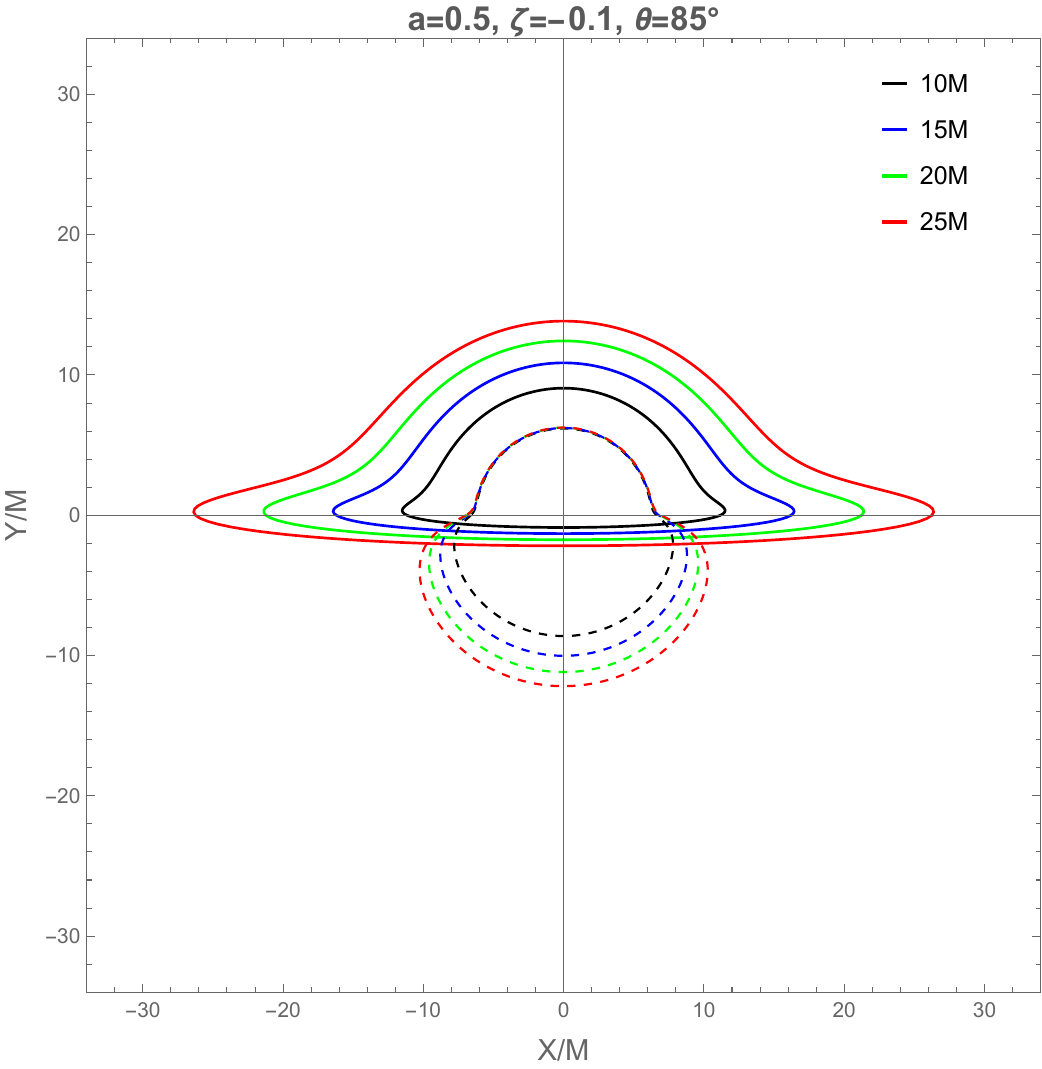}
	\includegraphics[width=5.1 cm]{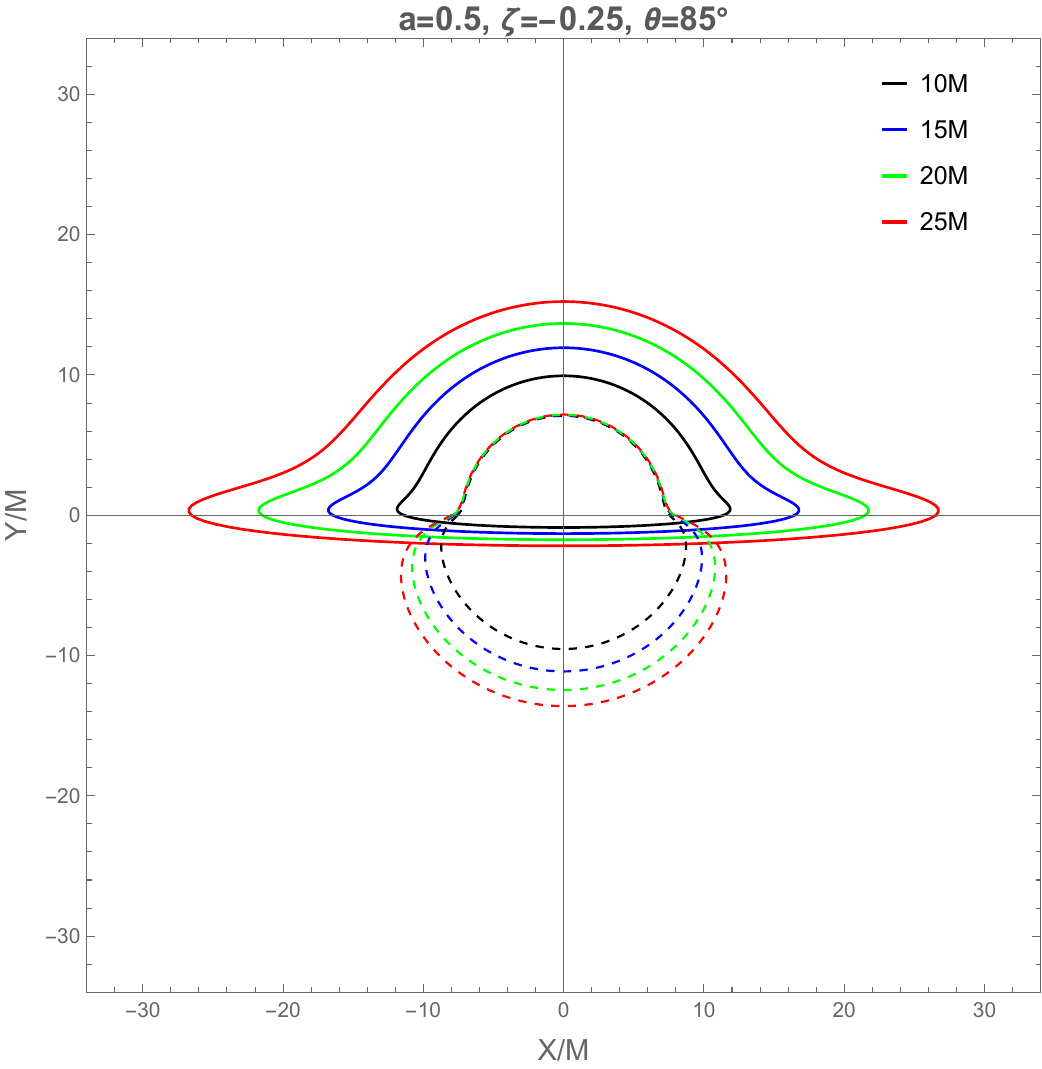}
	\caption{Direct and secondary images of thin accretion disks for Schwarzschild and charged-PFDM black holes. Rows: inclination angles \(17^\circ, 53^\circ, 85^\circ\) (top to bottom). Columns: Schwarzschild; \(a=0.5\), \(\zeta=-0.1\); \(a=0.5\), \(\zeta=-0.25\) (left to right).} \label{fig:14}
\end{figure*}

\begin{figure*}
	\includegraphics[width=5.1 cm]{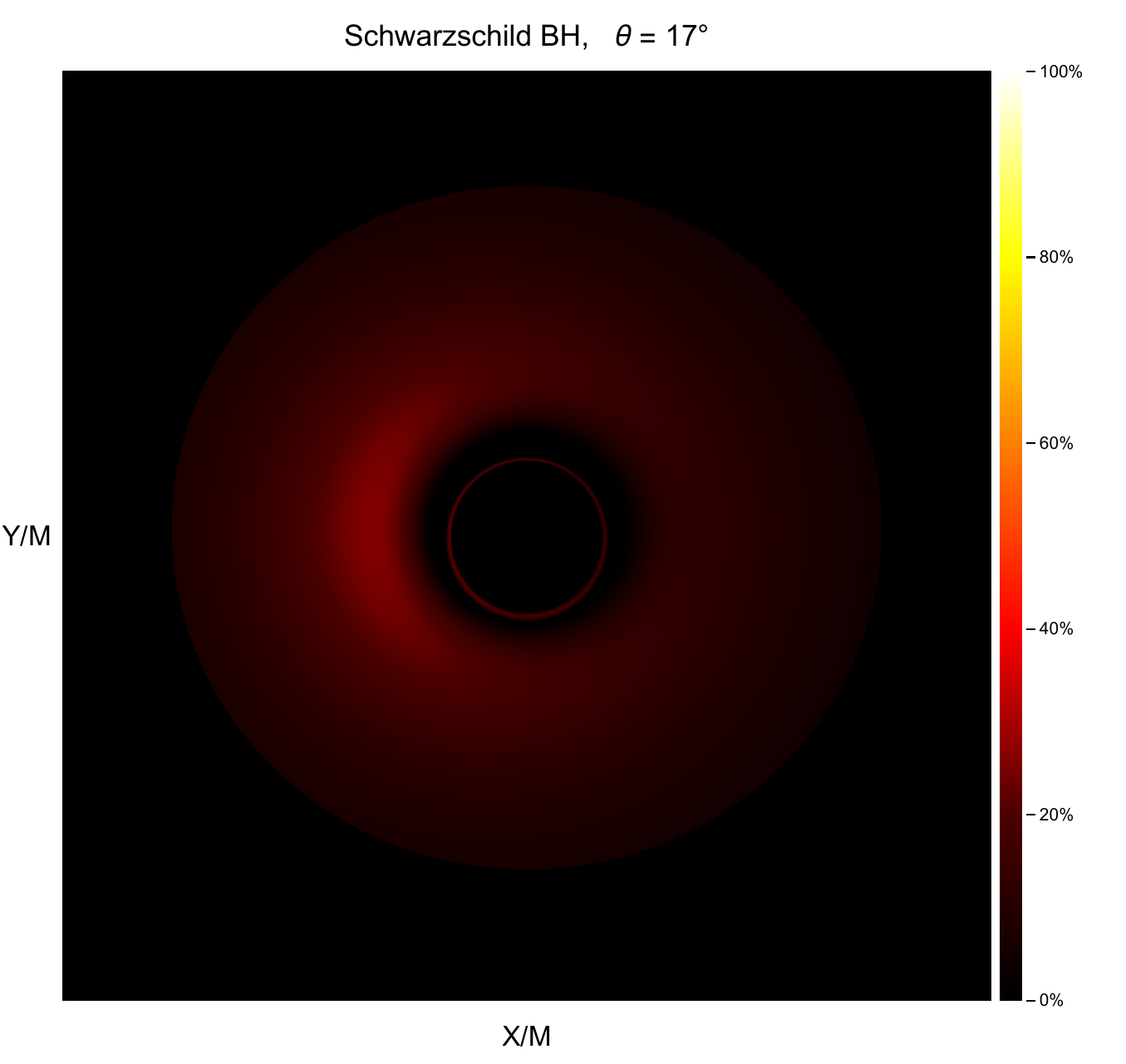}
	\includegraphics[width=5.1 cm]{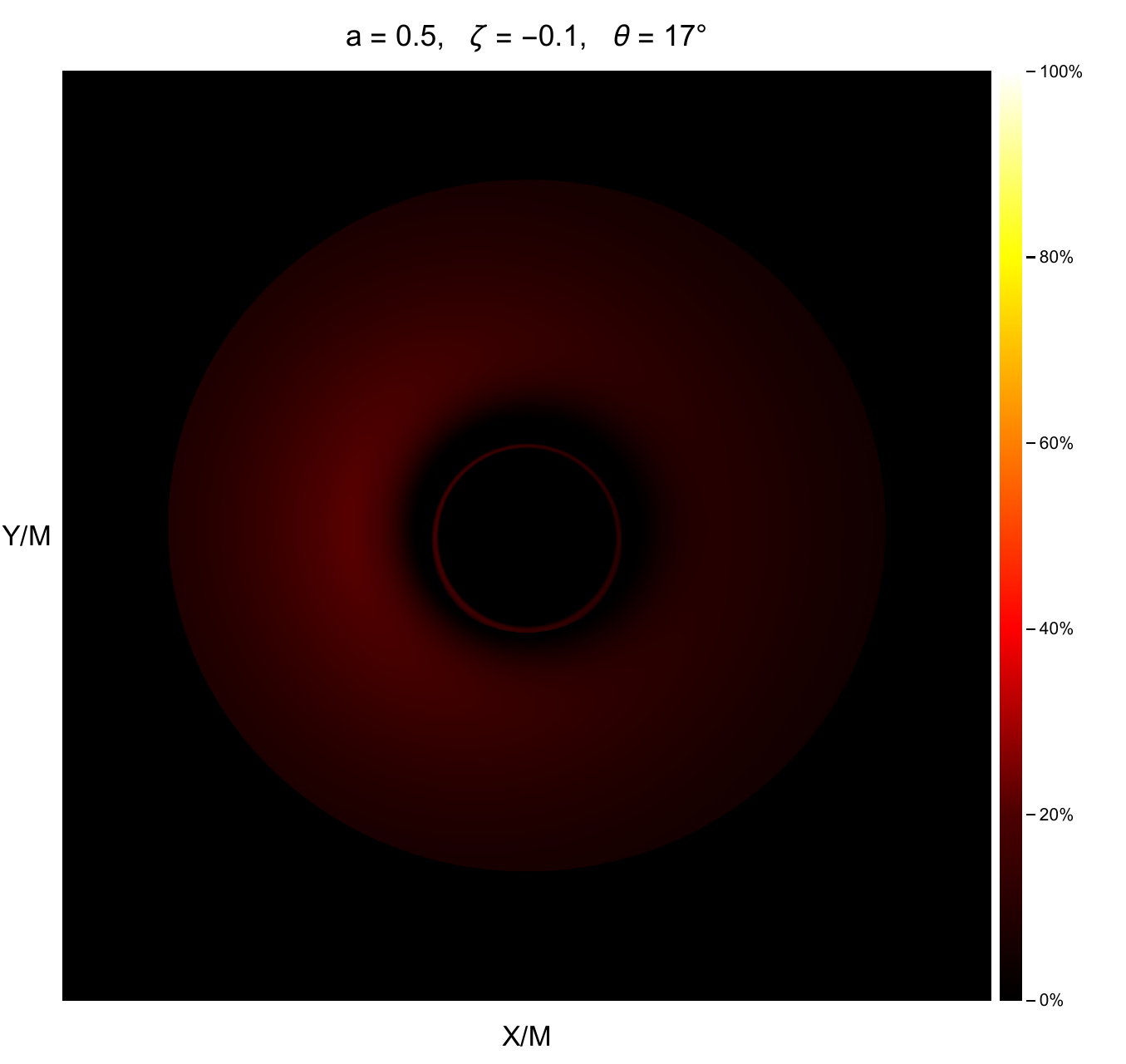}
	\includegraphics[width=5.1 cm]{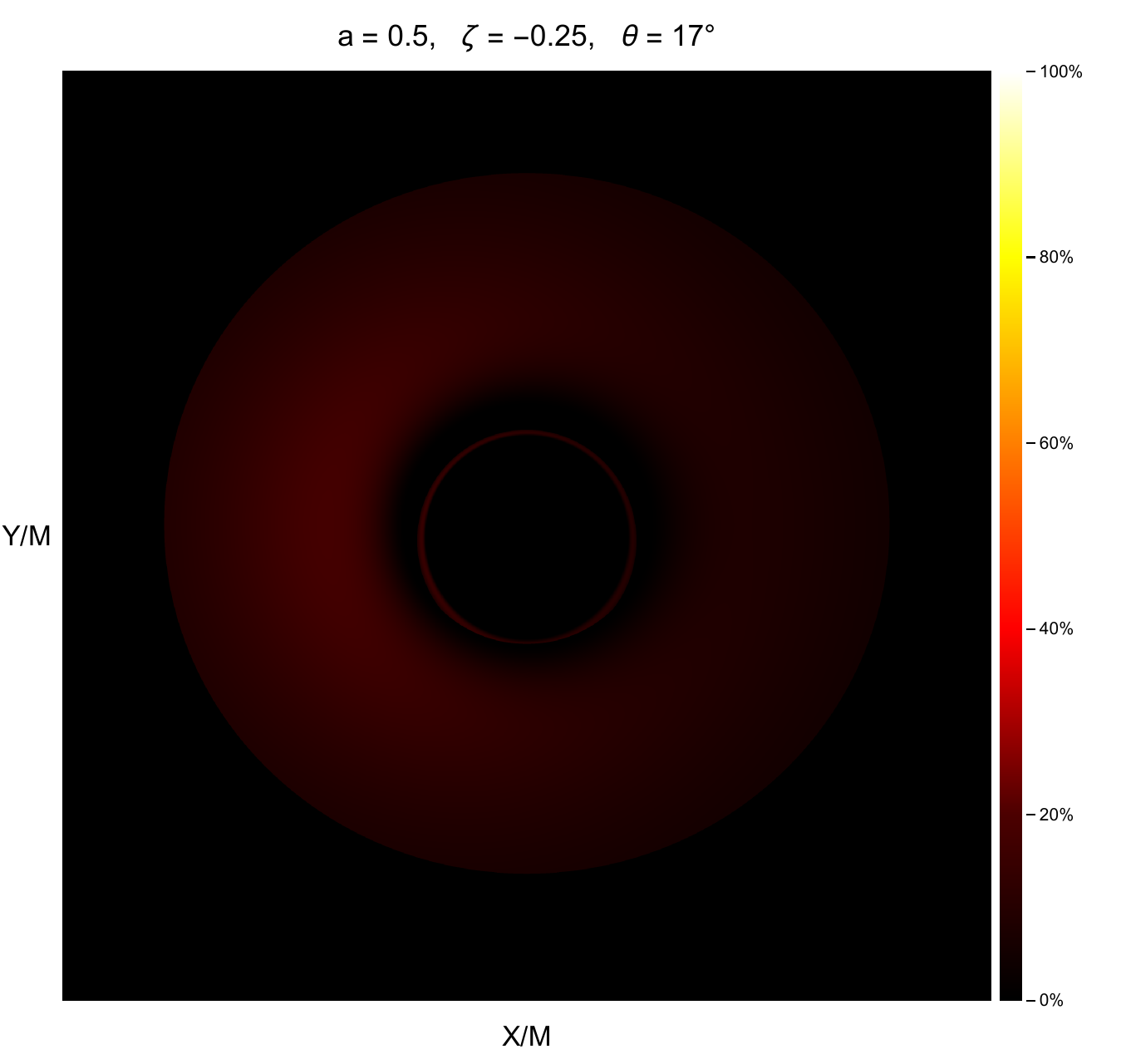}
	\includegraphics[width=5.1 cm]{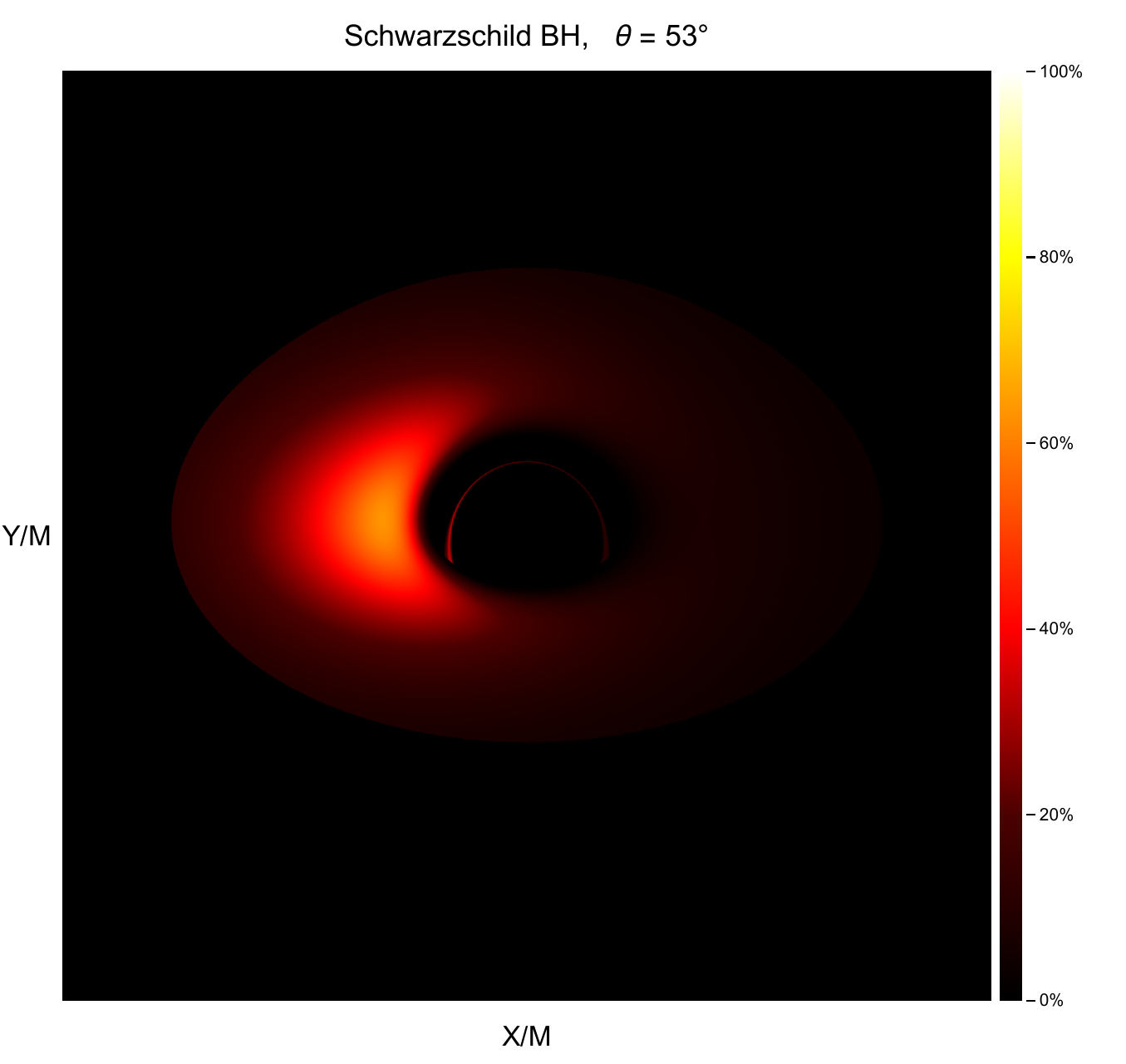}
	\includegraphics[width=5.1 cm]{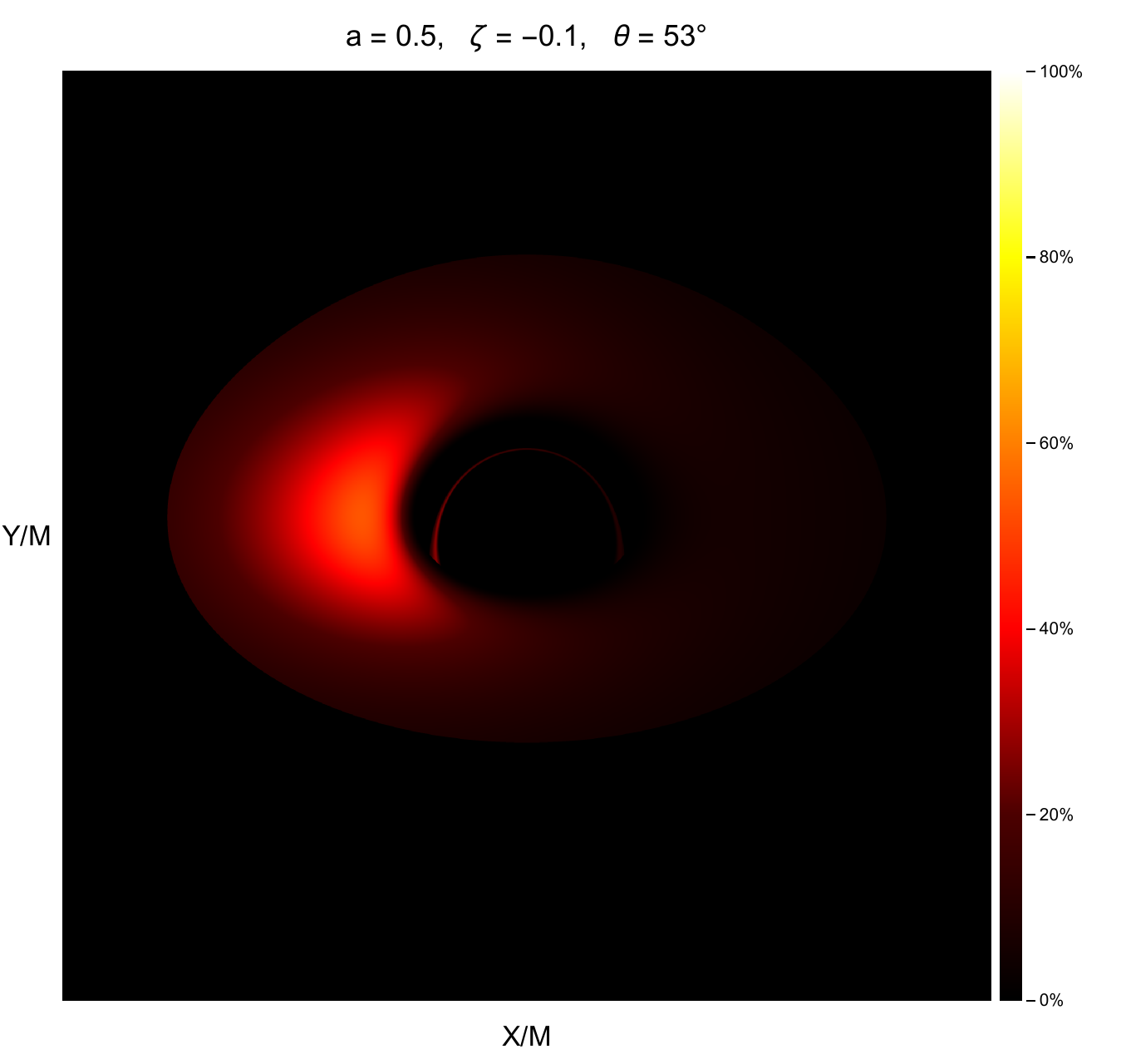}
	\includegraphics[width=5.1 cm]{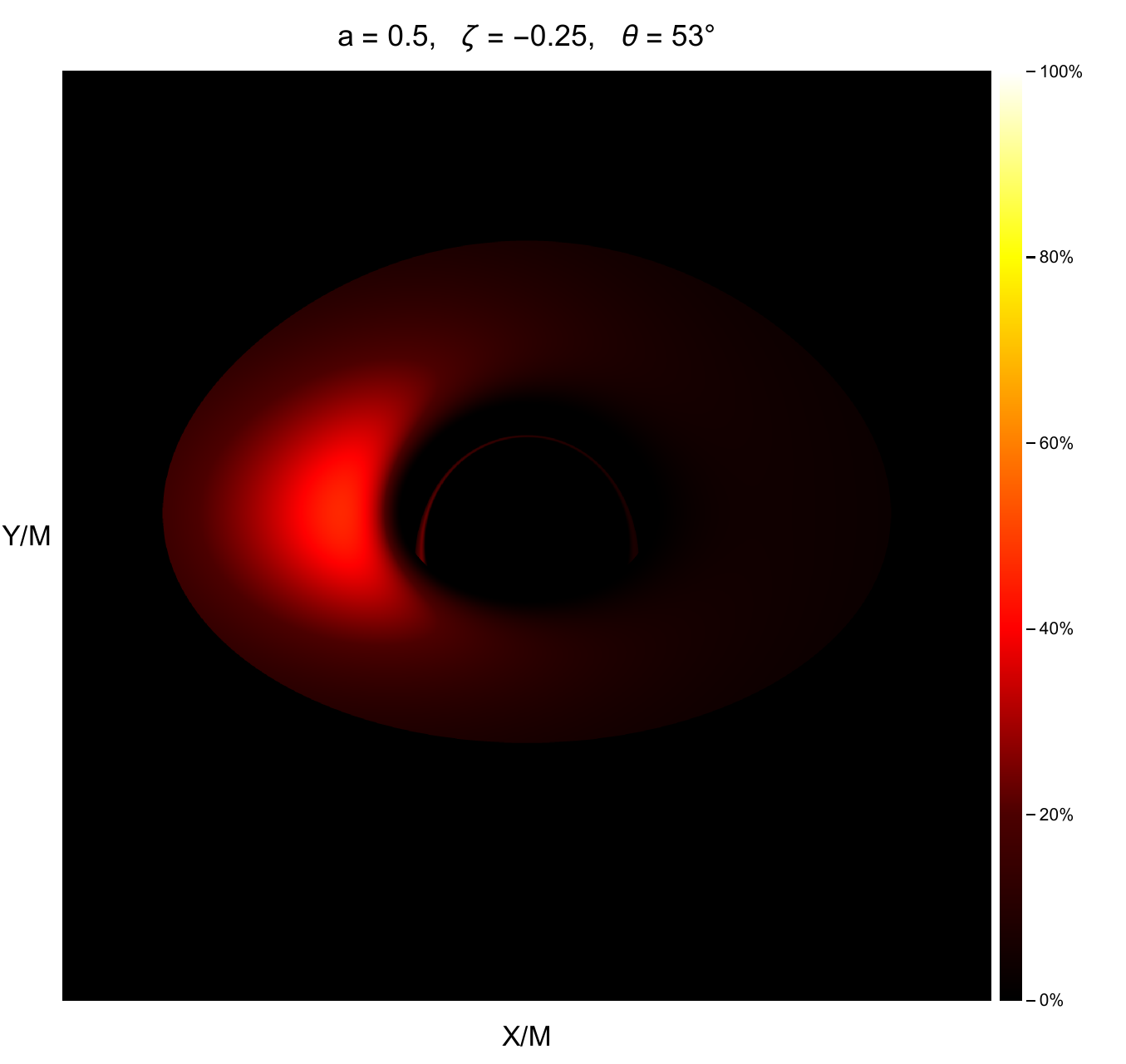}
	\includegraphics[width=5.1 cm]{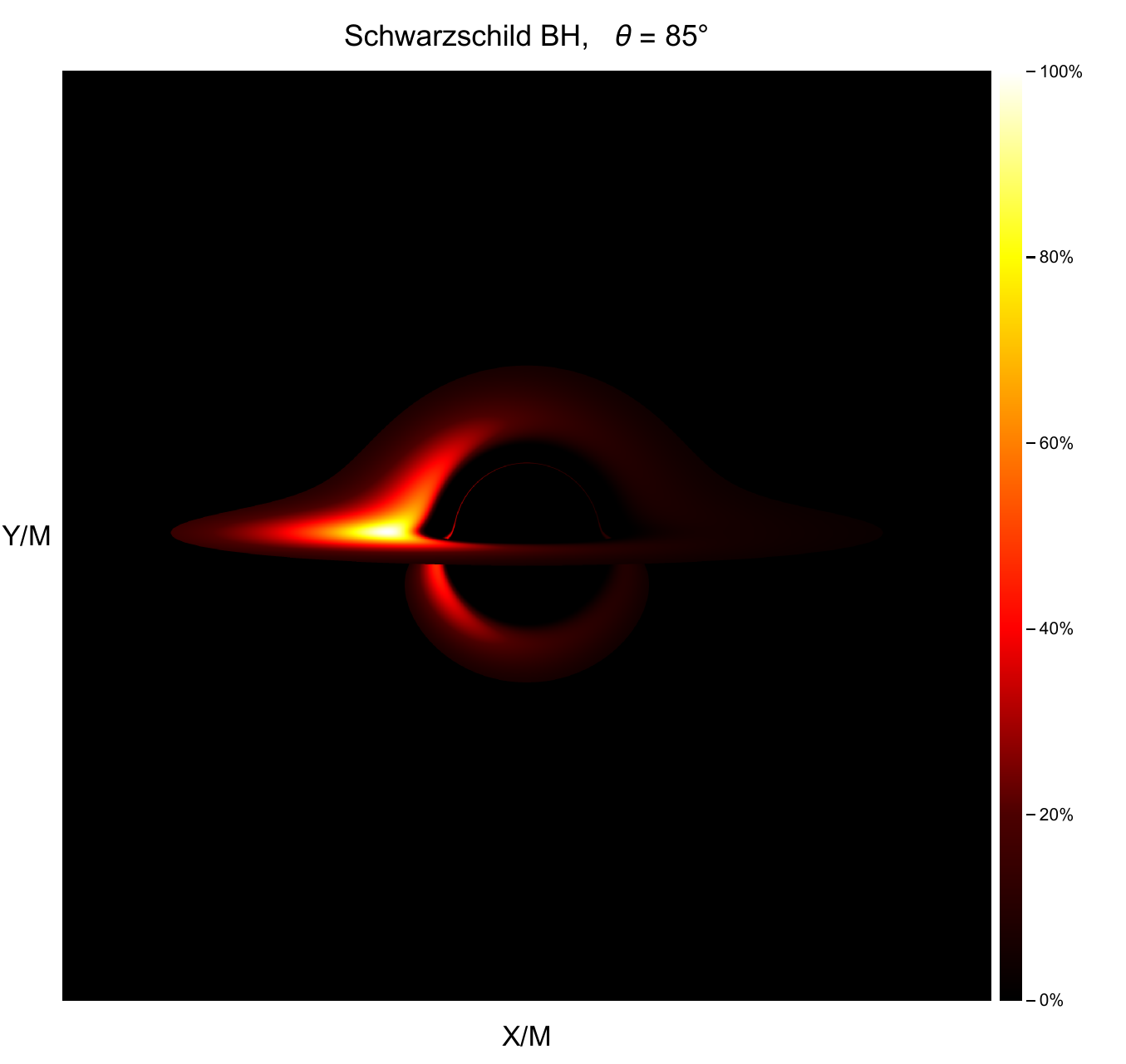}
	\includegraphics[width=5.1 cm]{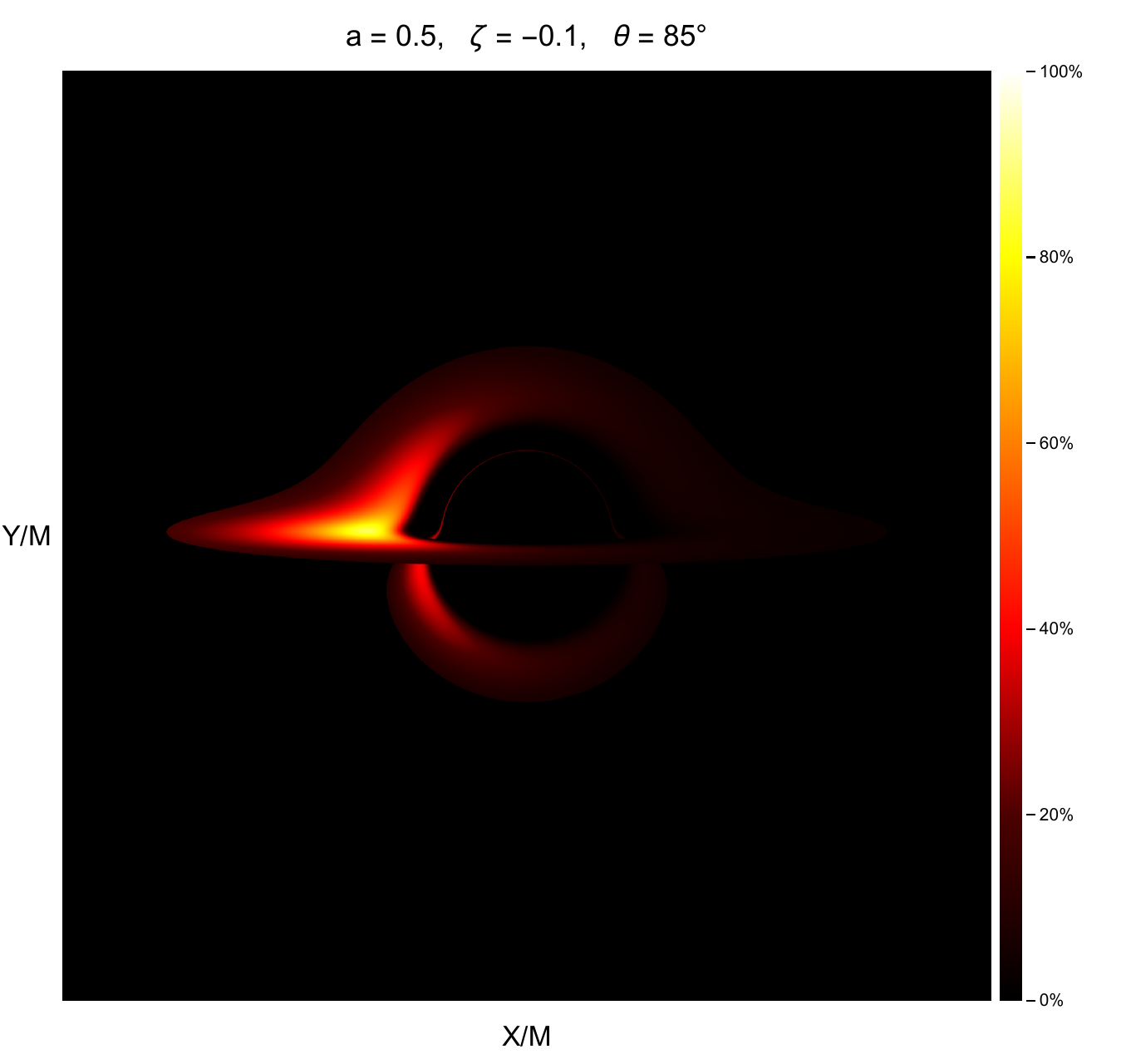}
	\includegraphics[width=5.1 cm]{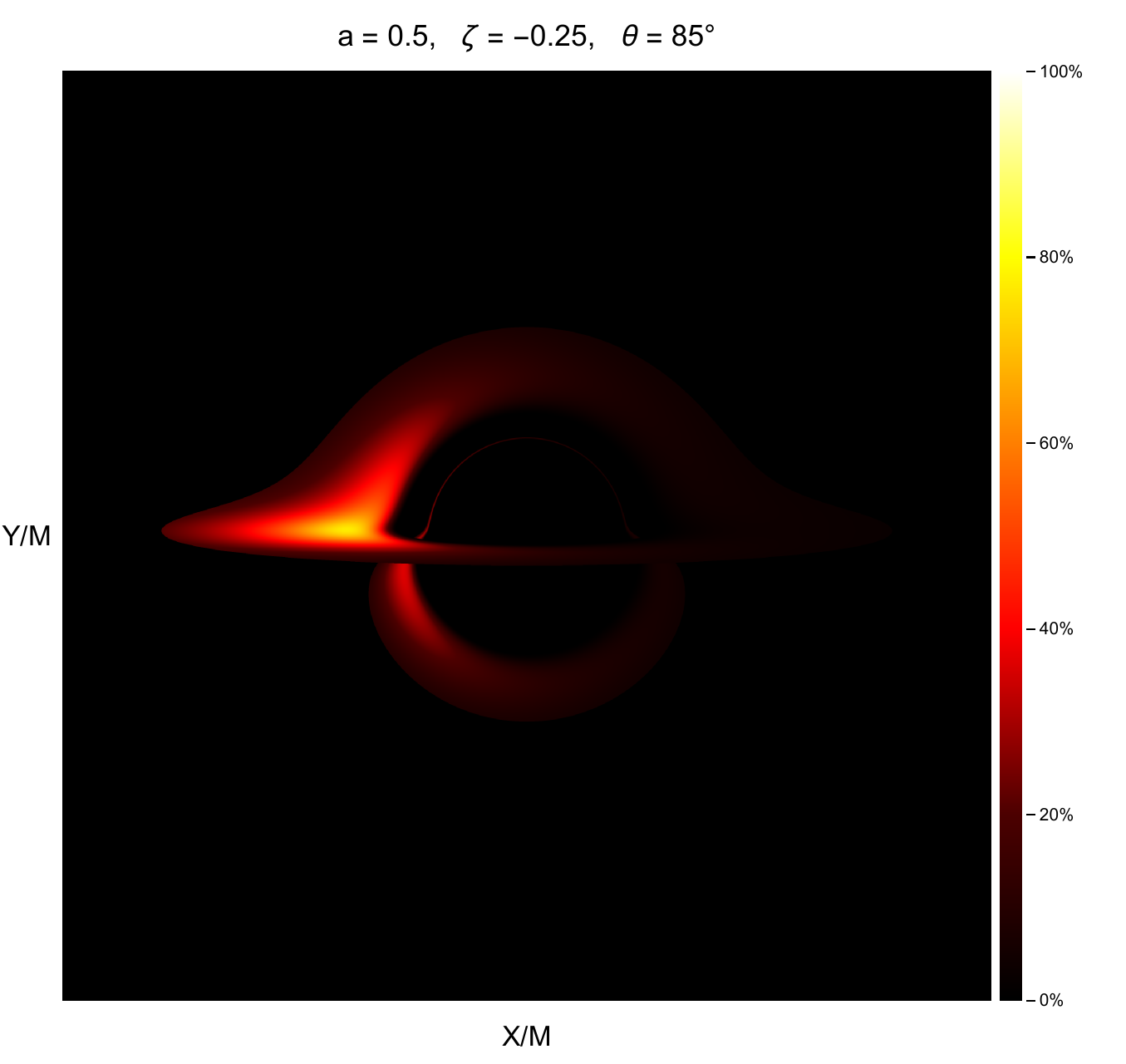}
	\caption{Observed flux \(F_{\mathrm{obs}}\) in direct and secondary images of thin accretion disks around Schwarzschild and charged-PFDM black holes, for different \(\zeta\) and inclination angles.} \label{fig:15}
\end{figure*}

\begin{figure*}
	\includegraphics[width=5.1 cm]{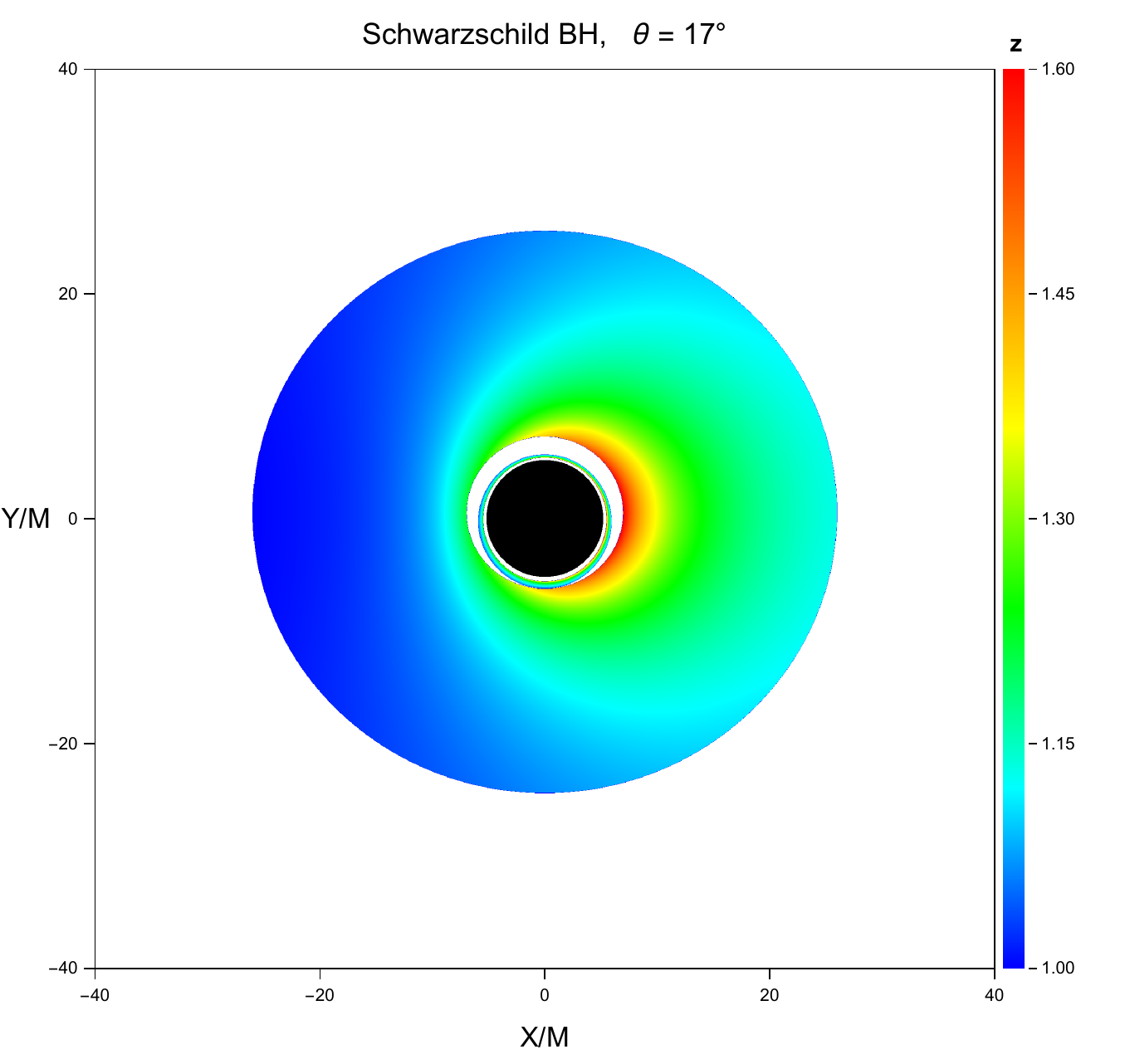}
	\includegraphics[width=5.1 cm]{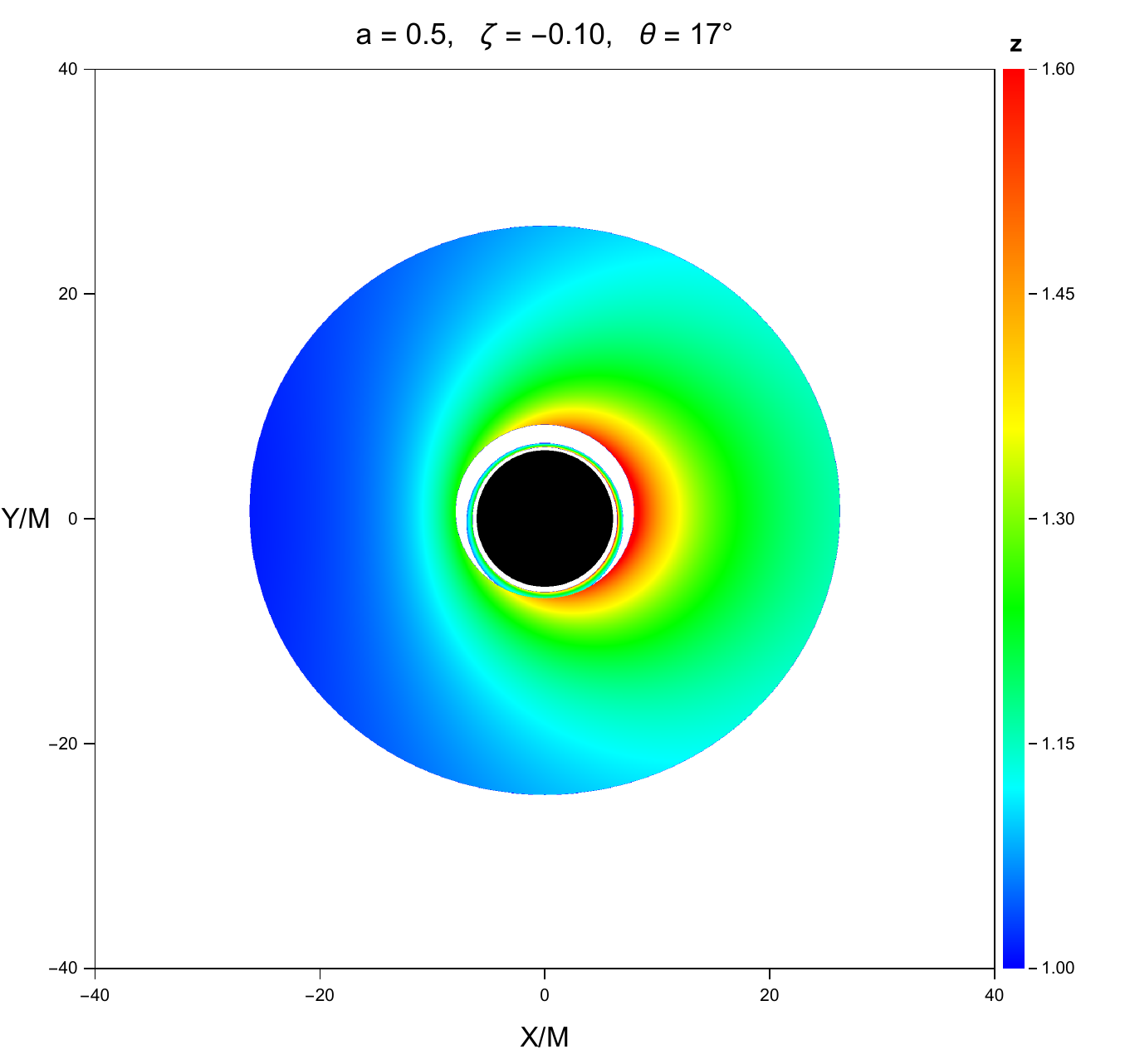}
	\includegraphics[width=5.1 cm]{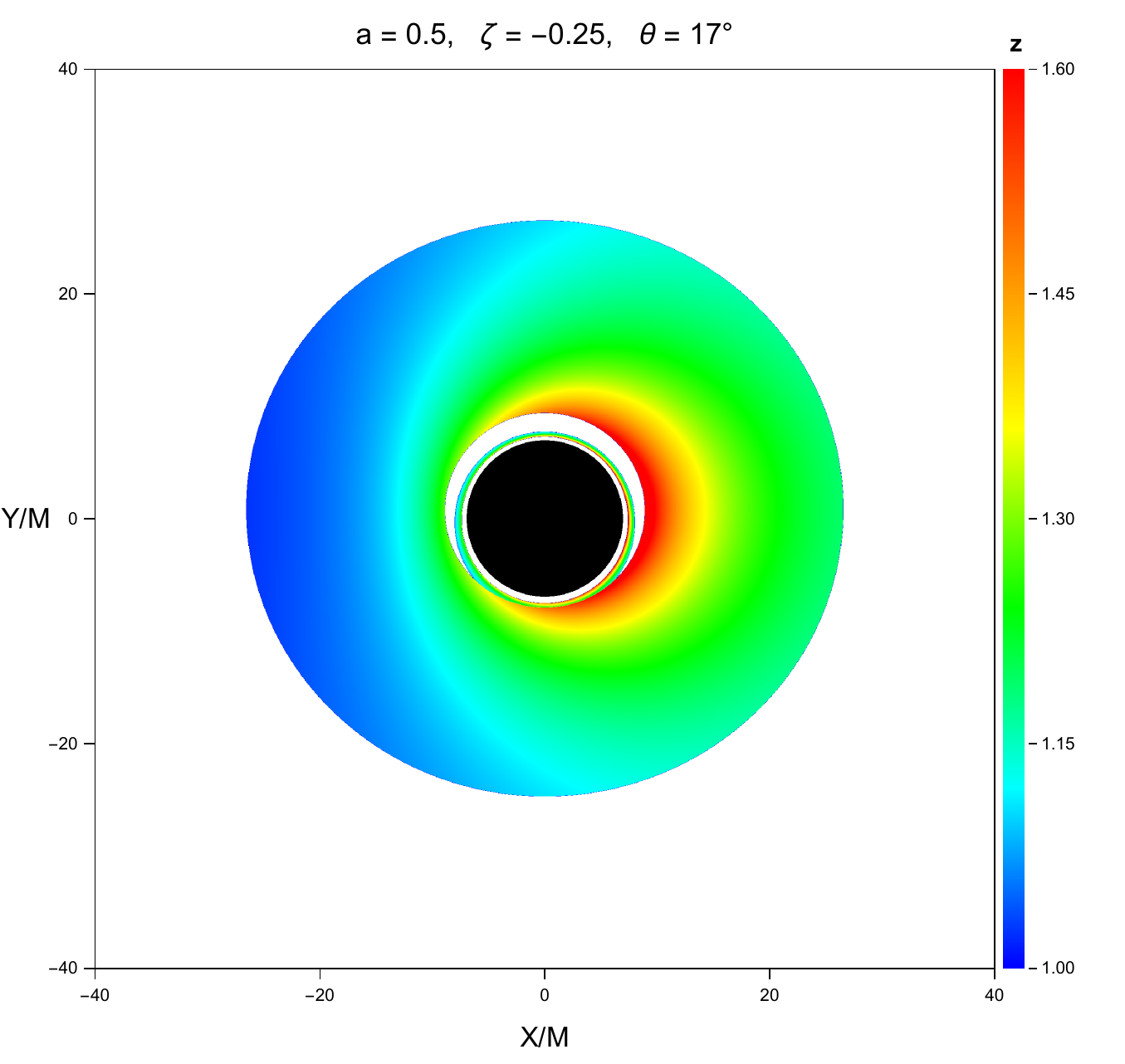}
	\includegraphics[width=5.1 cm]{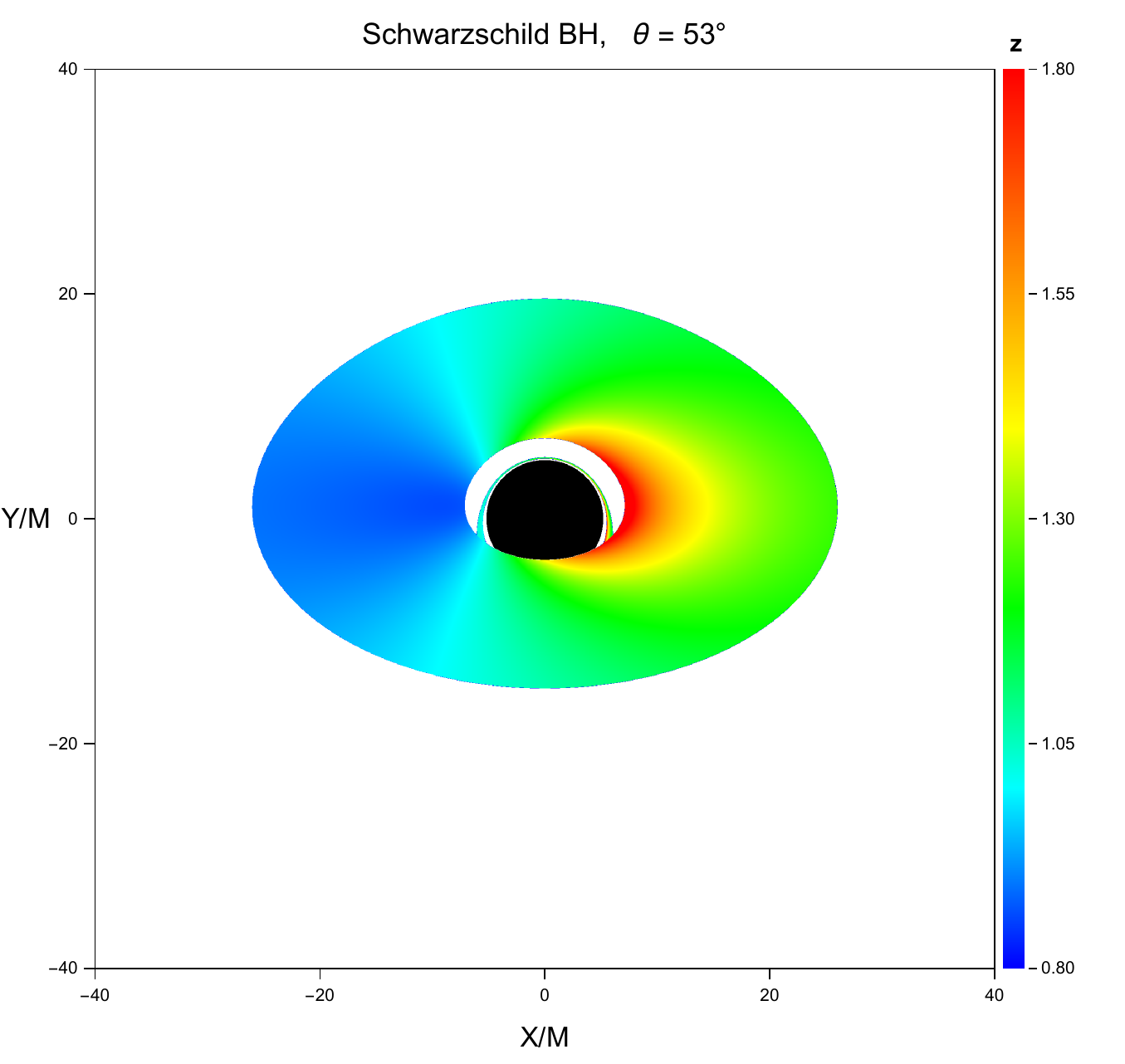}
	\includegraphics[width=5.1 cm]{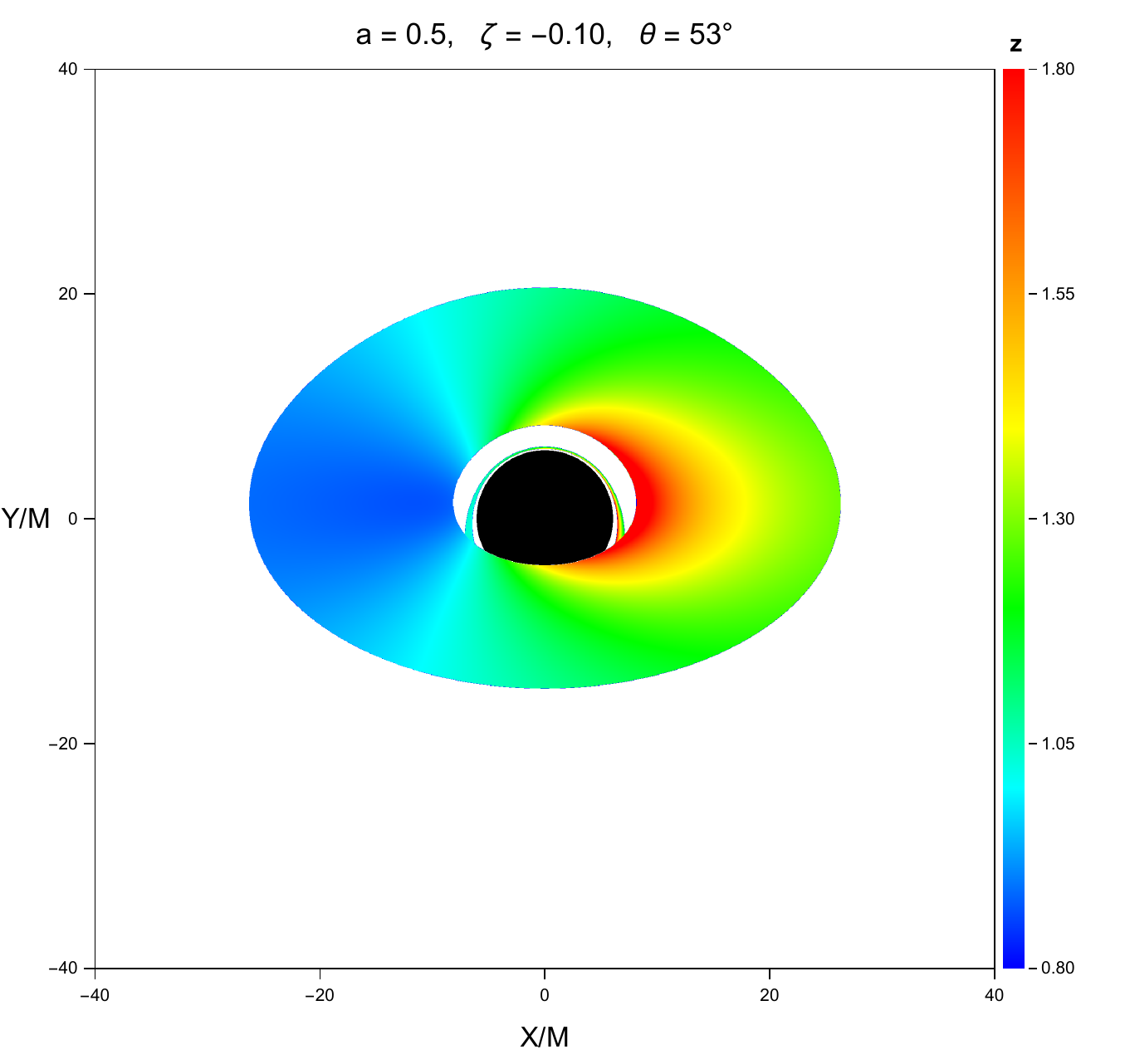}
	\includegraphics[width=5.1 cm]{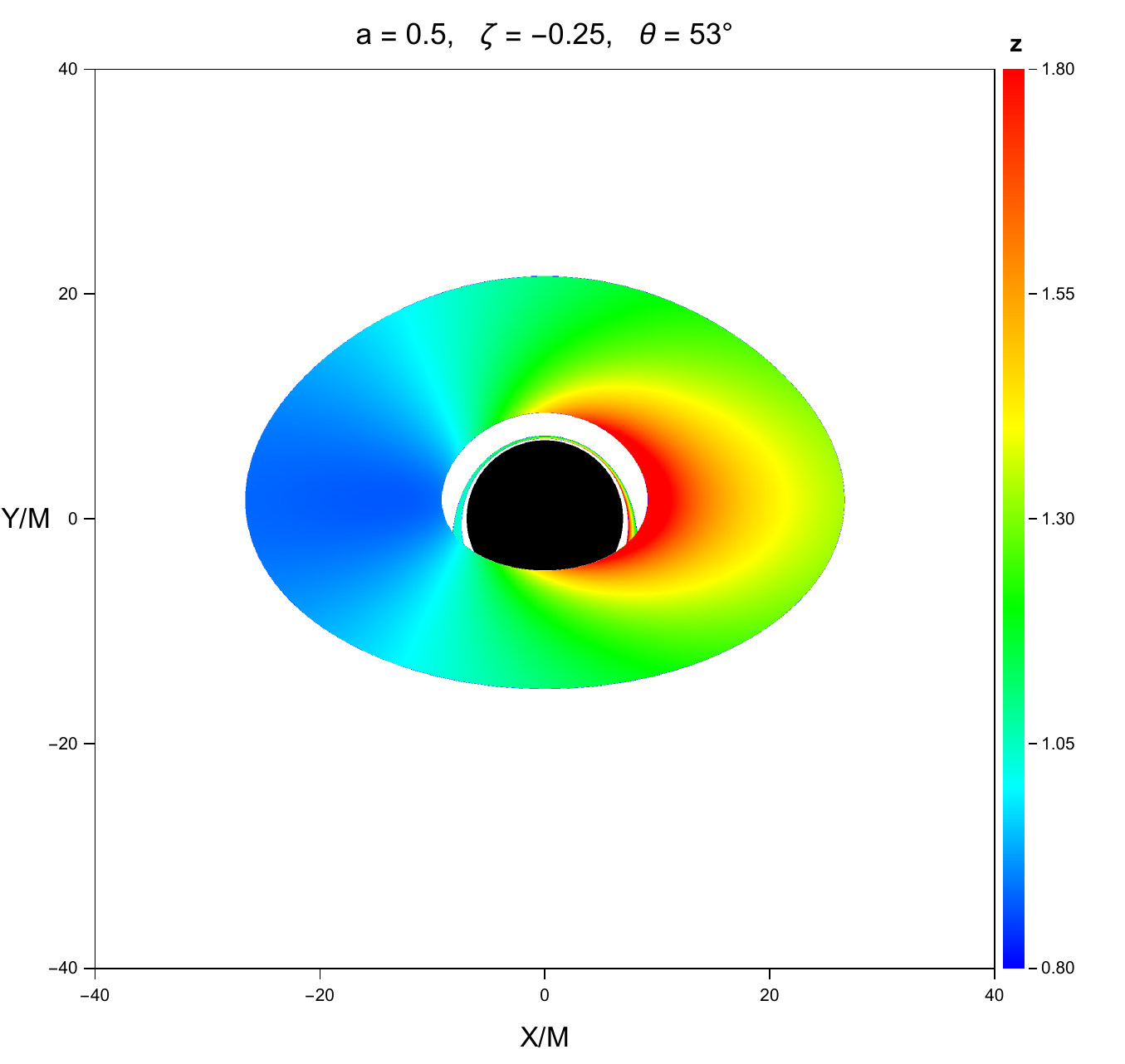}
	\includegraphics[width=5.1 cm]{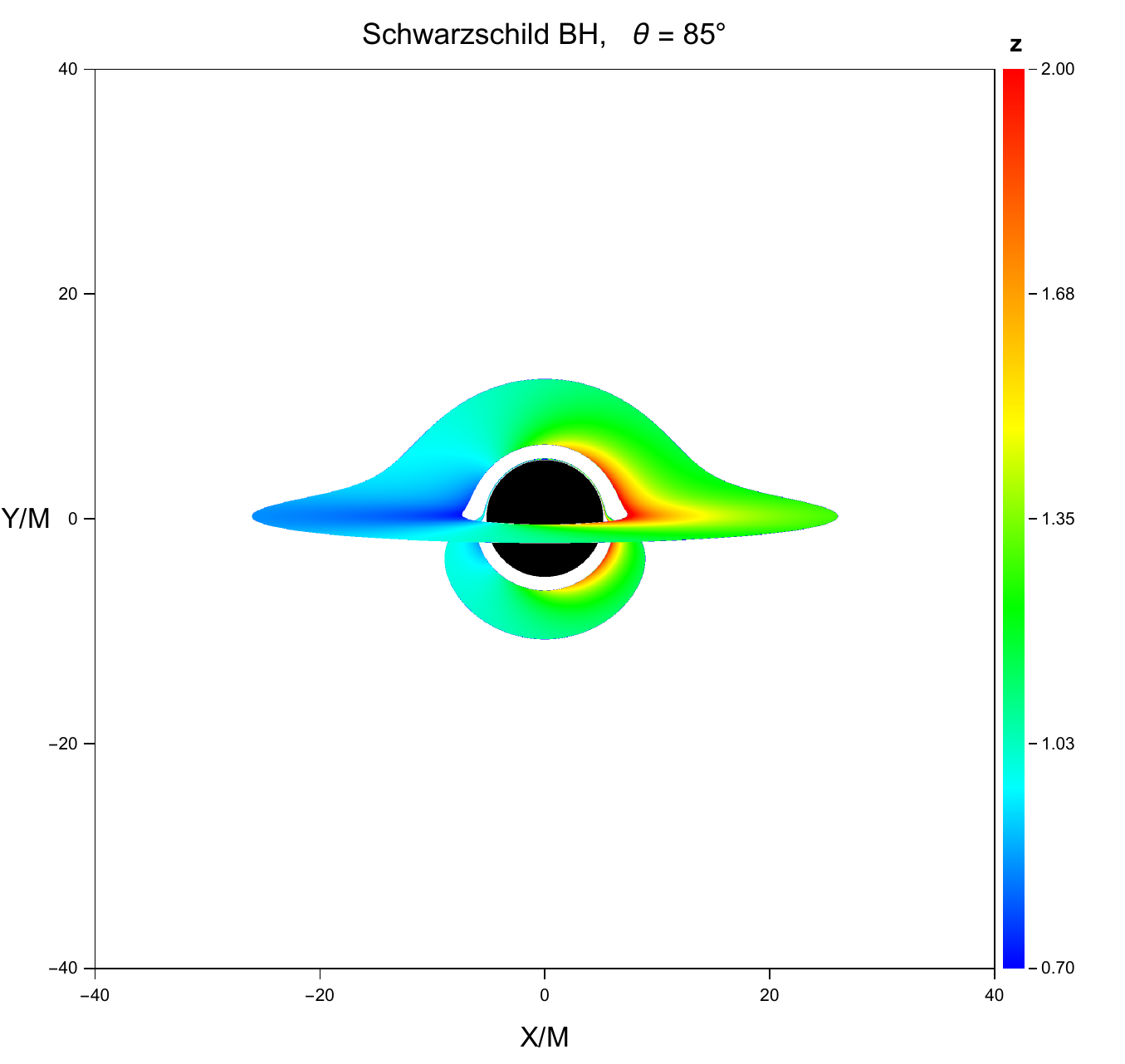}
	\includegraphics[width=5.1 cm]{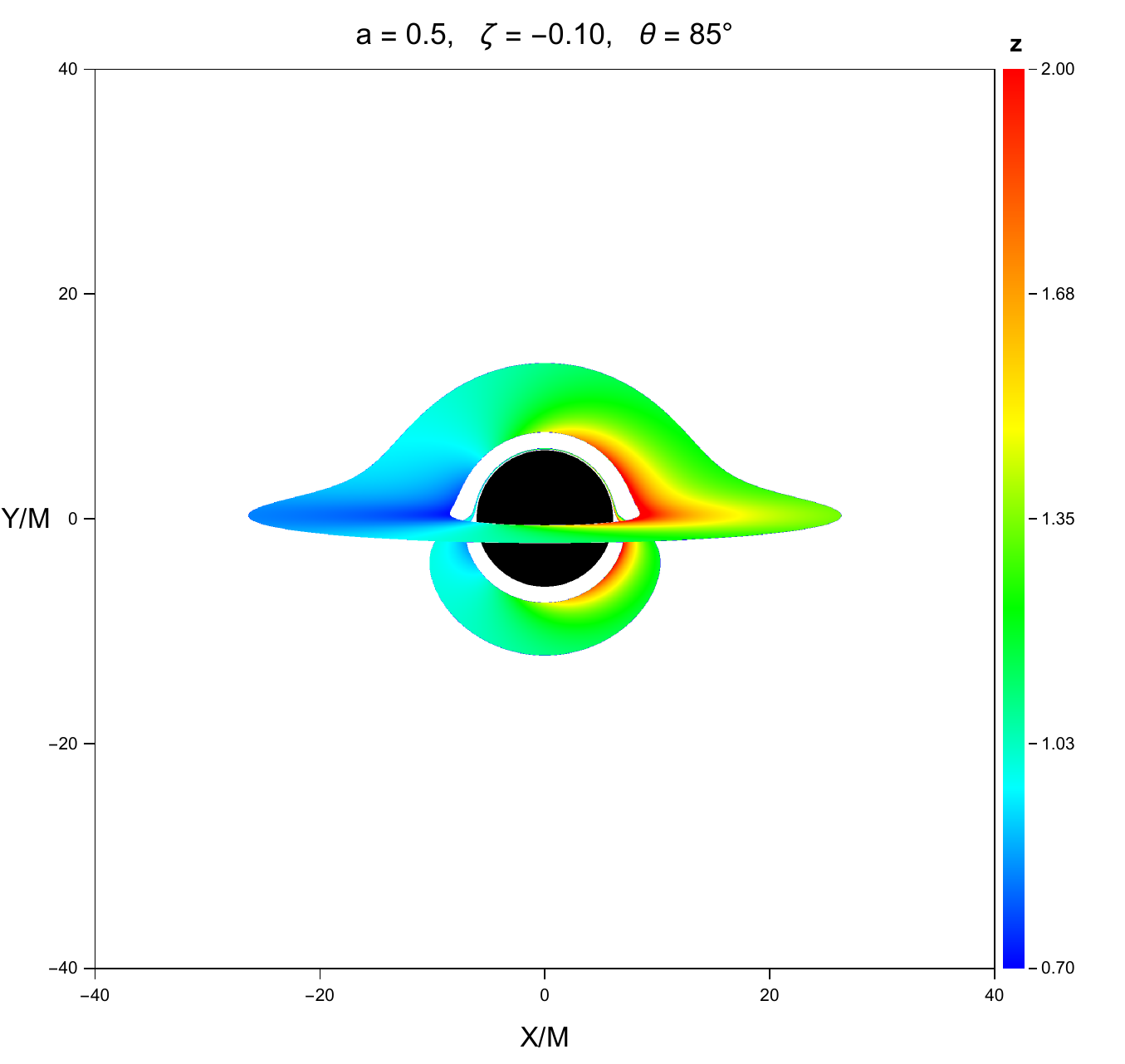}
	\includegraphics[width=5.1 cm]{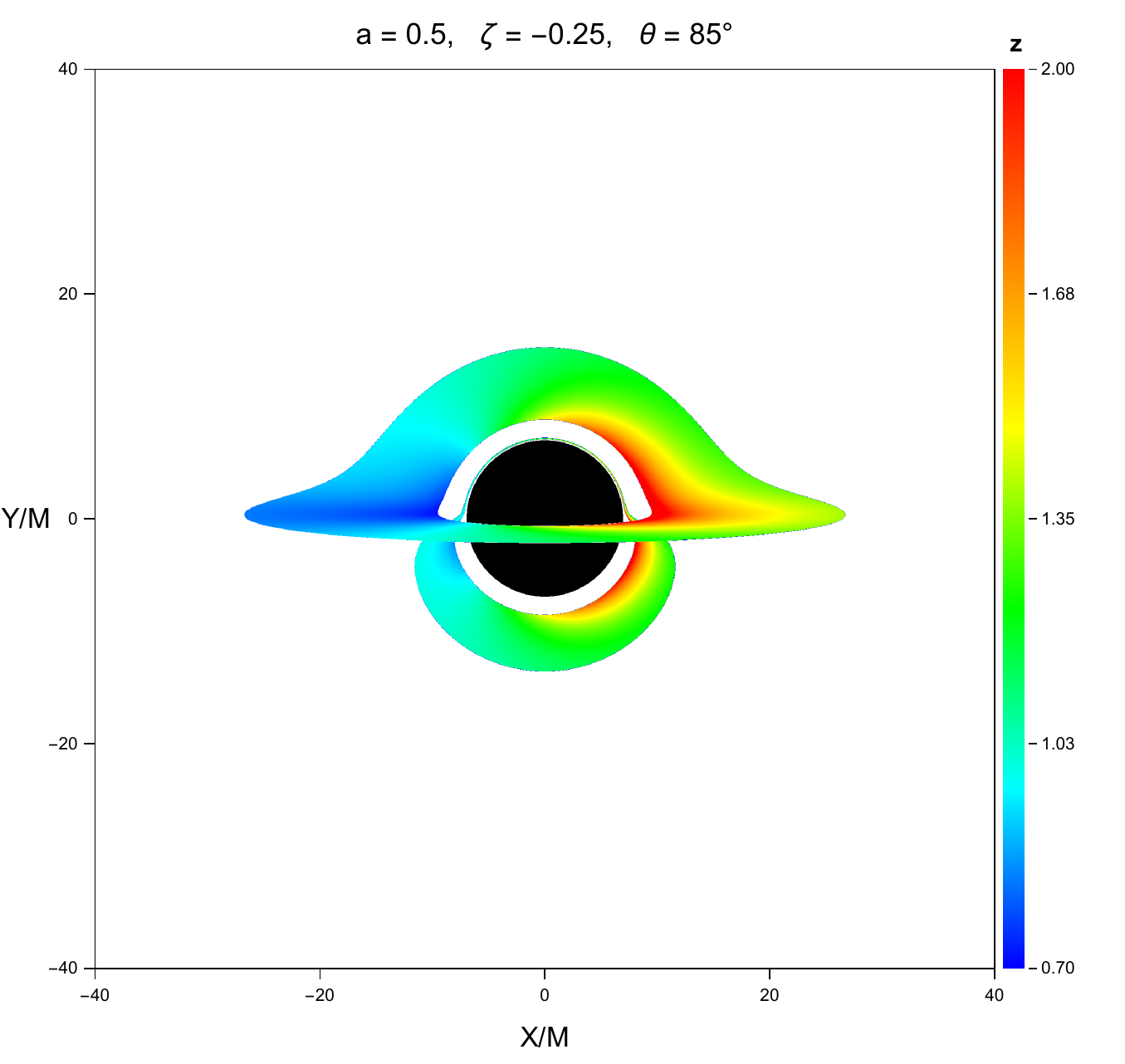}
	\caption{Redshift distributions for direct and secondary images of thin accretion disks around Schwarzschild and charged-PFDM black holes. Columns (left to right): Schwarzschild; \(a=0.5\), \(\zeta=-0.1\); \(a=0.5\), \(\zeta=-0.25\). Rows (top to bottom): \(\theta=17^\circ, 53^\circ, 85^\circ\). The radius of the central black shadow is $b_c$.} \label{fig:16}
\end{figure*}

\section{Conclusion}
\label{sec:conclusion}

This work presents a systematic investigation of accretion dynamics around static black holes endowed with a magnetic charge within a perfect fluid dark matter (PFDM) background. We first employed the shadow observations of M87* from the EHT to constrain the black hole’s magnetic charge parameter $a$ and the PFDM parameter $\zeta$. The observationally allowed parameter ranges were determined to be $0 < a \leq 1$ and $-0.25\leq \zeta<0$, thereby establishing the astrophysical viability of this model.

With the parameter space fixed, we analyzed the geodesic structure around the black hole. For massive particles (\(\varepsilon=1\)), the magnetic charge and the dark matter environment significantly modify the effective potential, the ISCO radius, and the specific energy and angular momentum: a more negative \(\zeta\) pushes the ISCO outward, whereas a larger \(a\) shifts it inward. For photons (\(\varepsilon=0\)), we derived the photon sphere radius \(r_{\mathrm{ph}}\), the critical impact parameter \(b_c\), the deflection angle \(\varphi(b)\), and the winding number \(n(b)=\varphi/(2\pi)\). As \(\lvert\zeta\rvert\) increases, \(b_c\), \(b_1^-\), \(b_2^{\pm}\), and \(b_3^{\pm}\) all shift outward, whereas the effect of \(a\) is marginal. Quantitatively, varying \(a\) from \(0.2\) to \(1\) changes \(b_c\) by less than \(4\%\), while changing \(\zeta\) from \(-0.1\) to \(-0.25\) shifts \(b_c\) by more than \(14\%\). The optical appearance of the charged-PFDM black hole is therefore predominantly governed by the dark matter parameter \(\zeta\).

Based on the Novikov‑Thorne model of a thin accretion disk, we computed the radiant energy flux, temperature distribution, radiative efficiency, direct and secondary images, redshift distribution, and observed flux. At a given radius, the local flux $F(r)$ and temperature $T(r)$ around a charged PFDM black hole are lower than those around a Schwarzschild black hole. Nevertheless, for the same accretion rate, its overall radiative efficiency $\eta^*$ and total observed luminosity are higher. This behaviour is attributed to two mechanisms: (i) a more negative $\zeta$ pushes the ISCO outward, increasing the radiating area; (ii) a larger $a$ pulls the ISCO inward, enhancing the release of gravitational energy per unit mass. Within the parameter space allowed by EHT constraints, the thin‑disk radiative efficiency of the charged PFDM black hole ($\sim 7.3\%$) lies between that of the Schwarzschild case ($\sim 6\%$) and the pure PFDM case ($\sim 20\%$). The opposite effects of the PFDM parameter $\zeta$ and the magnetic charge $a$ on the ISCO radius $\zeta$ pushes outward, $a$ pulls inward) lead to a competition in which the magnetic charge partially offsets the efficiency enhancement induced by dark matter. Consequently, the thin‑disk radiative efficiency of M87* is expected to lie in the range of $7\text{--}8\%$, a prediction that can be tested by future X‑ray/extreme‑ultraviolet continuum observations.

In terms of optical appearance, the direct and secondary images evolve from a nearly circular ring to a ``hat'' shape as the inclination angle increases. The distribution of the observed flux \(F_{\mathrm{obs}}\) shows that, for the same accretion rate, the overall brightness of the charged-PFDM black hole is lower than that in the Schwarzschild case, and this reduction becomes more pronounced as \(\lvert\zeta\rvert\) increases. The redshift distribution further indicates that, with growing inclination, the Doppler effect gradually dominates, producing a strong blueshift on the approaching side of the disk and a strong redshift on the receding side. A comparison of images for different parameters reveals that a larger \(\lvert\zeta\rvert\) leads to more significant outward expansion of the photon ring and lensing ring, as well as stronger redshift asymmetry. These features provide falsifiable theoretical predictions for future high-resolution observations, such as those by ngEHT, to distinguish a dark matter environment from the standard Schwarzschild spacetime.

\section*{Acknowledgments}
This research partly supported by the
National Natural Science Foundation of China (Grant No. 12265007) and the
Guizhou Provincial Major Scientific and Technological Program(Grant
No. XKBF(2025)010).

\newpage


\bibliographystyle{unsrt}  
\bibliography{ref1}

\begin{thebibliography}{79}%
\makeatletter
\providecommand \@ifxundefined [1]{%
 \@ifx{#1\undefined}
}%
\providecommand \@ifnum [1]{%
 \ifnum #1\expandafter \@firstoftwo
 \else \expandafter \@secondoftwo
 \fi
}%
\providecommand \@ifx [1]{%
 \ifx #1\expandafter \@firstoftwo
 \else \expandafter \@secondoftwo
 \fi
}%
\providecommand \natexlab [1]{#1}%
\providecommand \enquote  [1]{``#1''}%
\providecommand \bibnamefont  [1]{#1}%
\providecommand \bibfnamefont [1]{#1}%
\providecommand \citenamefont [1]{#1}%
\providecommand \href@noop [0]{\@secondoftwo}%
\providecommand \href [0]{\begingroup \@sanitize@url \@href}%
\providecommand \@href[1]{\@@startlink{#1}\@@href}%
\providecommand \@@href[1]{\endgroup#1\@@endlink}%
\providecommand \@sanitize@url [0]{\catcode `\\12\catcode `\$12\catcode
  `\&12\catcode `\#12\catcode `\^12\catcode `\_12\catcode `\%12\relax}%
\providecommand \@@startlink[1]{}%
\providecommand \@@endlink[0]{}%
\providecommand \url  [0]{\begingroup\@sanitize@url \@url }%
\providecommand \@url [1]{\endgroup\@href {#1}{\urlprefix }}%
\providecommand \urlprefix  [0]{URL }%
\providecommand \Eprint [0]{\href }%
\providecommand \doibase [0]{http://dx.doi.org/}%
\providecommand \selectlanguage [0]{\@gobble}%
\providecommand \bibinfo  [0]{\@secondoftwo}%
\providecommand \bibfield  [0]{\@secondoftwo}%
\providecommand \translation [1]{[#1]}%
\providecommand \BibitemOpen [0]{}%
\providecommand \bibitemStop [0]{}%
\providecommand \bibitemNoStop [0]{.\EOS\space}%
\providecommand \EOS [0]{\spacefactor3000\relax}%
\providecommand \BibitemShut  [1]{\csname bibitem#1\endcsname}%
\let\auto@bib@innerbib\@empty
\bibitem [{\citenamefont {Abbott}\ \emph {et~al.}(2016)\citenamefont {Abbott}
  \emph {et~al.}}]{LIGOScientific:2016emj}%
  \BibitemOpen
  \bibfield  {author} {\bibinfo {author} {\bibfnamefont {B.~P.}\ \bibnamefont
  {Abbott}} \emph {et~al.} (\bibinfo {collaboration} {LIGO Scientific,
  Virgo}),\ }\href {\doibase 10.1103/PhysRevLett.116.131103} {\bibfield
  {journal} {\bibinfo  {journal} {Phys. Rev. Lett.}\ }\textbf {\bibinfo
  {volume} {116}},\ \bibinfo {pages} {131103} (\bibinfo {year} {2016})},\
  \Eprint {http://arxiv.org/abs/1602.03838} {arXiv:1602.03838 [gr-qc]}
  \BibitemShut {NoStop}%
\bibitem [{\citenamefont {Akiyama}\ \emph
  {et~al.}(2019{\natexlab{a}})\citenamefont {Akiyama} \emph
  {et~al.}}]{EventHorizonTelescope:2019dse}%
  \BibitemOpen
  \bibfield  {author} {\bibinfo {author} {\bibfnamefont {K.}~\bibnamefont
  {Akiyama}} \emph {et~al.} (\bibinfo {collaboration} {Event Horizon
  Telescope}),\ }\href {\doibase 10.3847/2041-8213/ab0ec7} {\bibfield
  {journal} {\bibinfo  {journal} {Astrophys. J. Lett.}\ }\textbf {\bibinfo
  {volume} {875}},\ \bibinfo {pages} {L1} (\bibinfo {year}
  {2019}{\natexlab{a}})},\ \Eprint {http://arxiv.org/abs/1906.11238}
  {arXiv:1906.11238 [astro-ph.GA]} \BibitemShut {NoStop}%
\bibitem [{\citenamefont {Akiyama}\ \emph
  {et~al.}(2019{\natexlab{b}})\citenamefont {Akiyama} \emph
  {et~al.}}]{EventHorizonTelescope:2019ths}%
  \BibitemOpen
  \bibfield  {author} {\bibinfo {author} {\bibfnamefont {K.}~\bibnamefont
  {Akiyama}} \emph {et~al.} (\bibinfo {collaboration} {Event Horizon
  Telescope}),\ }\href {\doibase 10.3847/2041-8213/ab0e85} {\bibfield
  {journal} {\bibinfo  {journal} {Astrophys. J. Lett.}\ }\textbf {\bibinfo
  {volume} {875}},\ \bibinfo {pages} {L4} (\bibinfo {year}
  {2019}{\natexlab{b}})},\ \Eprint {http://arxiv.org/abs/1906.11241}
  {arXiv:1906.11241 [astro-ph.GA]} \BibitemShut {NoStop}%
\bibitem [{\citenamefont {Gibbons}(2001)}]{Gibbons:2001gy}%
  \BibitemOpen
  \bibfield  {author} {\bibinfo {author} {\bibfnamefont {G.~W.}\ \bibnamefont
  {Gibbons}},\ }\href {\doibase 10.1063/1.1419338} {\bibfield  {journal}
  {\bibinfo  {journal} {AIP Conf. Proc.}\ }\textbf {\bibinfo {volume} {589}},\
  \bibinfo {pages} {324} (\bibinfo {year} {2001})},\ \Eprint
  {http://arxiv.org/abs/hep-th/0106059} {arXiv:hep-th/0106059} \BibitemShut
  {NoStop}%
\bibitem [{\citenamefont {Born}(1934)}]{Born:1934ji}%
  \BibitemOpen
  \bibfield  {author} {\bibinfo {author} {\bibfnamefont {M.}~\bibnamefont
  {Born}},\ }\href {\doibase 10.1098/rspa.1934.0010} {\bibfield  {journal}
  {\bibinfo  {journal} {Proc. Roy. Soc. Lond. A}\ }\textbf {\bibinfo {volume}
  {143}},\ \bibinfo {pages} {410} (\bibinfo {year} {1934})}\BibitemShut
  {NoStop}%
\bibitem [{\citenamefont {Born}\ and\ \citenamefont
  {Infeld}(1934)}]{Born:1934gh}%
  \BibitemOpen
  \bibfield  {author} {\bibinfo {author} {\bibfnamefont {M.}~\bibnamefont
  {Born}}\ and\ \bibinfo {author} {\bibfnamefont {L.}~\bibnamefont {Infeld}},\
  }\href {\doibase 10.1098/rspa.1934.0059} {\bibfield  {journal} {\bibinfo
  {journal} {Proc. Roy. Soc. Lond. A}\ }\textbf {\bibinfo {volume} {144}},\
  \bibinfo {pages} {425} (\bibinfo {year} {1934})}\BibitemShut {NoStop}%
\bibitem [{\citenamefont {Gunasekaran}\ \emph {et~al.}(2012)\citenamefont
  {Gunasekaran}, \citenamefont {Mann},\ and\ \citenamefont
  {Kubiznak}}]{Gunasekaran:2012dq}%
  \BibitemOpen
  \bibfield  {author} {\bibinfo {author} {\bibfnamefont {S.}~\bibnamefont
  {Gunasekaran}}, \bibinfo {author} {\bibfnamefont {R.~B.}\ \bibnamefont
  {Mann}}, \ and\ \bibinfo {author} {\bibfnamefont {D.}~\bibnamefont
  {Kubiznak}},\ }\href {\doibase 10.1007/JHEP11(2012)110} {\bibfield  {journal}
  {\bibinfo  {journal} {JHEP}\ }\textbf {\bibinfo {volume} {11}},\ \bibinfo
  {pages} {110} (\bibinfo {year} {2012})},\ \Eprint
  {http://arxiv.org/abs/1208.6251} {arXiv:1208.6251 [hep-th]} \BibitemShut
  {NoStop}%
\bibitem [{\citenamefont {Bronnikov}(2001)}]{Bronnikov:2000vy}%
  \BibitemOpen
  \bibfield  {author} {\bibinfo {author} {\bibfnamefont {K.~A.}\ \bibnamefont
  {Bronnikov}},\ }\href {\doibase 10.1103/PhysRevD.63.044005} {\bibfield
  {journal} {\bibinfo  {journal} {Phys. Rev. D}\ }\textbf {\bibinfo {volume}
  {63}},\ \bibinfo {pages} {044005} (\bibinfo {year} {2001})},\ \Eprint
  {http://arxiv.org/abs/gr-qc/0006014} {arXiv:gr-qc/0006014} \BibitemShut
  {NoStop}%
\bibitem [{\citenamefont {Bronnikov}(2018)}]{Bronnikov:2017sgg}%
  \BibitemOpen
  \bibfield  {author} {\bibinfo {author} {\bibfnamefont {K.~A.}\ \bibnamefont
  {Bronnikov}},\ }\href {\doibase 10.1142/S0218271818410055} {\bibfield
  {journal} {\bibinfo  {journal} {Int. J. Mod. Phys. D}\ }\textbf {\bibinfo
  {volume} {27}},\ \bibinfo {pages} {1841005} (\bibinfo {year} {2018})},\
  \Eprint {http://arxiv.org/abs/1711.00087} {arXiv:1711.00087 [gr-qc]}
  \BibitemShut {NoStop}%
\bibitem [{\citenamefont {Rincon}\ \emph {et~al.}(2021)\citenamefont {Rincon},
  \citenamefont {Contreras}, \citenamefont {Bargueno}, \citenamefont {Koch},\
  and\ \citenamefont {Panotopoulos}}]{Rincon:2021hjj}%
  \BibitemOpen
  \bibfield  {author} {\bibinfo {author} {\bibfnamefont {A.}~\bibnamefont
  {Rincon}}, \bibinfo {author} {\bibfnamefont {E.}~\bibnamefont {Contreras}},
  \bibinfo {author} {\bibfnamefont {P.}~\bibnamefont {Bargueno}}, \bibinfo
  {author} {\bibfnamefont {B.}~\bibnamefont {Koch}}, \ and\ \bibinfo {author}
  {\bibfnamefont {G.}~\bibnamefont {Panotopoulos}},\ }\href {\doibase
  10.1016/j.dark.2021.100783} {\bibfield  {journal} {\bibinfo  {journal} {Phys.
  Dark Univ.}\ }\textbf {\bibinfo {volume} {31}},\ \bibinfo {pages} {100783}
  (\bibinfo {year} {2021})},\ \Eprint {http://arxiv.org/abs/2102.02426}
  {arXiv:2102.02426 [gr-qc]} \BibitemShut {NoStop}%
\bibitem [{\citenamefont {Bronnikov}(2022)}]{Bronnikov:2022ofk}%
  \BibitemOpen
  \bibfield  {author} {\bibinfo {author} {\bibfnamefont {K.~A.}\ \bibnamefont
  {Bronnikov}},\ }\href@noop {} {\  (\bibinfo {year} {2022})},\ \Eprint
  {http://arxiv.org/abs/2211.00743} {arXiv:2211.00743 [gr-qc]} \BibitemShut
  {NoStop}%
\bibitem [{\citenamefont {Fradkin}\ and\ \citenamefont
  {Tseytlin}(1985)}]{Fradkin:1985qd}%
  \BibitemOpen
  \bibfield  {author} {\bibinfo {author} {\bibfnamefont {E.~S.}\ \bibnamefont
  {Fradkin}}\ and\ \bibinfo {author} {\bibfnamefont {A.~A.}\ \bibnamefont
  {Tseytlin}},\ }\href {\doibase 10.1016/0370-2693(85)90205-9} {\bibfield
  {journal} {\bibinfo  {journal} {Phys. Lett. B}\ }\textbf {\bibinfo {volume}
  {163}},\ \bibinfo {pages} {123} (\bibinfo {year} {1985})}\BibitemShut
  {NoStop}%
\bibitem [{\citenamefont {Seiberg}\ and\ \citenamefont
  {Witten}(1999)}]{Seiberg:1999vs}%
  \BibitemOpen
  \bibfield  {author} {\bibinfo {author} {\bibfnamefont {N.}~\bibnamefont
  {Seiberg}}\ and\ \bibinfo {author} {\bibfnamefont {E.}~\bibnamefont
  {Witten}},\ }\href {\doibase 10.1088/1126-6708/1999/09/032} {\bibfield
  {journal} {\bibinfo  {journal} {JHEP}\ }\textbf {\bibinfo {volume} {09}},\
  \bibinfo {pages} {032} (\bibinfo {year} {1999})},\ \Eprint
  {http://arxiv.org/abs/hep-th/9908142} {arXiv:hep-th/9908142} \BibitemShut
  {NoStop}%
\bibitem [{\citenamefont {De~Lorenci}\ \emph {et~al.}(2002)\citenamefont
  {De~Lorenci}, \citenamefont {Klippert}, \citenamefont {Novello},\ and\
  \citenamefont {Salim}}]{DeLorenci:2002mi}%
  \BibitemOpen
  \bibfield  {author} {\bibinfo {author} {\bibfnamefont {V.~A.}\ \bibnamefont
  {De~Lorenci}}, \bibinfo {author} {\bibfnamefont {R.}~\bibnamefont
  {Klippert}}, \bibinfo {author} {\bibfnamefont {M.}~\bibnamefont {Novello}}, \
  and\ \bibinfo {author} {\bibfnamefont {J.~M.}\ \bibnamefont {Salim}},\ }\href
  {\doibase 10.1103/PhysRevD.65.063501} {\bibfield  {journal} {\bibinfo
  {journal} {Phys. Rev. D}\ }\textbf {\bibinfo {volume} {65}},\ \bibinfo
  {pages} {063501} (\bibinfo {year} {2002})}\BibitemShut {NoStop}%
\bibitem [{\citenamefont {Novello}\ \emph {et~al.}(2004)\citenamefont
  {Novello}, \citenamefont {Perez~Bergliaffa},\ and\ \citenamefont
  {Salim}}]{Novello:2003kh}%
  \BibitemOpen
  \bibfield  {author} {\bibinfo {author} {\bibfnamefont {M.}~\bibnamefont
  {Novello}}, \bibinfo {author} {\bibfnamefont {S.~E.}\ \bibnamefont
  {Perez~Bergliaffa}}, \ and\ \bibinfo {author} {\bibfnamefont
  {J.}~\bibnamefont {Salim}},\ }\href {\doibase 10.1103/PhysRevD.69.127301}
  {\bibfield  {journal} {\bibinfo  {journal} {Phys. Rev. D}\ }\textbf {\bibinfo
  {volume} {69}},\ \bibinfo {pages} {127301} (\bibinfo {year} {2004})},\
  \Eprint {http://arxiv.org/abs/astro-ph/0312093} {arXiv:astro-ph/0312093}
  \BibitemShut {NoStop}%
\bibitem [{\citenamefont {Novello}\ and\ \citenamefont
  {Bergliaffa}(2008)}]{Novello:2008ra}%
  \BibitemOpen
  \bibfield  {author} {\bibinfo {author} {\bibfnamefont {M.}~\bibnamefont
  {Novello}}\ and\ \bibinfo {author} {\bibfnamefont {S.~E.~P.}\ \bibnamefont
  {Bergliaffa}},\ }\href {\doibase 10.1016/j.physrep.2008.04.006} {\bibfield
  {journal} {\bibinfo  {journal} {Phys. Rept.}\ }\textbf {\bibinfo {volume}
  {463}},\ \bibinfo {pages} {127} (\bibinfo {year} {2008})},\ \Eprint
  {http://arxiv.org/abs/0802.1634} {arXiv:0802.1634 [astro-ph]} \BibitemShut
  {NoStop}%
\bibitem [{\citenamefont {Novello}\ \emph {et~al.}(2009)\citenamefont
  {Novello}, \citenamefont {Araujo},\ and\ \citenamefont
  {Salim}}]{Novello:2008xp}%
  \BibitemOpen
  \bibfield  {author} {\bibinfo {author} {\bibfnamefont {M.}~\bibnamefont
  {Novello}}, \bibinfo {author} {\bibfnamefont {A.~N.}\ \bibnamefont {Araujo}},
  \ and\ \bibinfo {author} {\bibfnamefont {J.~M.}\ \bibnamefont {Salim}},\
  }\href {\doibase 10.1142/S0217751X09046321} {\bibfield  {journal} {\bibinfo
  {journal} {Int. J. Mod. Phys. A}\ }\textbf {\bibinfo {volume} {24}},\
  \bibinfo {pages} {5639} (\bibinfo {year} {2009})},\ \Eprint
  {http://arxiv.org/abs/0802.1875} {arXiv:0802.1875 [astro-ph]} \BibitemShut
  {NoStop}%
\bibitem [{\citenamefont {Xiang}\ \emph {et~al.}(2013)\citenamefont {Xiang},
  \citenamefont {Ling},\ and\ \citenamefont {Shen}}]{Xiang:2013sza}%
  \BibitemOpen
  \bibfield  {author} {\bibinfo {author} {\bibfnamefont {L.}~\bibnamefont
  {Xiang}}, \bibinfo {author} {\bibfnamefont {Y.}~\bibnamefont {Ling}}, \ and\
  \bibinfo {author} {\bibfnamefont {Y.~G.}\ \bibnamefont {Shen}},\ }\href
  {\doibase 10.1142/S0218271813420169} {\bibfield  {journal} {\bibinfo
  {journal} {Int. J. Mod. Phys. D}\ }\textbf {\bibinfo {volume} {22}},\
  \bibinfo {pages} {1342016} (\bibinfo {year} {2013})},\ \Eprint
  {http://arxiv.org/abs/1305.3851} {arXiv:1305.3851 [gr-qc]} \BibitemShut
  {NoStop}%
\bibitem [{\citenamefont {Balart}\ and\ \citenamefont
  {Vagenas}(2014)}]{Balart:2014cga}%
  \BibitemOpen
  \bibfield  {author} {\bibinfo {author} {\bibfnamefont {L.}~\bibnamefont
  {Balart}}\ and\ \bibinfo {author} {\bibfnamefont {E.~C.}\ \bibnamefont
  {Vagenas}},\ }\href {\doibase 10.1103/PhysRevD.90.124045} {\bibfield
  {journal} {\bibinfo  {journal} {Phys. Rev. D}\ }\textbf {\bibinfo {volume}
  {90}},\ \bibinfo {pages} {124045} (\bibinfo {year} {2014})},\ \Eprint
  {http://arxiv.org/abs/1408.0306} {arXiv:1408.0306 [gr-qc]} \BibitemShut
  {NoStop}%
\bibitem [{\citenamefont {Culetu}(2015)}]{Culetu:2014lca}%
  \BibitemOpen
  \bibfield  {author} {\bibinfo {author} {\bibfnamefont {H.}~\bibnamefont
  {Culetu}},\ }\href {\doibase 10.1007/s10773-015-2521-6} {\bibfield  {journal}
  {\bibinfo  {journal} {Int. J. Theor. Phys.}\ }\textbf {\bibinfo {volume}
  {54}},\ \bibinfo {pages} {2855} (\bibinfo {year} {2015})},\ \Eprint
  {http://arxiv.org/abs/1408.3334} {arXiv:1408.3334 [gr-qc]} \BibitemShut
  {NoStop}%
\bibitem [{\citenamefont {Cirilo~Lombardo}(2004)}]{CiriloLombardo:2004qw}%
  \BibitemOpen
  \bibfield  {author} {\bibinfo {author} {\bibfnamefont {D.~J.}\ \bibnamefont
  {Cirilo~Lombardo}},\ }\href {\doibase 10.1088/0264-9381/21/6/009} {\bibfield
  {journal} {\bibinfo  {journal} {Class. Quant. Grav.}\ }\textbf {\bibinfo
  {volume} {21}},\ \bibinfo {pages} {1407} (\bibinfo {year} {2004})},\ \Eprint
  {http://arxiv.org/abs/gr-qc/0612063} {arXiv:gr-qc/0612063} \BibitemShut
  {NoStop}%
\bibitem [{\citenamefont {Newman}\ and\ \citenamefont
  {Janis}(1965)}]{Newman:1965tw}%
  \BibitemOpen
  \bibfield  {author} {\bibinfo {author} {\bibfnamefont {E.~T.}\ \bibnamefont
  {Newman}}\ and\ \bibinfo {author} {\bibfnamefont {A.~I.}\ \bibnamefont
  {Janis}},\ }\href {\doibase 10.1063/1.1704350} {\bibfield  {journal}
  {\bibinfo  {journal} {J. Math. Phys.}\ }\textbf {\bibinfo {volume} {6}},\
  \bibinfo {pages} {915} (\bibinfo {year} {1965})}\BibitemShut {NoStop}%
\bibitem [{\citenamefont
  {Azreg-Ainou}(2014{\natexlab{a}})}]{Azreg-Ainou:2014aqa}%
  \BibitemOpen
  \bibfield  {author} {\bibinfo {author} {\bibfnamefont {M.}~\bibnamefont
  {Azreg-Ainou}},\ }\href {\doibase 10.1140/epjc/s10052-014-2865-8} {\bibfield
  {journal} {\bibinfo  {journal} {Eur. Phys. J. C}\ }\textbf {\bibinfo {volume}
  {74}},\ \bibinfo {pages} {2865} (\bibinfo {year} {2014}{\natexlab{a}})},\
  \Eprint {http://arxiv.org/abs/1401.4292} {arXiv:1401.4292 [gr-qc]}
  \BibitemShut {NoStop}%
\bibitem [{\citenamefont
  {Azreg-Ainou}(2014{\natexlab{b}})}]{Azreg-Ainou:2014nra}%
  \BibitemOpen
  \bibfield  {author} {\bibinfo {author} {\bibfnamefont {M.}~\bibnamefont
  {Azreg-Ainou}},\ }\href {\doibase 10.1016/j.physletb.2014.01.041} {\bibfield
  {journal} {\bibinfo  {journal} {Phys. Lett. B}\ }\textbf {\bibinfo {volume}
  {730}},\ \bibinfo {pages} {95} (\bibinfo {year} {2014}{\natexlab{b}})},\
  \Eprint {http://arxiv.org/abs/1401.0787} {arXiv:1401.0787 [gr-qc]}
  \BibitemShut {NoStop}%
\bibitem [{\citenamefont
  {Azreg-Ainou}(2014{\natexlab{c}})}]{Azreg-Ainou:2014pra}%
  \BibitemOpen
  \bibfield  {author} {\bibinfo {author} {\bibfnamefont {M.}~\bibnamefont
  {Azreg-Ainou}},\ }\href {\doibase 10.1103/PhysRevD.90.064041} {\bibfield
  {journal} {\bibinfo  {journal} {Phys. Rev. D}\ }\textbf {\bibinfo {volume}
  {90}},\ \bibinfo {pages} {064041} (\bibinfo {year} {2014}{\natexlab{c}})},\
  \Eprint {http://arxiv.org/abs/1405.2569} {arXiv:1405.2569 [gr-qc]}
  \BibitemShut {NoStop}%
\bibitem [{\citenamefont {Ghosh}\ and\ \citenamefont
  {Walia}(2021)}]{Ghosh:2021clx}%
  \BibitemOpen
  \bibfield  {author} {\bibinfo {author} {\bibfnamefont {S.~G.}\ \bibnamefont
  {Ghosh}}\ and\ \bibinfo {author} {\bibfnamefont {R.~K.}\ \bibnamefont
  {Walia}},\ }\href {\doibase 10.1016/j.aop.2021.168619} {\bibfield  {journal}
  {\bibinfo  {journal} {Annals Phys.}\ }\textbf {\bibinfo {volume} {434}},\
  \bibinfo {pages} {168619} (\bibinfo {year} {2021})},\ \Eprint
  {http://arxiv.org/abs/2109.13031} {arXiv:2109.13031 [gr-qc]} \BibitemShut
  {NoStop}%
\bibitem [{\citenamefont {Hawking}\ and\ \citenamefont
  {Ellis}(2023)}]{Hawking:1973uf}%
  \BibitemOpen
  \bibfield  {author} {\bibinfo {author} {\bibfnamefont {S.~W.}\ \bibnamefont
  {Hawking}}\ and\ \bibinfo {author} {\bibfnamefont {G.~F.~R.}\ \bibnamefont
  {Ellis}},\ }\href {\doibase 10.1017/9781009253161} {\bibfield  {journal}
  {\bibinfo  {journal} {Cambridge University Press}\ } (\bibinfo {year}
  {2023}),\ 10.1017/9781009253161}\BibitemShut {NoStop}%
\bibitem [{\citenamefont {Kiselev}(2003{\natexlab{a}})}]{Kiselev:2003ah}%
  \BibitemOpen
  \bibfield  {author} {\bibinfo {author} {\bibfnamefont {V.~V.}\ \bibnamefont
  {Kiselev}},\ }\href@noop {} {\  (\bibinfo {year} {2003}{\natexlab{a}})},\
  \Eprint {http://arxiv.org/abs/gr-qc/0303031} {arXiv:gr-qc/0303031}
  \BibitemShut {NoStop}%
\bibitem [{\citenamefont {Kiselev}(2003{\natexlab{b}})}]{Kiselev:2002dx}%
  \BibitemOpen
  \bibfield  {author} {\bibinfo {author} {\bibfnamefont {V.~V.}\ \bibnamefont
  {Kiselev}},\ }\href {\doibase 10.1088/0264-9381/20/6/310} {\bibfield
  {journal} {\bibinfo  {journal} {Class. Quant. Grav.}\ }\textbf {\bibinfo
  {volume} {20}},\ \bibinfo {pages} {1187} (\bibinfo {year}
  {2003}{\natexlab{b}})},\ \Eprint {http://arxiv.org/abs/gr-qc/0210040}
  {arXiv:gr-qc/0210040} \BibitemShut {NoStop}%
\bibitem [{\citenamefont {Guzman}\ \emph {et~al.}(2003)\citenamefont {Guzman},
  \citenamefont {Matos}, \citenamefont {Nunez},\ and\ \citenamefont
  {Ramirez}}]{Guzman:2000zba}%
  \BibitemOpen
  \bibfield  {author} {\bibinfo {author} {\bibfnamefont {F.~S.}\ \bibnamefont
  {Guzman}}, \bibinfo {author} {\bibfnamefont {T.}~\bibnamefont {Matos}},
  \bibinfo {author} {\bibfnamefont {D.}~\bibnamefont {Nunez}}, \ and\ \bibinfo
  {author} {\bibfnamefont {E.}~\bibnamefont {Ramirez}},\ }\href@noop {}
  {\bibfield  {journal} {\bibinfo  {journal} {Rev. Mex. Fis.}\ }\textbf
  {\bibinfo {volume} {49}},\ \bibinfo {pages} {203} (\bibinfo {year} {2003})},\
  \Eprint {http://arxiv.org/abs/astro-ph/0003105} {arXiv:astro-ph/0003105}
  \BibitemShut {NoStop}%
\bibitem [{\citenamefont {Rahaman}\ \emph {et~al.}(2011)\citenamefont
  {Rahaman}, \citenamefont {Nandi}, \citenamefont {Bhadra}, \citenamefont
  {Kalam},\ and\ \citenamefont {Chakraborty}}]{Rahaman:2010xs}%
  \BibitemOpen
  \bibfield  {author} {\bibinfo {author} {\bibfnamefont {F.}~\bibnamefont
  {Rahaman}}, \bibinfo {author} {\bibfnamefont {K.~K.}\ \bibnamefont {Nandi}},
  \bibinfo {author} {\bibfnamefont {A.}~\bibnamefont {Bhadra}}, \bibinfo
  {author} {\bibfnamefont {M.}~\bibnamefont {Kalam}}, \ and\ \bibinfo {author}
  {\bibfnamefont {K.}~\bibnamefont {Chakraborty}},\ }\href {\doibase
  10.1016/j.physletb.2010.09.038} {\bibfield  {journal} {\bibinfo  {journal}
  {Phys. Lett. B}\ }\textbf {\bibinfo {volume} {694}},\ \bibinfo {pages} {10}
  (\bibinfo {year} {2011})},\ \Eprint {http://arxiv.org/abs/1009.3572}
  {arXiv:1009.3572 [gr-qc]} \BibitemShut {NoStop}%
\bibitem [{\citenamefont {Potapov}\ \emph {et~al.}(2016)\citenamefont
  {Potapov}, \citenamefont {Garipova},\ and\ \citenamefont
  {Nandi}}]{Potapov:2016obe}%
  \BibitemOpen
  \bibfield  {author} {\bibinfo {author} {\bibfnamefont {A.~A.}\ \bibnamefont
  {Potapov}}, \bibinfo {author} {\bibfnamefont {G.~M.}\ \bibnamefont
  {Garipova}}, \ and\ \bibinfo {author} {\bibfnamefont {K.~K.}\ \bibnamefont
  {Nandi}},\ }\href {\doibase 10.1016/j.physletb.2015.11.057} {\bibfield
  {journal} {\bibinfo  {journal} {Phys. Lett. B}\ }\textbf {\bibinfo {volume}
  {753}},\ \bibinfo {pages} {140} (\bibinfo {year} {2016})},\ \Eprint
  {http://arxiv.org/abs/1606.07733} {arXiv:1606.07733 [gr-qc]} \BibitemShut
  {NoStop}%
\bibitem [{\citenamefont {Rizwan}\ \emph {et~al.}(2019)\citenamefont {Rizwan},
  \citenamefont {Jamil},\ and\ \citenamefont {Jusufi}}]{Rizwan:2018rgs}%
  \BibitemOpen
  \bibfield  {author} {\bibinfo {author} {\bibfnamefont {M.}~\bibnamefont
  {Rizwan}}, \bibinfo {author} {\bibfnamefont {M.}~\bibnamefont {Jamil}}, \
  and\ \bibinfo {author} {\bibfnamefont {K.}~\bibnamefont {Jusufi}},\ }\href
  {\doibase 10.1103/PhysRevD.99.024050} {\bibfield  {journal} {\bibinfo
  {journal} {Phys. Rev. D}\ }\textbf {\bibinfo {volume} {99}},\ \bibinfo
  {pages} {024050} (\bibinfo {year} {2019})},\ \Eprint
  {http://arxiv.org/abs/1812.01331} {arXiv:1812.01331 [gr-qc]} \BibitemShut
  {NoStop}%
\bibitem [{\citenamefont {Ndongmo}\ \emph {et~al.}(2023)\citenamefont
  {Ndongmo}, \citenamefont {Mahamat}, \citenamefont {Tabi}, \citenamefont
  {Bouetou},\ and\ \citenamefont {Kofane}}]{Ndongmo:2021how}%
  \BibitemOpen
  \bibfield  {author} {\bibinfo {author} {\bibfnamefont {R.~B.}\ \bibnamefont
  {Ndongmo}}, \bibinfo {author} {\bibfnamefont {S.}~\bibnamefont {Mahamat}},
  \bibinfo {author} {\bibfnamefont {C.~B.}\ \bibnamefont {Tabi}}, \bibinfo
  {author} {\bibfnamefont {T.~B.}\ \bibnamefont {Bouetou}}, \ and\ \bibinfo
  {author} {\bibfnamefont {T.~C.}\ \bibnamefont {Kofane}},\ }\href {\doibase
  10.1016/j.dark.2023.101299} {\bibfield  {journal} {\bibinfo  {journal} {Phys.
  Dark Univ.}\ }\textbf {\bibinfo {volume} {42}},\ \bibinfo {pages} {101299}
  (\bibinfo {year} {2023})},\ \Eprint {http://arxiv.org/abs/2111.05045}
  {arXiv:2111.05045 [gr-qc]} \BibitemShut {NoStop}%
\bibitem [{\citenamefont {Xu}\ \emph {et~al.}(2018)\citenamefont {Xu},
  \citenamefont {Hou}, \citenamefont {Gong},\ and\ \citenamefont
  {Wang}}]{Xu:2018wow}%
  \BibitemOpen
  \bibfield  {author} {\bibinfo {author} {\bibfnamefont {Z.}~\bibnamefont
  {Xu}}, \bibinfo {author} {\bibfnamefont {X.}~\bibnamefont {Hou}}, \bibinfo
  {author} {\bibfnamefont {X.}~\bibnamefont {Gong}}, \ and\ \bibinfo {author}
  {\bibfnamefont {J.}~\bibnamefont {Wang}},\ }\href {\doibase
  10.1088/1475-7516/2018/09/038} {\bibfield  {journal} {\bibinfo  {journal}
  {JCAP}\ }\textbf {\bibinfo {volume} {09}},\ \bibinfo {pages} {038} (\bibinfo
  {year} {2018})},\ \Eprint {http://arxiv.org/abs/1803.00767} {arXiv:1803.00767
  [gr-qc]} \BibitemShut {NoStop}%
\bibitem [{\citenamefont {Vachher}\ \emph {et~al.}(2024)\citenamefont
  {Vachher}, \citenamefont {Baboolal},\ and\ \citenamefont
  {Ghosh}}]{Vachher:2024ldc}%
  \BibitemOpen
  \bibfield  {author} {\bibinfo {author} {\bibfnamefont {A.}~\bibnamefont
  {Vachher}}, \bibinfo {author} {\bibfnamefont {D.}~\bibnamefont {Baboolal}}, \
  and\ \bibinfo {author} {\bibfnamefont {S.~G.}\ \bibnamefont {Ghosh}},\ }\href
  {\doibase 10.1016/j.dark.2024.101493} {\bibfield  {journal} {\bibinfo
  {journal} {Phys. Dark Univ.}\ }\textbf {\bibinfo {volume} {44}},\ \bibinfo
  {pages} {101493} (\bibinfo {year} {2024})}\BibitemShut {NoStop}%
\bibitem [{\citenamefont {Tan}\ \emph {et~al.}(2025)\citenamefont {Tan},
  \citenamefont {Liu}, \citenamefont {Liang},\ and\ \citenamefont
  {Long}}]{Tan:2025usr}%
  \BibitemOpen
  \bibfield  {author} {\bibinfo {author} {\bibfnamefont {Q.}~\bibnamefont
  {Tan}}, \bibinfo {author} {\bibfnamefont {D.}~\bibnamefont {Liu}}, \bibinfo
  {author} {\bibfnamefont {J.}~\bibnamefont {Liang}}, \ and\ \bibinfo {author}
  {\bibfnamefont {Z.~W.}\ \bibnamefont {Long}},\ }\href {\doibase
  10.1140/epjc/s10052-025-14407-3} {\bibfield  {journal} {\bibinfo  {journal}
  {Eur. Phys. J. C}\ }\textbf {\bibinfo {volume} {85}},\ \bibinfo {pages} {687}
  (\bibinfo {year} {2025})},\ \Eprint {http://arxiv.org/abs/2504.05641}
  {arXiv:2504.05641 [gr-qc]} \BibitemShut {NoStop}%
\bibitem [{\citenamefont {Akiyama}\ \emph
  {et~al.}(2022{\natexlab{a}})\citenamefont {Akiyama} \emph
  {et~al.}}]{EventHorizonTelescope:2022wkp}%
  \BibitemOpen
  \bibfield  {author} {\bibinfo {author} {\bibfnamefont {K.}~\bibnamefont
  {Akiyama}} \emph {et~al.} (\bibinfo {collaboration} {Event Horizon
  Telescope}),\ }\href {\doibase 10.3847/2041-8213/ac6674} {\bibfield
  {journal} {\bibinfo  {journal} {Astrophys. J. Lett.}\ }\textbf {\bibinfo
  {volume} {930}},\ \bibinfo {pages} {L12} (\bibinfo {year}
  {2022}{\natexlab{a}})},\ \Eprint {http://arxiv.org/abs/2311.08680}
  {arXiv:2311.08680 [astro-ph.HE]} \BibitemShut {NoStop}%
\bibitem [{\citenamefont {Kocherlakota}\ \emph {et~al.}(2021)\citenamefont
  {Kocherlakota} \emph {et~al.}}]{EventHorizonTelescope:2021dqv}%
  \BibitemOpen
  \bibfield  {author} {\bibinfo {author} {\bibfnamefont {P.}~\bibnamefont
  {Kocherlakota}} \emph {et~al.} (\bibinfo {collaboration} {Event Horizon
  Telescope}),\ }\href {\doibase 10.1103/PhysRevD.103.104047} {\bibfield
  {journal} {\bibinfo  {journal} {Phys. Rev. D}\ }\textbf {\bibinfo {volume}
  {103}},\ \bibinfo {pages} {104047} (\bibinfo {year} {2021})},\ \Eprint
  {http://arxiv.org/abs/2105.09343} {arXiv:2105.09343 [gr-qc]} \BibitemShut
  {NoStop}%
\bibitem [{\citenamefont {Shakura}\ and\ \citenamefont
  {Sunyaev}(1973)}]{Shakura:1972te}%
  \BibitemOpen
  \bibfield  {author} {\bibinfo {author} {\bibfnamefont {N.~I.}\ \bibnamefont
  {Shakura}}\ and\ \bibinfo {author} {\bibfnamefont {R.~A.}\ \bibnamefont
  {Sunyaev}},\ }\href@noop {} {\bibfield  {journal} {\bibinfo  {journal}
  {Astron. Astrophys.}\ }\textbf {\bibinfo {volume} {24}},\ \bibinfo {pages}
  {337} (\bibinfo {year} {1973})}\BibitemShut {NoStop}%
\bibitem [{\citenamefont {Thorne}(1974)}]{Thorne:1974ve}%
  \BibitemOpen
  \bibfield  {author} {\bibinfo {author} {\bibfnamefont {K.~S.}\ \bibnamefont
  {Thorne}},\ }\href {\doibase 10.1086/152991} {\bibfield  {journal} {\bibinfo
  {journal} {Astrophys. J.}\ }\textbf {\bibinfo {volume} {191}},\ \bibinfo
  {pages} {507} (\bibinfo {year} {1974})}\BibitemShut {NoStop}%
\bibitem [{\citenamefont {Page}\ and\ \citenamefont
  {Thorne}(1974)}]{Page:1974he}%
  \BibitemOpen
  \bibfield  {author} {\bibinfo {author} {\bibfnamefont {D.~N.}\ \bibnamefont
  {Page}}\ and\ \bibinfo {author} {\bibfnamefont {K.~S.}\ \bibnamefont
  {Thorne}},\ }\href {\doibase 10.1086/152990} {\bibfield  {journal} {\bibinfo
  {journal} {Astrophys. J.}\ }\textbf {\bibinfo {volume} {191}},\ \bibinfo
  {pages} {499} (\bibinfo {year} {1974})}\BibitemShut {NoStop}%
\bibitem [{\citenamefont {Novikov}\ and\ \citenamefont
  {Thorne}(1973)}]{Novikov:1973kta}%
  \BibitemOpen
  \bibfield  {author} {\bibinfo {author} {\bibfnamefont {I.~D.}\ \bibnamefont
  {Novikov}}\ and\ \bibinfo {author} {\bibfnamefont {K.~S.}\ \bibnamefont
  {Thorne}},\ }\href@noop {} {\  (\bibinfo {year} {1973})}\BibitemShut
  {NoStop}%
\bibitem [{\citenamefont {Li}\ \emph {et~al.}(2005)\citenamefont {Li},
  \citenamefont {Zimmerman}, \citenamefont {Narayan},\ and\ \citenamefont
  {McClintock}}]{Li:2004aq}%
  \BibitemOpen
  \bibfield  {author} {\bibinfo {author} {\bibfnamefont {L.~X.}\ \bibnamefont
  {Li}}, \bibinfo {author} {\bibfnamefont {E.~R.}\ \bibnamefont {Zimmerman}},
  \bibinfo {author} {\bibfnamefont {R.}~\bibnamefont {Narayan}}, \ and\
  \bibinfo {author} {\bibfnamefont {J.~E.}\ \bibnamefont {McClintock}},\ }\href
  {\doibase 10.1086/428089} {\bibfield  {journal} {\bibinfo  {journal}
  {Astrophys. J. Suppl.}\ }\textbf {\bibinfo {volume} {157}},\ \bibinfo {pages}
  {335} (\bibinfo {year} {2005})},\ \Eprint
  {http://arxiv.org/abs/astro-ph/0411583} {arXiv:astro-ph/0411583} \BibitemShut
  {NoStop}%
\bibitem [{\citenamefont {Pun}\ \emph {et~al.}(2008)\citenamefont {Pun},
  \citenamefont {Kovacs},\ and\ \citenamefont {Harko}}]{Pun:2008ua}%
  \BibitemOpen
  \bibfield  {author} {\bibinfo {author} {\bibfnamefont {C.~S.~J.}\
  \bibnamefont {Pun}}, \bibinfo {author} {\bibfnamefont {Z.}~\bibnamefont
  {Kovacs}}, \ and\ \bibinfo {author} {\bibfnamefont {T.}~\bibnamefont
  {Harko}},\ }\href {\doibase 10.1103/PhysRevD.78.084015} {\bibfield  {journal}
  {\bibinfo  {journal} {Phys. Rev. D}\ }\textbf {\bibinfo {volume} {78}},\
  \bibinfo {pages} {084015} (\bibinfo {year} {2008})},\ \Eprint
  {http://arxiv.org/abs/0809.1284} {arXiv:0809.1284 [gr-qc]} \BibitemShut
  {NoStop}%
\bibitem [{\citenamefont {Harko}\ \emph {et~al.}(2011)\citenamefont {Harko},
  \citenamefont {Kovacs},\ and\ \citenamefont {Lobo}}]{Harko:2010ua}%
  \BibitemOpen
  \bibfield  {author} {\bibinfo {author} {\bibfnamefont {T.}~\bibnamefont
  {Harko}}, \bibinfo {author} {\bibfnamefont {Z.}~\bibnamefont {Kovacs}}, \
  and\ \bibinfo {author} {\bibfnamefont {F.~S.~N.}\ \bibnamefont {Lobo}},\
  }\href {\doibase 10.1088/0264-9381/28/16/165001} {\bibfield  {journal}
  {\bibinfo  {journal} {Class. Quant. Grav.}\ }\textbf {\bibinfo {volume}
  {28}},\ \bibinfo {pages} {165001} (\bibinfo {year} {2011})},\ \Eprint
  {http://arxiv.org/abs/1009.1958} {arXiv:1009.1958 [gr-qc]} \BibitemShut
  {NoStop}%
\bibitem [{\citenamefont {Chakraborty}(2015)}]{Chakraborty:2014eha}%
  \BibitemOpen
  \bibfield  {author} {\bibinfo {author} {\bibfnamefont {S.}~\bibnamefont
  {Chakraborty}},\ }\href {\doibase 10.1088/0264-9381/32/7/075007} {\bibfield
  {journal} {\bibinfo  {journal} {Class. Quant. Grav.}\ }\textbf {\bibinfo
  {volume} {32}},\ \bibinfo {pages} {075007} (\bibinfo {year} {2015})},\
  \Eprint {http://arxiv.org/abs/1406.0417} {arXiv:1406.0417 [gr-qc]}
  \BibitemShut {NoStop}%
\bibitem [{\citenamefont {Bambi}(2017)}]{Bambi:2015kza}%
  \BibitemOpen
  \bibfield  {author} {\bibinfo {author} {\bibfnamefont {C.}~\bibnamefont
  {Bambi}},\ }\href {\doibase 10.1103/RevModPhys.89.025001} {\bibfield
  {journal} {\bibinfo  {journal} {Rev. Mod. Phys.}\ }\textbf {\bibinfo {volume}
  {89}},\ \bibinfo {pages} {025001} (\bibinfo {year} {2017})},\ \Eprint
  {http://arxiv.org/abs/1509.03884} {arXiv:1509.03884 [gr-qc]} \BibitemShut
  {NoStop}%
\bibitem [{\citenamefont {He}\ \emph {et~al.}(2022)\citenamefont {He},
  \citenamefont {Cai},\ and\ \citenamefont {Yang}}]{He:2022lrc}%
  \BibitemOpen
  \bibfield  {author} {\bibinfo {author} {\bibfnamefont {T.~Y.}\ \bibnamefont
  {He}}, \bibinfo {author} {\bibfnamefont {Z.}~\bibnamefont {Cai}}, \ and\
  \bibinfo {author} {\bibfnamefont {R.~J.}\ \bibnamefont {Yang}},\ }\href
  {\doibase 10.1140/epjc/s10052-022-11037-x} {\bibfield  {journal} {\bibinfo
  {journal} {Eur. Phys. J. C}\ }\textbf {\bibinfo {volume} {82}},\ \bibinfo
  {pages} {1067} (\bibinfo {year} {2022})},\ \Eprint
  {http://arxiv.org/abs/2208.03723} {arXiv:2208.03723 [gr-qc]} \BibitemShut
  {NoStop}%
\bibitem [{\citenamefont {Zheng}\ \emph {et~al.}(2025)\citenamefont {Zheng},
  \citenamefont {Wu}, \citenamefont {Li},\ and\ \citenamefont
  {Jiang}}]{Zheng:2024ftk}%
  \BibitemOpen
  \bibfield  {author} {\bibinfo {author} {\bibfnamefont {H.-B.}\ \bibnamefont
  {Zheng}}, \bibinfo {author} {\bibfnamefont {M.-Q.}\ \bibnamefont {Wu}},
  \bibinfo {author} {\bibfnamefont {G.-P.}\ \bibnamefont {Li}}, \ and\ \bibinfo
  {author} {\bibfnamefont {Q.-Q.}\ \bibnamefont {Jiang}},\ }\href {\doibase
  10.1140/epjc/s10052-025-13791-0} {\bibfield  {journal} {\bibinfo  {journal}
  {Eur. Phys. J. C}\ }\textbf {\bibinfo {volume} {85}},\ \bibinfo {pages} {46}
  (\bibinfo {year} {2025})},\ \Eprint {http://arxiv.org/abs/2411.10315}
  {arXiv:2411.10315 [gr-qc]} \BibitemShut {NoStop}%
\bibitem [{\citenamefont {Wang}\ \emph {et~al.}(2025)\citenamefont {Wang},
  \citenamefont {Guo}, \citenamefont {Li}, \citenamefont {Mai}, \citenamefont
  {Wu}, \citenamefont {Deng},\ and\ \citenamefont {Jiang}}]{Wang:2025buh}%
  \BibitemOpen
  \bibfield  {author} {\bibinfo {author} {\bibfnamefont {P.}~\bibnamefont
  {Wang}}, \bibinfo {author} {\bibfnamefont {S.}~\bibnamefont {Guo}}, \bibinfo
  {author} {\bibfnamefont {L.-F.}\ \bibnamefont {Li}}, \bibinfo {author}
  {\bibfnamefont {Z.-F.}\ \bibnamefont {Mai}}, \bibinfo {author} {\bibfnamefont
  {B.-F.}\ \bibnamefont {Wu}}, \bibinfo {author} {\bibfnamefont {W.-H.}\
  \bibnamefont {Deng}}, \ and\ \bibinfo {author} {\bibfnamefont {Q.-Q.}\
  \bibnamefont {Jiang}},\ }\href {\doibase 10.1140/epjc/s10052-025-14441-1}
  {\bibfield  {journal} {\bibinfo  {journal} {Eur. Phys. J. C}\ }\textbf
  {\bibinfo {volume} {85}},\ \bibinfo {pages} {747} (\bibinfo {year} {2025})},\
  \Eprint {http://arxiv.org/abs/2507.17217} {arXiv:2507.17217 [gr-qc]}
  \BibitemShut {NoStop}%
\bibitem [{\citenamefont {Chen}\ and\ \citenamefont
  {Yang}(2026)}]{Chen:2025wut}%
  \BibitemOpen
  \bibfield  {author} {\bibinfo {author} {\bibfnamefont {J.}~\bibnamefont
  {Chen}}\ and\ \bibinfo {author} {\bibfnamefont {J.}~\bibnamefont {Yang}},\
  }\href {\doibase 10.1140/epjc/s10052-025-15183-w} {\bibfield  {journal}
  {\bibinfo  {journal} {Eur. Phys. J. C}\ }\textbf {\bibinfo {volume} {86}},\
  \bibinfo {pages} {9} (\bibinfo {year} {2026})},\ \Eprint
  {http://arxiv.org/abs/2506.22891} {arXiv:2506.22891 [gr-qc]} \BibitemShut
  {NoStop}%
\bibitem [{\citenamefont {Nieto}\ \emph {et~al.}(2025)\citenamefont {Nieto},
  \citenamefont {Hosseinifar}, \citenamefont {Boshkayev}, \citenamefont
  {Zare},\ and\ \citenamefont {Hassanabadi}}]{Nieto:2025apz}%
  \BibitemOpen
  \bibfield  {author} {\bibinfo {author} {\bibfnamefont {L.~M.}\ \bibnamefont
  {Nieto}}, \bibinfo {author} {\bibfnamefont {F.}~\bibnamefont {Hosseinifar}},
  \bibinfo {author} {\bibfnamefont {K.}~\bibnamefont {Boshkayev}}, \bibinfo
  {author} {\bibfnamefont {S.}~\bibnamefont {Zare}}, \ and\ \bibinfo {author}
  {\bibfnamefont {H.}~\bibnamefont {Hassanabadi}},\ }\href {\doibase
  10.1016/j.dark.2025.102151} {\bibfield  {journal} {\bibinfo  {journal} {Phys.
  Dark Univ.}\ }\textbf {\bibinfo {volume} {50}},\ \bibinfo {pages} {102151}
  (\bibinfo {year} {2025})},\ \Eprint {http://arxiv.org/abs/2507.14305}
  {arXiv:2507.14305 [gr-qc]} \BibitemShut {NoStop}%
\bibitem [{\citenamefont {Darvishi}\ \emph {et~al.}(2025)\citenamefont
  {Darvishi}, \citenamefont {Heydari-Fard},\ and\ \citenamefont
  {Mohseni}}]{Darvishi:2024ndu}%
  \BibitemOpen
  \bibfield  {author} {\bibinfo {author} {\bibfnamefont {M.}~\bibnamefont
  {Darvishi}}, \bibinfo {author} {\bibfnamefont {M.}~\bibnamefont
  {Heydari-Fard}}, \ and\ \bibinfo {author} {\bibfnamefont {M.}~\bibnamefont
  {Mohseni}},\ }\href {\doibase 10.1016/j.cjph.2024.12.018} {\bibfield
  {journal} {\bibinfo  {journal} {Chin. J. Phys.}\ }\textbf {\bibinfo {volume}
  {93}},\ \bibinfo {pages} {632} (\bibinfo {year} {2025})},\ \Eprint
  {http://arxiv.org/abs/2405.13079} {arXiv:2405.13079 [gr-qc]} \BibitemShut
  {NoStop}%
\bibitem [{\citenamefont {Yang}\ \emph {et~al.}(2025)\citenamefont {Yang},
  \citenamefont {Aslam}, \citenamefont {Zeng},\ and\ \citenamefont
  {Saleem}}]{Yang:2024nin}%
  \BibitemOpen
  \bibfield  {author} {\bibinfo {author} {\bibfnamefont {C.-Y.}\ \bibnamefont
  {Yang}}, \bibinfo {author} {\bibfnamefont {M.~I.}\ \bibnamefont {Aslam}},
  \bibinfo {author} {\bibfnamefont {X.-X.}\ \bibnamefont {Zeng}}, \ and\
  \bibinfo {author} {\bibfnamefont {R.}~\bibnamefont {Saleem}},\ }\href
  {\doibase 10.1016/j.jheap.2025.01.017} {\bibfield  {journal} {\bibinfo
  {journal} {JHEAp}\ }\textbf {\bibinfo {volume} {46}},\ \bibinfo {pages} {345}
  (\bibinfo {year} {2025})},\ \Eprint {http://arxiv.org/abs/2411.11807}
  {arXiv:2411.11807 [astro-ph.HE]} \BibitemShut {NoStop}%
\bibitem [{\citenamefont {Zeng}\ \emph {et~al.}(2026)\citenamefont {Zeng},
  \citenamefont {Yang}, \citenamefont {Aslam},\ and\ \citenamefont
  {Saleem}}]{Zeng:2025pch}%
  \BibitemOpen
  \bibfield  {author} {\bibinfo {author} {\bibfnamefont {X.-X.}\ \bibnamefont
  {Zeng}}, \bibinfo {author} {\bibfnamefont {C.-Y.}\ \bibnamefont {Yang}},
  \bibinfo {author} {\bibfnamefont {M.~I.}\ \bibnamefont {Aslam}}, \ and\
  \bibinfo {author} {\bibfnamefont {R.}~\bibnamefont {Saleem}},\ }\href
  {\doibase 10.1016/j.jheap.2025.100540} {\bibfield  {journal} {\bibinfo
  {journal} {JHEAp}\ }\textbf {\bibinfo {volume} {51}},\ \bibinfo {pages}
  {100540} (\bibinfo {year} {2026})},\ \Eprint
  {http://arxiv.org/abs/2509.05803} {arXiv:2509.05803 [gr-qc]} \BibitemShut
  {NoStop}%
\bibitem [{\citenamefont {Cai}\ \emph {et~al.}(2025)\citenamefont {Cai},
  \citenamefont {Ban}, \citenamefont {Wang}, \citenamefont {Feng},\ and\
  \citenamefont {Long}}]{Cai:2025rst}%
  \BibitemOpen
  \bibfield  {author} {\bibinfo {author} {\bibfnamefont {Z.}~\bibnamefont
  {Cai}}, \bibinfo {author} {\bibfnamefont {Z.}~\bibnamefont {Ban}}, \bibinfo
  {author} {\bibfnamefont {L.}~\bibnamefont {Wang}}, \bibinfo {author}
  {\bibfnamefont {H.}~\bibnamefont {Feng}}, \ and\ \bibinfo {author}
  {\bibfnamefont {Z.-W.}\ \bibnamefont {Long}},\ }\href {\doibase
  10.1088/1475-7516/2025/10/101} {\bibfield  {journal} {\bibinfo  {journal}
  {JCAP}\ }\textbf {\bibinfo {volume} {10}},\ \bibinfo {pages} {101} (\bibinfo
  {year} {2025})},\ \Eprint {http://arxiv.org/abs/2503.08424} {arXiv:2503.08424
  [gr-qc]} \BibitemShut {NoStop}%
\bibitem [{\citenamefont {Bozza}(2002)}]{Bozza:2002zj}%
  \BibitemOpen
  \bibfield  {author} {\bibinfo {author} {\bibfnamefont {V.}~\bibnamefont
  {Bozza}},\ }\href {\doibase 10.1103/PhysRevD.66.103001} {\bibfield  {journal}
  {\bibinfo  {journal} {Phys. Rev. D}\ }\textbf {\bibinfo {volume} {66}},\
  \bibinfo {pages} {103001} (\bibinfo {year} {2002})},\ \Eprint
  {http://arxiv.org/abs/gr-qc/0208075} {arXiv:gr-qc/0208075} \BibitemShut
  {NoStop}%
\bibitem [{\citenamefont {Perlick}\ and\ \citenamefont
  {Tsupko}(2022)}]{Perlick:2021aok}%
  \BibitemOpen
  \bibfield  {author} {\bibinfo {author} {\bibfnamefont {V.}~\bibnamefont
  {Perlick}}\ and\ \bibinfo {author} {\bibfnamefont {O.~Y.}\ \bibnamefont
  {Tsupko}},\ }\href {\doibase 10.1016/j.physrep.2021.10.004} {\bibfield
  {journal} {\bibinfo  {journal} {Phys. Rept.}\ }\textbf {\bibinfo {volume}
  {947}},\ \bibinfo {pages} {1} (\bibinfo {year} {2022})},\ \Eprint
  {http://arxiv.org/abs/2105.07101} {arXiv:2105.07101 [gr-qc]} \BibitemShut
  {NoStop}%
\bibitem [{\citenamefont {Akiyama}\ \emph
  {et~al.}(2019{\natexlab{c}})\citenamefont {Akiyama} \emph
  {et~al.}}]{EventHorizonTelescope:2019ggy}%
  \BibitemOpen
  \bibfield  {author} {\bibinfo {author} {\bibfnamefont {K.}~\bibnamefont
  {Akiyama}} \emph {et~al.} (\bibinfo {collaboration} {Event Horizon
  Telescope}),\ }\href {\doibase 10.3847/2041-8213/ab1141} {\bibfield
  {journal} {\bibinfo  {journal} {Astrophys. J. Lett.}\ }\textbf {\bibinfo
  {volume} {875}},\ \bibinfo {pages} {L6} (\bibinfo {year}
  {2019}{\natexlab{c}})},\ \Eprint {http://arxiv.org/abs/1906.11243}
  {arXiv:1906.11243 [astro-ph.GA]} \BibitemShut {NoStop}%
\bibitem [{\citenamefont {Akiyama}\ \emph
  {et~al.}(2022{\natexlab{b}})\citenamefont {Akiyama} \emph
  {et~al.}}]{EventHorizonTelescope:2022urf}%
  \BibitemOpen
  \bibfield  {author} {\bibinfo {author} {\bibfnamefont {K.}~\bibnamefont
  {Akiyama}} \emph {et~al.} (\bibinfo {collaboration} {Event Horizon
  Telescope}),\ }\href {\doibase 10.3847/2041-8213/ac6672} {\bibfield
  {journal} {\bibinfo  {journal} {Astrophys. J. Lett.}\ }\textbf {\bibinfo
  {volume} {930}},\ \bibinfo {pages} {L16} (\bibinfo {year}
  {2022}{\natexlab{b}})},\ \Eprint {http://arxiv.org/abs/2311.09478}
  {arXiv:2311.09478 [astro-ph.HE]} \BibitemShut {NoStop}%
\bibitem [{\citenamefont {Akiyama}\ \emph
  {et~al.}(2022{\natexlab{c}})\citenamefont {Akiyama} \emph
  {et~al.}}]{EventHorizonTelescope:2022xqj}%
  \BibitemOpen
  \bibfield  {author} {\bibinfo {author} {\bibfnamefont {K.}~\bibnamefont
  {Akiyama}} \emph {et~al.} (\bibinfo {collaboration} {Event Horizon
  Telescope}),\ }\href {\doibase 10.3847/2041-8213/ac6756} {\bibfield
  {journal} {\bibinfo  {journal} {Astrophys. J. Lett.}\ }\textbf {\bibinfo
  {volume} {930}},\ \bibinfo {pages} {L17} (\bibinfo {year}
  {2022}{\natexlab{c}})},\ \Eprint {http://arxiv.org/abs/2311.09484}
  {arXiv:2311.09484 [astro-ph.HE]} \BibitemShut {NoStop}%
\bibitem [{\citenamefont {Saurabh}\ and\ \citenamefont
  {Jusufi}(2021)}]{Saurabh:2020zqg}%
  \BibitemOpen
  \bibfield  {author} {\bibinfo {author} {\bibfnamefont {K.}~\bibnamefont
  {Saurabh}}\ and\ \bibinfo {author} {\bibfnamefont {K.}~\bibnamefont
  {Jusufi}},\ }\href {\doibase 10.1140/epjc/s10052-021-09280-9} {\bibfield
  {journal} {\bibinfo  {journal} {Eur. Phys. J. C}\ }\textbf {\bibinfo {volume}
  {81}},\ \bibinfo {pages} {490} (\bibinfo {year} {2021})},\ \Eprint
  {http://arxiv.org/abs/2009.10599} {arXiv:2009.10599 [gr-qc]} \BibitemShut
  {NoStop}%
\bibitem [{\citenamefont {Vagnozzi}\ \emph {et~al.}(2023)\citenamefont
  {Vagnozzi}, \citenamefont {Roy}, \citenamefont {Tsai}, \citenamefont
  {Visinelli}, \citenamefont {Afrin}, \citenamefont {Allahyari}, \citenamefont
  {Bambhaniya}, \citenamefont {Dey}, \citenamefont {Ghosh}, \citenamefont
  {Joshi} \emph {et~al.}}]{Vagnozzi:2022moj}%
  \BibitemOpen
  \bibfield  {author} {\bibinfo {author} {\bibfnamefont {S.}~\bibnamefont
  {Vagnozzi}}, \bibinfo {author} {\bibfnamefont {R.}~\bibnamefont {Roy}},
  \bibinfo {author} {\bibfnamefont {Y.~D.}\ \bibnamefont {Tsai}}, \bibinfo
  {author} {\bibfnamefont {L.}~\bibnamefont {Visinelli}}, \bibinfo {author}
  {\bibfnamefont {M.}~\bibnamefont {Afrin}}, \bibinfo {author} {\bibfnamefont
  {A.}~\bibnamefont {Allahyari}}, \bibinfo {author} {\bibfnamefont
  {P.}~\bibnamefont {Bambhaniya}}, \bibinfo {author} {\bibfnamefont
  {D.}~\bibnamefont {Dey}}, \bibinfo {author} {\bibfnamefont {S.~G.}\
  \bibnamefont {Ghosh}}, \bibinfo {author} {\bibfnamefont {P.~S.}\ \bibnamefont
  {Joshi}},  \emph {et~al.},\ }\href {\doibase 10.1088/1361-6382/acd97b}
  {\bibfield  {journal} {\bibinfo  {journal} {Class. Quant. Grav.}\ }\textbf
  {\bibinfo {volume} {40}},\ \bibinfo {pages} {165007} (\bibinfo {year}
  {2023})},\ \Eprint {http://arxiv.org/abs/2205.07787} {arXiv:2205.07787
  [gr-qc]} \BibitemShut {NoStop}%
\bibitem [{\citenamefont {Collodel}\ \emph {et~al.}(2021)\citenamefont
  {Collodel}, \citenamefont {Doneva},\ and\ \citenamefont
  {Yazadjiev}}]{Collodel:2021gxu}%
  \BibitemOpen
  \bibfield  {author} {\bibinfo {author} {\bibfnamefont {L.~G.}\ \bibnamefont
  {Collodel}}, \bibinfo {author} {\bibfnamefont {D.~D.}\ \bibnamefont
  {Doneva}}, \ and\ \bibinfo {author} {\bibfnamefont {S.~S.}\ \bibnamefont
  {Yazadjiev}},\ }\href {\doibase 10.3847/1538-4357/abe305} {\bibfield
  {journal} {\bibinfo  {journal} {Astrophys. J.}\ }\textbf {\bibinfo {volume}
  {910}},\ \bibinfo {pages} {52} (\bibinfo {year} {2021})},\ \Eprint
  {http://arxiv.org/abs/2101.05073} {arXiv:2101.05073 [astro-ph.HE]}
  \BibitemShut {NoStop}%
\bibitem [{\citenamefont {Luminet}(1979)}]{Luminet:1979nyg}%
  \BibitemOpen
  \bibfield  {author} {\bibinfo {author} {\bibfnamefont {J.~P.}\ \bibnamefont
  {Luminet}},\ }\href@noop {} {\bibfield  {journal} {\bibinfo  {journal}
  {Astron. Astrophys.}\ }\textbf {\bibinfo {volume} {75}},\ \bibinfo {pages}
  {228} (\bibinfo {year} {1979})}\BibitemShut {NoStop}%
\bibitem [{\citenamefont {Bhattacharyya}\ \emph {et~al.}(2001)\citenamefont
  {Bhattacharyya}, \citenamefont {Misra},\ and\ \citenamefont
  {Thampan}}]{Bhattacharyya:2000kt}%
  \BibitemOpen
  \bibfield  {author} {\bibinfo {author} {\bibfnamefont {S.}~\bibnamefont
  {Bhattacharyya}}, \bibinfo {author} {\bibfnamefont {R.}~\bibnamefont
  {Misra}}, \ and\ \bibinfo {author} {\bibfnamefont {A.~V.}\ \bibnamefont
  {Thampan}},\ }\href {\doibase 10.1086/319807} {\bibfield  {journal} {\bibinfo
   {journal} {Astrophys. J.}\ }\textbf {\bibinfo {volume} {550}},\ \bibinfo
  {pages} {841} (\bibinfo {year} {2001})},\ \Eprint
  {http://arxiv.org/abs/astro-ph/0011519} {arXiv:astro-ph/0011519} \BibitemShut
  {NoStop}%
\bibitem [{\citenamefont {Salahshoor}\ and\ \citenamefont
  {Nozari}(2018)}]{Salahshoor:2018plr}%
  \BibitemOpen
  \bibfield  {author} {\bibinfo {author} {\bibfnamefont {K.}~\bibnamefont
  {Salahshoor}}\ and\ \bibinfo {author} {\bibfnamefont {K.}~\bibnamefont
  {Nozari}},\ }\href {\doibase 10.1140/epjc/s10052-018-5946-2} {\bibfield
  {journal} {\bibinfo  {journal} {Eur. Phys. J. C}\ }\textbf {\bibinfo {volume}
  {78}},\ \bibinfo {pages} {486} (\bibinfo {year} {2018})},\ \Eprint
  {http://arxiv.org/abs/1806.08949} {arXiv:1806.08949 [gr-qc]} \BibitemShut
  {NoStop}%
\bibitem [{\citenamefont {Narzilloev}\ and\ \citenamefont
  {Ahmedov}(2022)}]{Narzilloev:2022avv}%
  \BibitemOpen
  \bibfield  {author} {\bibinfo {author} {\bibfnamefont {B.}~\bibnamefont
  {Narzilloev}}\ and\ \bibinfo {author} {\bibfnamefont {B.}~\bibnamefont
  {Ahmedov}},\ }\href {\doibase 10.3390/sym14091765} {\bibfield  {journal}
  {\bibinfo  {journal} {Symmetry}\ }\textbf {\bibinfo {volume} {14}},\ \bibinfo
  {pages} {1765} (\bibinfo {year} {2022})}\BibitemShut {NoStop}%
\bibitem [{\citenamefont {Gralla}\ \emph {et~al.}(2019)\citenamefont {Gralla},
  \citenamefont {Holz},\ and\ \citenamefont {Wald}}]{Gralla:2019xty}%
  \BibitemOpen
  \bibfield  {author} {\bibinfo {author} {\bibfnamefont {S.~E.}\ \bibnamefont
  {Gralla}}, \bibinfo {author} {\bibfnamefont {D.~E.}\ \bibnamefont {Holz}}, \
  and\ \bibinfo {author} {\bibfnamefont {R.~M.}\ \bibnamefont {Wald}},\ }\href
  {\doibase 10.1103/PhysRevD.100.024018} {\bibfield  {journal} {\bibinfo
  {journal} {Phys. Rev. D}\ }\textbf {\bibinfo {volume} {100}},\ \bibinfo
  {pages} {024018} (\bibinfo {year} {2019})},\ \Eprint
  {http://arxiv.org/abs/1906.00873} {arXiv:1906.00873 [astro-ph.HE]}
  \BibitemShut {NoStop}%
\bibitem [{\citenamefont {Peng}\ \emph {et~al.}(2021)\citenamefont {Peng},
  \citenamefont {Guo},\ and\ \citenamefont {Feng}}]{Peng:2020wun}%
  \BibitemOpen
  \bibfield  {author} {\bibinfo {author} {\bibfnamefont {J.}~\bibnamefont
  {Peng}}, \bibinfo {author} {\bibfnamefont {M.}~\bibnamefont {Guo}}, \ and\
  \bibinfo {author} {\bibfnamefont {X.-H.}\ \bibnamefont {Feng}},\ }\href
  {\doibase 10.1088/1674-1137/ac06bb} {\bibfield  {journal} {\bibinfo
  {journal} {Chin. Phys. C}\ }\textbf {\bibinfo {volume} {45}},\ \bibinfo
  {pages} {085103} (\bibinfo {year} {2021})},\ \Eprint
  {http://arxiv.org/abs/2008.00657} {arXiv:2008.00657 [gr-qc]} \BibitemShut
  {NoStop}%
\bibitem [{\citenamefont {Liu}\ \emph {et~al.}(2022)\citenamefont {Liu},
  \citenamefont {Tang},\ and\ \citenamefont {Jing}}]{Liu:2021lvk}%
  \BibitemOpen
  \bibfield  {author} {\bibinfo {author} {\bibfnamefont {C.}~\bibnamefont
  {Liu}}, \bibinfo {author} {\bibfnamefont {L.}~\bibnamefont {Tang}}, \ and\
  \bibinfo {author} {\bibfnamefont {J.}~\bibnamefont {Jing}},\ }\href {\doibase
  10.1142/S0218271822500419} {\bibfield  {journal} {\bibinfo  {journal} {Int.
  J. Mod. Phys. D}\ }\textbf {\bibinfo {volume} {31}},\ \bibinfo {pages}
  {2250041} (\bibinfo {year} {2022})},\ \Eprint
  {http://arxiv.org/abs/2109.01867} {arXiv:2109.01867 [gr-qc]} \BibitemShut
  {NoStop}%
\bibitem [{\citenamefont {Xamidov}\ \emph {et~al.}(2025)\citenamefont
  {Xamidov}, \citenamefont {Sharipov},\ and\ \citenamefont
  {Shaymatov}}]{Xamidov:2025gcs}%
  \BibitemOpen
  \bibfield  {author} {\bibinfo {author} {\bibfnamefont {T.}~\bibnamefont
  {Xamidov}}, \bibinfo {author} {\bibfnamefont {J.}~\bibnamefont {Sharipov}}, \
  and\ \bibinfo {author} {\bibfnamefont {S.}~\bibnamefont {Shaymatov}},\
  }\href@noop {} {\  (\bibinfo {year} {2025})},\ \Eprint
  {http://arxiv.org/abs/2509.10953} {arXiv:2509.10953 [gr-qc]} \BibitemShut
  {NoStop}%
\bibitem [{\citenamefont {You}\ \emph {et~al.}(2024{\natexlab{a}})\citenamefont
  {You}, \citenamefont {Feng}, \citenamefont {Wang}, \citenamefont {Hu},\ and\
  \citenamefont {Deng}}]{You:2024jeu}%
  \BibitemOpen
  \bibfield  {author} {\bibinfo {author} {\bibfnamefont {L.}~\bibnamefont
  {You}}, \bibinfo {author} {\bibfnamefont {Y.-H.}\ \bibnamefont {Feng}},
  \bibinfo {author} {\bibfnamefont {R.-B.}\ \bibnamefont {Wang}}, \bibinfo
  {author} {\bibfnamefont {X.-R.}\ \bibnamefont {Hu}}, \ and\ \bibinfo {author}
  {\bibfnamefont {J.-B.}\ \bibnamefont {Deng}},\ }\href {\doibase
  10.3390/universe10100393} {\bibfield  {journal} {\bibinfo  {journal}
  {Universe}\ }\textbf {\bibinfo {volume} {10}},\ \bibinfo {pages} {393}
  (\bibinfo {year} {2024}{\natexlab{a}})},\ \Eprint
  {http://arxiv.org/abs/2404.01418} {arXiv:2404.01418 [gr-qc]} \BibitemShut
  {NoStop}%
\bibitem [{\citenamefont {Sharipov}\ \emph {et~al.}(2026)\citenamefont
  {Sharipov}, \citenamefont {Xamidov}, \citenamefont {Wu}, \citenamefont
  {Shaymatov},\ and\ \citenamefont {Zhu}}]{Sharipov:2025yfw}%
  \BibitemOpen
  \bibfield  {author} {\bibinfo {author} {\bibfnamefont {J.}~\bibnamefont
  {Sharipov}}, \bibinfo {author} {\bibfnamefont {T.}~\bibnamefont {Xamidov}},
  \bibinfo {author} {\bibfnamefont {Q.}~\bibnamefont {Wu}}, \bibinfo {author}
  {\bibfnamefont {S.}~\bibnamefont {Shaymatov}}, \ and\ \bibinfo {author}
  {\bibfnamefont {T.}~\bibnamefont {Zhu}},\ }\href {\doibase
  10.1088/1674-1137/ae6b2f} {\bibfield  {journal} {\bibinfo  {journal} {Chin.
  Phys. C}\ }\textbf {\bibinfo {volume} {50}},\ \bibinfo {pages} {085105}
  (\bibinfo {year} {2026})},\ \Eprint {http://arxiv.org/abs/2511.10043}
  {arXiv:2511.10043 [gr-qc]} \BibitemShut {NoStop}%
\bibitem [{\citenamefont {You}\ \emph {et~al.}(2024{\natexlab{b}})\citenamefont
  {You}, \citenamefont {Wang}, \citenamefont {Ma}, \citenamefont {Deng},\ and\
  \citenamefont {Hu}}]{You:2024uql}%
  \BibitemOpen
  \bibfield  {author} {\bibinfo {author} {\bibfnamefont {L.}~\bibnamefont
  {You}}, \bibinfo {author} {\bibfnamefont {R.-b.}\ \bibnamefont {Wang}},
  \bibinfo {author} {\bibfnamefont {S.-J.}\ \bibnamefont {Ma}}, \bibinfo
  {author} {\bibfnamefont {J.-B.}\ \bibnamefont {Deng}}, \ and\ \bibinfo
  {author} {\bibfnamefont {X.-R.}\ \bibnamefont {Hu}},\ }\href@noop {} {\
  (\bibinfo {year} {2024}{\natexlab{b}})},\ \Eprint
  {http://arxiv.org/abs/2403.12840} {arXiv:2403.12840 [gr-qc]} \BibitemShut
  {NoStop}%
\bibitem [{\citenamefont {Gyulchev}\ \emph {et~al.}(2020)\citenamefont
  {Gyulchev}, \citenamefont {Kunz}, \citenamefont {Nedkova}, \citenamefont
  {Vetsov},\ and\ \citenamefont {Yazadjiev}}]{Gyulchev:2020cvo}%
  \BibitemOpen
  \bibfield  {author} {\bibinfo {author} {\bibfnamefont {G.}~\bibnamefont
  {Gyulchev}}, \bibinfo {author} {\bibfnamefont {J.}~\bibnamefont {Kunz}},
  \bibinfo {author} {\bibfnamefont {P.}~\bibnamefont {Nedkova}}, \bibinfo
  {author} {\bibfnamefont {T.}~\bibnamefont {Vetsov}}, \ and\ \bibinfo {author}
  {\bibfnamefont {S.}~\bibnamefont {Yazadjiev}},\ }\href {\doibase
  10.1140/epjc/s10052-020-08575-7} {\bibfield  {journal} {\bibinfo  {journal}
  {Eur. Phys. J. C}\ }\textbf {\bibinfo {volume} {80}},\ \bibinfo {pages}
  {1017} (\bibinfo {year} {2020})},\ \Eprint {http://arxiv.org/abs/2003.06943}
  {arXiv:2003.06943 [gr-qc]} \BibitemShut {NoStop}%
\bibitem [{\citenamefont {Guo}\ \emph {et~al.}(2024)\citenamefont {Guo},
  \citenamefont {Huang}, \citenamefont {Liang}, \citenamefont {Liang},
  \citenamefont {Jiang},\ and\ \citenamefont {Lin}}]{Guo:2024mij}%
  \BibitemOpen
  \bibfield  {author} {\bibinfo {author} {\bibfnamefont {S.}~\bibnamefont
  {Guo}}, \bibinfo {author} {\bibfnamefont {Y.-X.}\ \bibnamefont {Huang}},
  \bibinfo {author} {\bibfnamefont {E.-W.}\ \bibnamefont {Liang}}, \bibinfo
  {author} {\bibfnamefont {Y.}~\bibnamefont {Liang}}, \bibinfo {author}
  {\bibfnamefont {Q.-Q.}\ \bibnamefont {Jiang}}, \ and\ \bibinfo {author}
  {\bibfnamefont {K.}~\bibnamefont {Lin}},\ }\href {\doibase
  10.3847/1538-4357/ad7d85} {\bibfield  {journal} {\bibinfo  {journal}
  {Astrophys. J.}\ }\textbf {\bibinfo {volume} {975}},\ \bibinfo {pages} {237}
  (\bibinfo {year} {2024})},\ \Eprint {http://arxiv.org/abs/2411.07914}
  {arXiv:2411.07914 [astro-ph.HE]} \BibitemShut {NoStop}%
\bibitem [{\citenamefont {Ellis}(1971)}]{Ellis:1971pg}%
  \BibitemOpen
  \bibfield  {author} {\bibinfo {author} {\bibfnamefont {G.~F.~R.}\
  \bibnamefont {Ellis}},\ }\href {\doibase 10.1007/s10714-009-0760-7}
  {\bibfield  {journal} {\bibinfo  {journal} {Proc. Int. Sch. Phys. Fermi}\
  }\textbf {\bibinfo {volume} {47}},\ \bibinfo {pages} {104} (\bibinfo {year}
  {1971})}\BibitemShut {NoStop}%
\end{thebibliography}%
\bibliographystyle{apsrev4-1}

\end{document}